\documentclass[10pt,pra,aps,twocolumn]{revtex4-2}
\usepackage{color,ifthen,amsthm,amsmath,amsxtra,amsfonts,dsfont,graphicx,bm,tikz,scalerel,wasysym,bbm,graphicx,amsthm,braket,physics,empheq}
\usepackage[inline]{enumitem}
\usepackage{CircuitTikz}
\usepackage{MnSymbol}%
\usepackage{wasysym}%
\usepackage[colorlinks=true, linkcolor=purple, citecolor=violet, urlcolor=violet, bookmarks]{hyperref}

\newcommand{\fl}{-}
\newcommand{\pprime}{\phantom{\prime}}

\newcommand{\Ue}{U^{\mathrm{e}}}
\newcommand{\Uo}{U^{\mathrm{o}}}
\newcommand{\tU}{\tilde{U}}
\newcommand{\tUe}{\tilde{U}^{\mathrm{e}}}
\newcommand{\tUo}{\tilde{U}^{\mathrm{o}}}

\newcommand{\n}{{\bm n}}
\newcommand{\tu}{\tilde{u}}

\newcommand{\change}[1]{#1}

\newcommand{\er}[1]{Eq.~\eqref{#1}}

\newcommand{\era}[2]{Eqs.~(\ref{#1}) and (\ref{#2})}

\newcommand{\gatens}[4]{
  \begin{tikzpicture}
    [baseline={([yshift=-0.6ex]current bounding box.center)},scale=0.5]
    \node at (-0.875,0.875) {#1};
    \node at (0.875,0.875) {#2};
    \node at (-0.875,-0.875) {#3};
    \node at (0.875,-0.875) {#4};
    \prop{0}{0}{colU}
  \end{tikzpicture}}

\newcommand{\gatefsd}{
\begin{tikzpicture}[baseline={([yshift=-0.6ex]current bounding box.center)},scale=0.5]
    \prop{0}{0}{colU}
    \MEld{0.5}{-0.5}
    \MErd{-0.5}{-0.5}
  \end{tikzpicture}}
\newcommand{\gatefsu}{
\begin{tikzpicture}[baseline={([yshift=-0.6ex]current bounding box.center)},scale=0.5]
    \prop{0}{0}{colU}
    \MErd{0.5}{0.5}
    \MEld{-0.5}{0.5}
  \end{tikzpicture}}
\newcommand{\gatefsr}{
\begin{tikzpicture}[baseline={([yshift=-0.6ex]current bounding box.center)},scale=0.5]
    \prop{0}{0}{colU}
    \MErd{0.5}{0.5}
    \MEld{0.5}{-0.5}
  \end{tikzpicture}}
\newcommand{\gatefsl}{
\begin{tikzpicture}[baseline={([yshift=-0.6ex]current bounding box.center)},scale=0.5]
    \prop{0}{0}{colU}
    \MEld{-0.5}{0.5}
    \MErd{-0.5}{-0.5}
  \end{tikzpicture}}
\newcommand{\gateyzz}{
  \begin{tikzpicture}[baseline={([yshift=-0.6ex]current bounding box.center)},scale=0.5]
    \nctgridLine{0}{0}{0.75}{0.75}
    \nctgridLine{0}{0}{0.75}{-0.75}
    \prop{0}{0}{colUinactive}
    \obsZero{0.5}{0.5}
    \obsZero{0.5}{-0.5}
  \end{tikzpicture}}      
\newcommand{\gatey}{
  \begin{tikzpicture}[baseline={([yshift=-0.6ex]current bounding box.center)},scale=0.5]
    \prop{0}{0}{colUinactive}
  \end{tikzpicture}}      
\newcommand{\gateyfs}[1]{
  \begin{tikzpicture}[baseline={([yshift=-0.6ex]current bounding box.center)},scale=0.5]
    \prop{0}{0}{colUinactive}
    \ifthenelse{#1>0}{
      \MEld{0.5}{-#1*0.5}
    }{
      \MErd{0.5}{-#1*0.5}
    }
  \end{tikzpicture}}      
\newcommand{\gatezz}{
  \begin{tikzpicture}[baseline={([yshift=-0.6ex]current bounding box.center)},scale=0.5]
    \nctgridLine{0}{0}{0.75}{0.75}
    \nctgridLine{0}{0}{0.75}{-0.75}
    \prop{0}{0}{colU}
    \obsZero{0.5}{0.5}
    \obsZero{0.5}{-0.5}
  \end{tikzpicture}}

\newcommand{\fsd}{\!
  \begin{tikzpicture}[baseline={([yshift=-0.6ex]current bounding box.center)},scale=0.5]
    \nctgridLine{0}{0}{0}{-0.75}
    \MEh{0}{-0.75}
  \end{tikzpicture}\!}
\newcommand{\fsu}{\!
  \begin{tikzpicture}[baseline={([yshift=-0.6ex]current bounding box.center)},scale=0.5]
    \nctgridLine{0}{0}{0}{0.75}
    \MEh{0}{0.75}
  \end{tikzpicture}\!}
\newcommand{\fsl}{
  \begin{tikzpicture}[baseline={([yshift=-0.6ex]current bounding box.center)},scale=0.5]
    \nctgridLine{0}{0}{-0.75}{0}
    \MEv{-0.75}{0}
  \end{tikzpicture}}
\newcommand{\fsr}{
  \begin{tikzpicture}[baseline={([yshift=-0.6ex]current bounding box.center)},scale=0.5]
    \nctgridLine{0}{0}{0.75}{0}
    \MEv{0.75}{0}
  \end{tikzpicture}}

\newcommand{\ds}{
  \begin{tikzpicture}[baseline={([yshift=-0.6ex]current bounding box.center)},scale=0.5]
    \nctgridLine{-0.5}{0}{0}{0}
    \nctgridLine{-0.5}{-0.5}{0}{-0.5}
    \bendL{-0.5}{-0.5}{0}
  \end{tikzpicture}}

\newcommand{\dsr}{
  \begin{tikzpicture}
    [baseline={([yshift=-0.6ex]current bounding box.center)},scale=0.5]
    \nctgridLine{-0.25}{0}{0.25}{0}
    \nctgridLine{-0.25}{-0.5}{0.25}{-0.5}
    \bendR{0.25}{-0.5}{0}
  \end{tikzpicture}}
\newcommand{\gatedsfsd}{
  \begin{tikzpicture}[baseline={([yshift=-0.6ex]current bounding box.center)},scale=0.5]
    \prop{0}{0}{colU}
    \nctgridLine{-0.5}{1}{0.5}{1}
    \MErd{-0.5}{-0.5}
    \bendLu{-0.5}{0.75-0.25}{0.75+0.25}
  \end{tikzpicture}}
\newcommand{\gatedsfsu}{
  \begin{tikzpicture}[baseline={([yshift=-0.6ex]current bounding box.center)},scale=0.5]
    \prop{0}{0}{colU}
    \nctgridLine{-0.5}{-1}{0.5}{-1}
    \MEld{-0.5}{0.5}
    \bendLd{-0.5}{-0.75-0.25}{-0.75+0.25}
  \end{tikzpicture}}

\newcommand{\fsdotfs}{\!
\begin{tikzpicture}[baseline={([yshift=-0.6ex]current bounding box.center)},scale=0.5]
    \nctgridLine{0}{-0.3}{0}{0.3}
    \MEh{0}{-0.3}
    \MEh{0}{0.3}
  \end{tikzpicture}\!}

\newcommand{\projz}{
  \begin{tikzpicture}[baseline={([yshift=-0.6ex]current bounding box.center)},scale=0.5]
    \nctgridLine{0}{-0.35}{0}{0.35}
    \obsZero{0}{0}
  \end{tikzpicture}}

\newcommand{\projo}{
  \begin{tikzpicture}[baseline={([yshift=-0.6ex]current bounding box.center)},scale=0.5]
    \nctgridLine{0}{-0.35}{0}{0.35}
    \obsOne{0}{0}
  \end{tikzpicture}}

\newcommand{\projzfs}[1]{\!
  \begin{tikzpicture}[baseline={([yshift=-0.6ex]current bounding box.center)},scale=0.5]
    \nctgridLine{0}{-0.5}{0}{0.5}
    \obsZero{0}{0}
    \MEh{0}{-0.5*#1}
  \end{tikzpicture}\!}

\begin{document}

\title{
  Exact 
  \change{pre-transition effects} 
  in kinetically constrained circuits: dynamical fluctuations in the Floquet-East model
  }

\author{Katja Klobas}

\affiliation{School of Physics and Astronomy, University of Nottingham, Nottingham, NG7 2RD, UK}
\affiliation{Centre for the Mathematics and Theoretical Physics of Quantum Non-Equilibrium Systems, University of Nottingham, Nottingham, NG7 2RD, UK}

\author{Cecilia De Fazio}

\affiliation{School of Physics and Astronomy, University of Nottingham, Nottingham, NG7 2RD, UK}
\affiliation{Centre for the Mathematics and Theoretical Physics of Quantum Non-Equilibrium Systems, University of Nottingham, Nottingham, NG7 2RD, UK}

\author{Juan P. Garrahan}

\affiliation{School of Physics and Astronomy, University of Nottingham, Nottingham, NG7 2RD, UK}
\affiliation{Centre for the Mathematics and Theoretical Physics of Quantum Non-Equilibrium Systems, University of Nottingham, Nottingham, NG7 2RD, UK}

\date{\today}

\begin{abstract}
We study the dynamics of a classical circuit corresponding to a discrete-time version of the kinetically constrained East model. We show that this classical ``Floquet-East'' model displays pre-transition behaviour which is a dynamical equivalent of the hydrophobic effect in water. 
For the deterministic version of the modelwe prove exactly: (i) a change in scaling with size in the probability of inactive space-time regions (akin to the ``energy-entropy'' crossover of the solvation free energy in water), (ii) a first-order phase transition in the dynamical large deviations, (iii) the existence of the optimal geometry for local phase separation to accommodate space-time solutes, and (iv) a dynamical analog of ``hydrophobic collapse''. 
\end{abstract}

\maketitle

\noindent
{\bf \em Introduction.---}
In thermodynamics, proximity to a phase transition gives rise to pre-transition effects in the presence of surfaces or solutes. For systems near criticality this leads to the Casimir effect 
\cite{casimir1948attraction, fisher1978wall, hertlein2008direct, gambassi2009the-casimir}. 
The equivalent for first-order transitions is the {\em hydrophobic effect} as occurs in water
\cite{lipowsky1982critical, lipowsky1984surface-induced, lum1999hydrophobicity, chandler2005interfaces}: near liquid-vapour coexistence, the free-energy to accommodate a solute \cite{lum1999hydrophobicity,chandler2005interfaces} displays an entropic-energetic crossover, from scaling with volume for small solutes to scaling with area for larger ones, since it becomes favourable to create a vapour domain around the solute paying only an interface cost (see also \cite{lipowsky1982critical,lipowsky1984surface-induced}). 
Hydrophobicity generalises to other systems with first-order transitions, known as the ``orderphobic effect'' \cite{katira2016pre-transition}.

Hydrophobic-like physics appears, at least numerically~\cite{katira2018solvation}, to also manifest 
in the dynamical fluctuations of kinetically constrained models (KCMs)~\cite{fredrickson1984kinetic,jackle1991a-hierarchically,ritort2003glassy}. These are models with explicit constraints in the dynamics that help explain~\cite{chandler2010dynamics,garrahan2018aspects,speck2019dynamic} many features relating to the glass transition~\cite{berthier2011theoretical,biroli2013perspective}. Specifically, in terms of KCMs~\cite{garrahan2002geometrical} dynamic heterogeneity can be understood as mesoscopic space-time fluctuations related to a nearby active/inactive first-order transition in the space of trajectories~\cite{merolle2005space-time,garrahan2007dynamical, garrahan2009first-order} found via dynamical large deviation methods~\cite{lecomte2007thermodynamic, touchette2009the-large,jack2020ergodicity} (cf.\ full-counting statistics~\cite{esposito2009nonequilibrium}). 
Beyond the glass problem, constrained dynamics is of interest in several other areas. These include: 
cellular automata (CA)~\cite{wolfram1983statistical,bobenko1993two,buca2021rule}, whose dynamics can also be understood in terms of local constraints;
quantum many-body systems 
such as driven Rydberg atoms which can behave like KCMs~\cite{lesanovsky2011many-body,lesanovsky2013kinetic,naik2023quantum,zhang2023quantum};
slow quantum thermalisation \cite{horssen2015dynamics, smith2017disorder-free, lan2018quantum,pancotti2020quantum, roy2020strong,valencia-tortora2022kinetically}; quantum scars~\cite{turner2018weak,moudgalya2022quantum}; and fractonic systems~\cite{nandkishore2019fractons,pretko2020fracton}. 

In this paper we prove dynamical hydrophobicity (i.e., pre-transition effects) analytically beyond just numerics~\cite{katira2018solvation}
for deterministic KCMs. We do so by studying a classical system with discrete-space/discrete-time ``circuit'' dynamics with the same local constraint as the (continuous-time) stochastic East model~\cite{jackle1991a-hierarchically,ritort2003glassy}. For the {\em deterministic} version of this ``Floquet-East'' model---where the circuit gates are unitary (i.e., permutations)---
we demonstrate by means of exact calculations all the relevant features of hydrophobicity in the dynamics: a crossover in the scaling of the probability of inactive space-time regions, a first-order phase transition in the dynamical large deviations, an optimal geometry to accommodate space-time solutes, and the dynamical analog of ``hydrophobic collapse''. Furthermore, the relevance of the deterministic results is that they bound properties of the dynamics in the presence of stochastic gates with the same constraint, which allows to prove dynamical hydrophobicity exactly also in the stochastic Floquet East model~\cite{defazio2024exact}. Below, we present our main results while the Supplemental Material (SM)~\cite{SM} contains additional proofs and details of calculations.

\begin{figure}[t]
    \centering
    \includegraphics[width=\columnwidth]{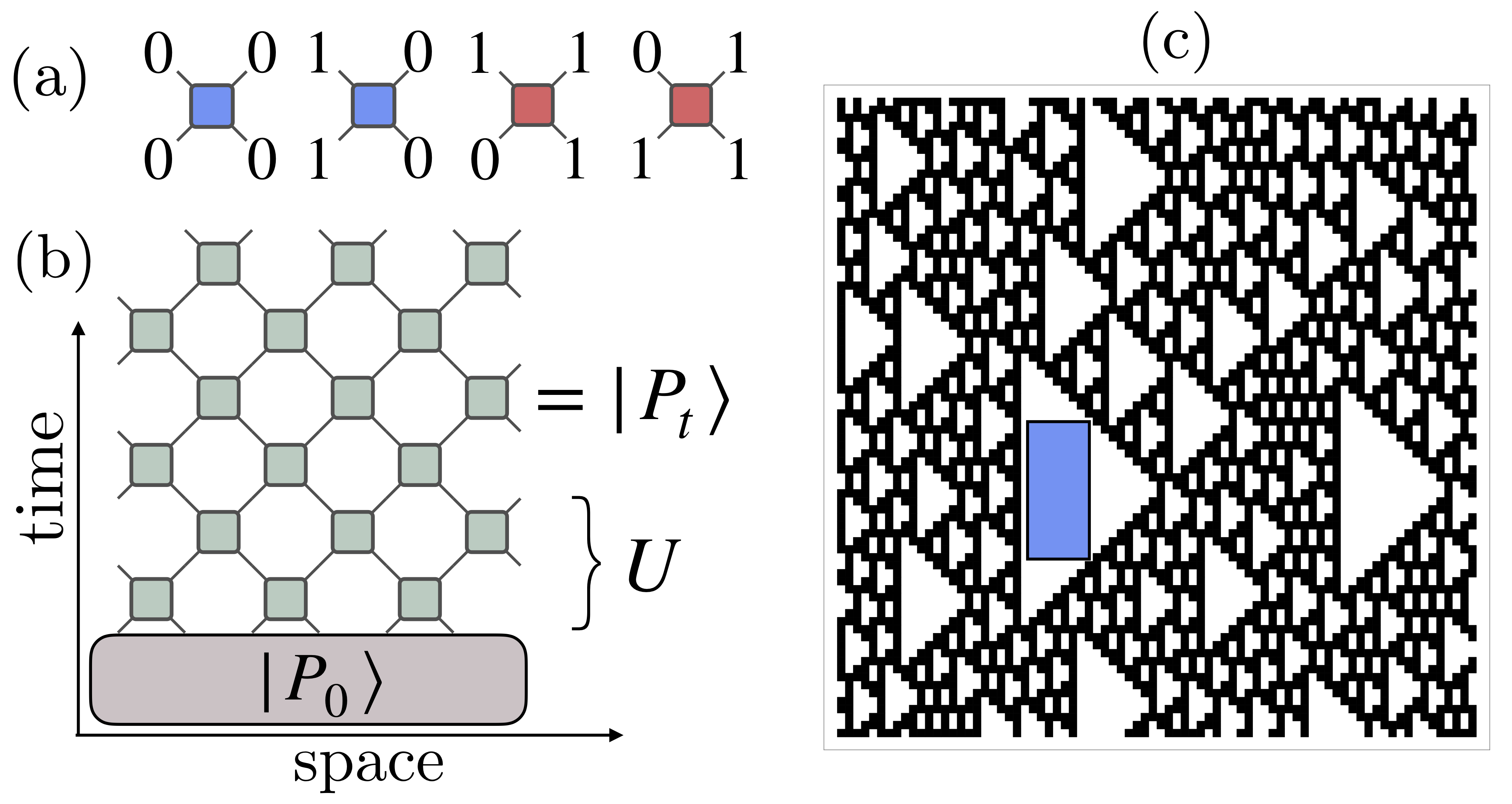}
    \caption{\label{fig1}
      \textbf{Floquet-East model.} 
      (a) Allowed local gates: the left spin can flip only if its right neighbour is 1; spin-flip gates are shaded red, no-flip gates shaded blue.
      (b) Tensor-network representation of an initial state $\ket{P_0}$ evolved under the Floquet dynamics, $\ket{P_t} = U^t \ket{P_0}$.
      (c) Sample trajectory from a random initial configuration. The blue box represents the condition of having only $0$s inside that space time region. 
    }
\end{figure}

\smallskip

\begin{figure*}[th!]
    \centering
    \includegraphics[width=\textwidth]{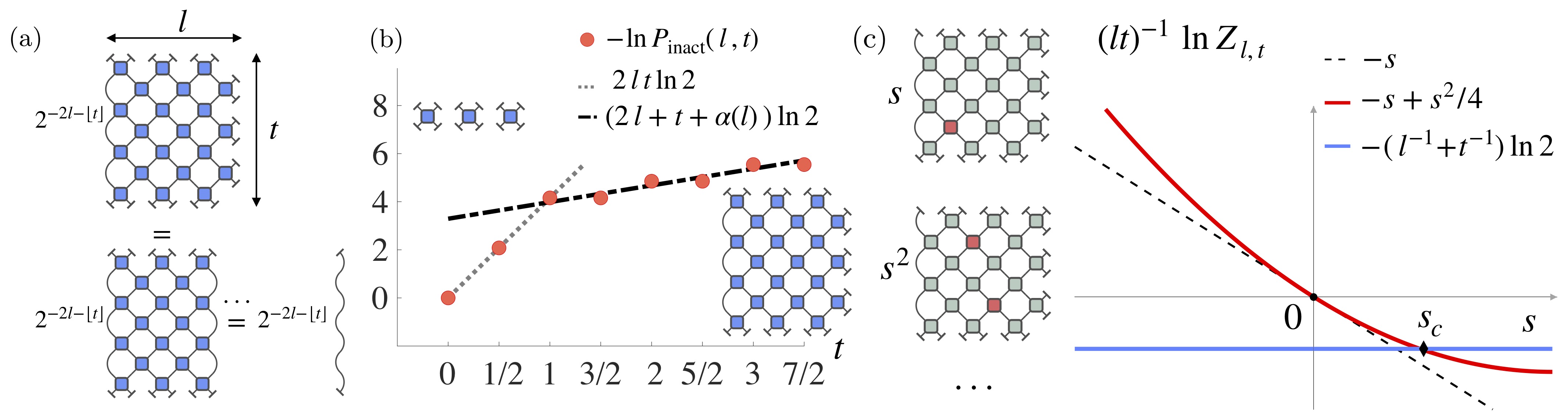}
    \caption{
    \label{fig2}
        {\bf Hydrophobic crossover and LD transition.}
        (a) Graphical representation of $P_{\rm inact}(l,t)$. The space-time region $l \times t$ is conditioned to only inactive gates (blue). This tensor network can be contracted to obtain \er{xover}. 
        \change{The pre-factor $2^{-2l-\lfloor t \rfloor}$ ensures that in the absence of conditioning the contraction is equal to  $1$ as required by probability conservation.}
        (b) Corresponding dynamical free-energy $- \ln P_{\rm inact}(l,t)$ as a function of time at $l = 3$, 
        displaying a crossover from area (dotted grey line) to perimeter (dashed black line) scaling. The coefficient $\alpha (l)$ is $5/4$ for integer $l$.
        (c) The scaled CGF for the activity, $\ln Z_{l,t}(s) / lt$, has 
        an active branch perturbatively connected to $s=0$ (red), and an inactive branch coming from $s \to \infty$ (blue). Their crossing at $s_c \approx t^{-1} + l^{-1}$ indicates a first-order transition in trajectory space in the large $l,t$ limit.  
    }
\end{figure*}

\noindent
{\bf \em Floquet-East model.---} 
We consider a chain of $2 L$ sites with a binary variable (or classical spin) per site $n_i \in \{ 0,1 \}$, with the site labels $i$ taking half-integer values, $i\in\{\frac{1}{2},1,\frac{3}{2},\ldots,L\}$.  Relevant statistical states are probability vectors $\ket{P} := \sum_{\n} \ket{\n} P(\n)$, where $\{ \ket{\n} := \ket*{n_{\frac{1}{2}}} \otimes \ket{n_{1}} \otimes \cdots \otimes \ket{n_L} \}$ is the configurational basis, $P(\n) \geq 0 \; \forall \n$,  and $\braket{\fl}{P}=\sum_{\n} P(\n) = 1$ (with $\bra{\fl} := \sum_{\n} \bra{\n}$ the ``flat state''). The dynamics is discrete, staggered in terms of two half time-steps given by the deterministic maps $\Ue$ and $\Uo$,
\begin{equation}\label{eq:Pt}
  \ket{P_{t+1}} =
  \Ue \ket*{P_{t+\frac{1}{2}}} = \Ue\Uo\ket{P_t},\qquad t\in\mathbb{N}.
\end{equation}
The maps $\Ue$ and $\Uo$ consist of two-site gates $u$ applied either to even or odd pairs of neighbouring sites,
\begin{equation}\label{eq:UeUo}
  \Ue = u^{\otimes L},\qquad
  \Uo = \Pi_{L} u^{\otimes L} \Pi_{L}^{\dagger},
\end{equation}
where $\Pi_L$ is a one-site shift operator for a chain of $2L$ sites with periodic boundaries. The local gate $u$ implements the deterministic East model rule: a spin flips if its neighbour to the right --- i.e., the one to the \emph{east} --- is in the state $1$, or stays the same otherwise,
\begin{equation*}
  u = 
  \begin{bmatrix}
    1 & 0 & 0 & 0 \\
    0 & 0 & 0 & 1 \\
    0 & 0 & 1 & 0 \\
    0 & 1 & 0 & 0 
  \end{bmatrix} , 
  \quad
 \mel{n^{\prime} \, m^{\prime}}{u}{n \; m} 
=:\mkern-16mu
  \gatens{$n^{\prime}$}{$m^{\prime}$}{$n$}{$m$}
\end{equation*}
where our convention is $\ket{0} = [1~0]^T$, $\ket{1} = [0~1]^T$, \change{and $\ket{n \, m}=\ket{n}\otimes \ket{m}$.} The graphical representation of $u$ allows to interpret the dynamics as a tensor network~\cite{TensorNetwork}, see Fig.~\ref{fig1}: panel (a) shows the allowed gates, while panel (b) gives the evolved state.

The Floquet-East model can also be thought of as a cellular automaton (CA), specifically Rule 60 in the classification of Refs.~\cite{wolfram1983statistical,bobenko1993two}. In contrast to other recently studied CAs such as Rule 54~\cite{prosen2016integrability,inoue2018two,prosen2017exact,gopalakrishnan2018operator,buca2019exact,klobas2020space,klobas2019time,alba2019operator,klobas2020matrix}, Rule 201~\cite{wilkinson2020exact,iadecola2020nonergodic} and Rule 150~\cite{gopalakrishnan2018facilitated,gombor2021integrable,wilkinson2022exact}, Rule 60 
\change{appears to be non-integrable}. 
Fig.~\ref{fig1}(c) shows a trajectory starting from a random initial configuration: the 
\change{mixing} 
nature of the dynamics is apparent, with  fluctuations highly reminiscent of the dynamic heterogeneity and ``space-time bubbles'' of the stochastic (and continuous-time) East model \cite{garrahan2002geometrical}. 
\change{This in turn suggests that much of the interesting behaviour of stochastic KCMs might manifest in deterministic KCMs with the same constraints (see also~\cite{deger2022arresting,deger2022constrained}). 
}

\begin{figure*}[th]
  \centering
  \includegraphics[width=\textwidth]{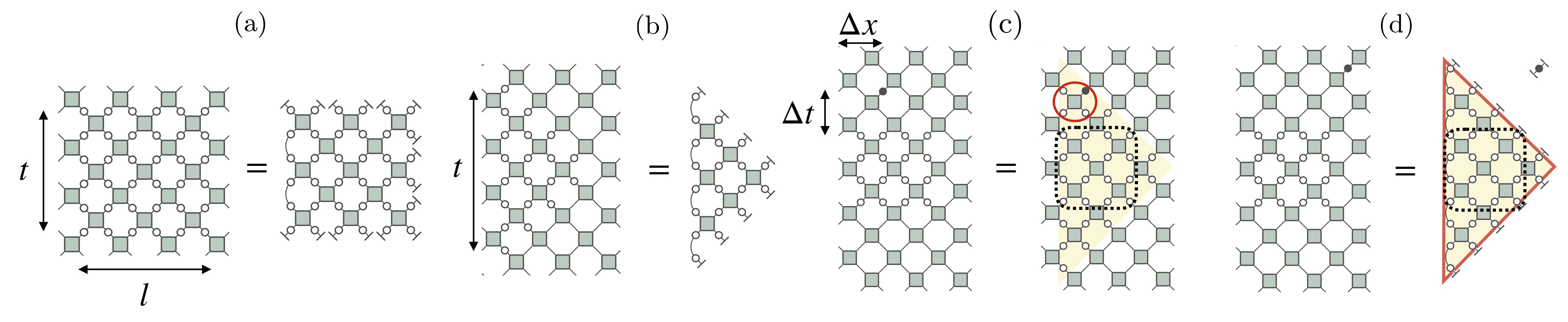}
  \caption{\label{fig3}
      \textbf{\bf Solvation and optimal void geometry.} 
      (a) Network representation of the probability $P_\square(l, t)$ of a square region conditioned to all zero spins, embedded in a 
      much larger trajectory. 
      (b) Same for $P_\rhd(t)$ for the optimal triangular geometry. 
      (c) Spin density outside from the condition: when the  site falls inside the triangle (shaded area) enclosing the condition (dashed square) the density is zero (due to the impossible gate, circled in red). 
      (d) Same but now the probed site falls outside the enclosing triangle: in this case the density is the average one. 
 }
\end{figure*}

\smallskip
\noindent
{\bf \em Propagation in space and invariant states.---} 
The definitions above describe the time evolution (down to up in Fig.~\ref{fig1}) of configurations.  Alternatively, one can consider also how a row pseudo-probability vector~\footnote{Note that the vector $\bra*{\tilde{P}_x}$ is not necessarily a valid probability vector since it does not need to be normalized.} $\bra*{\tilde{P}_x}$ over spins at a fixed point in space $x$ and all times is propagated {\em in space}, 
$\bra*{\tilde{P}_{x+1}} = \bra*{\tilde{P}_x}\tU$ (from left to right in Fig.~\ref{fig1}) under the dual operator $\tU = \tUo \tUe$, through the composition of the local gate 
\begin{equation}
  \tu := 
  \begin{bmatrix}
    1 & 0 & 0 & 0 \\
    0 & 0 & 0 & 1 \\
    0 & 0 & 0 & 1 \\
    1 & 0 & 0 & 0 
  \end{bmatrix} ,
  \quad
  \mel{n \; n^{\prime}}{\tu}{m \; m^{\prime}} =: \mkern-16mu
  \gatens{$n^{\prime}$}{$m^{\prime}$}{$n$}{$m$}\mkern-16mu,
\end{equation}
and similarly defined evolution of a column vector $\ket*{\tilde{P}_x}$ as
$\ket*{\tilde{P}_{x-1}}=\tUo\tUe \ket*{\tilde{P}_x}$.

The time-dynamics is deterministic and reversible (i.e., a special case of \emph{bi-stochastic} dynamics), which implies that the flat state is invariant under both the time-evolution and its inverse. This is a local property of the gate $u$ and can be stated graphically (see \cite{SM} for the details) as
\begin{equation}
    \fsd := \begin{bmatrix} 1 \\ 1 \end{bmatrix}
    , \mkern15mu
    \fsu := \begin{bmatrix} 1 & 1 \end{bmatrix}
    , \mkern15mu
    \gatefsd = \fsd~\fsd\,
    , \mkern15mu
    \gatefsu = \fsu~\fsu\,, 
    \label{ist}
\end{equation}
where e.g.\ the last relation is $[1~1~1~1]u=[1~1~1~1]$. Similarly, the invariant states of the space dynamics~\footnote{In the quantum setup, analogous objects --- the left and right fixed points of the space transfer-matrix --- are referred-to also as \emph{influence matrices}~\cite{lerose2021influence}. These have recently been understood to provide convenient numerical~\cite{banuls2009matrix,muller2012tensor,lerose2021influence,sonner2021influence,lerose2021scaling,friasperez2022lightcone}, and analytical tools~\cite{bertini2019exact,bertini2019entanglement,piroli2020exact,klobas2021exact,klobas2021exactrelaxation,klobas2021entanglement,giudice2022temporal,bertini2022entanglement,bertini2022growth,foligno2023temporal} to study quantum many-body dynamics.} follow from a set of local algebraic relations satisfied by the local gates,
\begin{gather}
    \label{iss} 
    \fsl := \begin{bmatrix} 1 & 1 \end{bmatrix}
    , \quad
    \fsr := \begin{bmatrix} 1 \\ 1 \end{bmatrix}
    , \quad
    \ds := \begin{bmatrix} 1&0&0&1\end{bmatrix},
    \\
    \gatefsl = 2 \, \ds
    , \quad
    \gatedsfsd = \begin{matrix} \fsl \\[-0.25em] \ds \end{matrix}
    , \quad
    \gatedsfsu = \begin{matrix} \ds \\[-0.25em] \fsl \end{matrix} 
    , \quad
    \gatefsr = \begin{matrix} \fsr \\[-0.25em] \fsr \end{matrix}\,.
    \nonumber
\end{gather}
From here we get that invariant states in space are, in the forward direction (left to right) the ``dimerised state'' obtained from the tensor product of $\ds$ (with appropriate boundaries, see below), and in the backward direction (right to left) the flat state. This is consistent with the space dynamics under $\tU$ being (right) stochastic, and is reminiscent of dual-unitarity~\cite{bertini2019exact,kos2021correlations,kos2022circuits} (see also Refs.~\cite{bertini2018exact,bertini2019entanglement,bertini2020operatorI,bertini2020operatorII,piroli2020exact,claeys2020maximum,claeys2021ergodic,bertini2021random,jonay2021triunitary,kasim2022dual,suzuki2022computational,foligno2022growth,foligno2023temporal,rampp2023dual}) in the quantum setting.

\smallskip

\noindent
{\bf \em Probability of inactive space-time regions.---} 
Relations \era{ist}{iss} allow to compute exactly several dynamical properties. Even though the dynamics is deterministic, if we consider a space-time region of size $l\times t$ inside a large box of $L\times T$, the dynamics in the region is probabilistic as the rest acts as an environment, and for $L,T\to\infty$ the circuit can be contracted to the boundary of the region in terms of the invariant states introduced above~\cite{lerose2021influence}, see Fig.~\ref{fig2}(a). As a first question, we consider the probability $P_{\rm inact}(l,t)$ of having no spin-flips in that space-time region. The probability $P_{\rm inact}(l,t)$ is then obtained by contracting the region of inactive gates shown in Fig.~\ref{fig2}(a), while the pre-factor $2^{-(2l+\lfloor t \rfloor)}$ is determined as the factor needed to normalise to $1$ the same region in the absence of conditioning. The inactive gates obey 
\begin{equation}\label{eq:relInactive}
  \gateyfs{1} = \dsr~\projzfs{1},\quad \gateyfs{-1} = \dsr~\projzfs{-1},
\end{equation}
with $\dsr = \ds^T$ and $\projz$ denoting the projector onto the 0 state, 
$\projz=[\begin{smallmatrix} 1 & 0 \\ 0 & 0 \end{smallmatrix}]$.
This means that the flat state on the right can be repeatedly propagated leftwards [see Fig.~\ref{fig2}(a)], until we are left with 
\begin{equation}
  P_{\rm inact}(l,t)=2^{-(2l+\lfloor t \rfloor)} \times \fsdotfs
  = 2^{-(2l+\lfloor t \rfloor-1)}.
\end{equation}
The calculation is different for the special case of $t=1/2$ (single row of inactive gates), as it reduces to the product of the expectation value of each gate~\cite{SM}. Overall,
\begin{equation}
  P_{\rm inact}(l,t)
  =
  \begin{cases}
    2^{-( 2l + \lfloor t \rfloor-1)},& t\ge 1,\\
    2^{- l },& t=\frac{1}{2},
  \end{cases}
  \label{xover}
\end{equation}
see Fig.~\ref{fig2}(b). This is similar to the crossover observed numerically in the stochastic East model \cite{katira2018solvation}, and analogous to that in the  free-energy (i.e., minus log probability) of solvation in the hydrophobic effect --- from a regime dominated by entropy for small solutes, to one dominated by energy for large ones~\cite{lum1999hydrophobicity,chandler2005interfaces,katira2016pre-transition}.

\smallskip
\noindent
{\bf \em Phase transition in dynamical large deviations.---} 
The result above suggests that in the limit of $l,t \to \infty$ (with $l/L,t/T \to 0$) the Floquet-East model will have a phase transition in the space of trajectories \cite{garrahan2007dynamical,touchette2009the-large,jack2020dynamical}. This can be shown by considering the statistics of the dynamical activity (total number of spin-flips) \cite{garrahan2007dynamical,lecomte2007thermodynamic,maes2020frenesy} in a space-time volume $l \times t$.
Given the set of trajectories $\{ \omega \}$ in the region, the moment generating function (MGF) of the activity is 
\begin{equation}
  Z_{l,t}(s) = \sum_{\omega} \pi(\omega) e^{-s K(\omega)},
\end{equation}
where $\pi(\omega)$ is the probability of the trajectory and $K(\omega)$ its activity. The MGF is obtained from a calculation similar to that of Fig.~\ref{fig2}(a), but where the active (spin-flip) gates carry an extra factor of $e^{-s}$.

Two limits are easy to calculate. In the limit of $s \to \infty$ all active gates are suppressed, and $Z_{l,t}(s)$ is given by \er{xover}.
Conversely, for $s \approx 0$ we can express $Z_{l,t}(s)$ as a series in $s$ with the moments of the activity as coefficients, 
\begin{equation}
  Z_{l,t}(s) = 1 - s \langle K \rangle + \frac{1}{2} s^2 \langle K^2 \rangle + \cdots.
\end{equation}
The first two moments are obtained straightforwardly,
\begin{equation}
  \langle K \rangle = l t,\qquad\langle K^2 \rangle = (l t)^2/2,
\end{equation}
the latter result given by the fact that all two-point correlators are disconnected in the dynamics \cite{SM}. The difference in scaling for small and large $s$ implies a crossover in the cumulant generating function (CGF), $\ln Z_{l,t}(s)$, which in the limit of $l,t \to \infty$ becomes singular, see Fig.~\ref{fig2}(c). Since the CGF is convex and non-increasing \cite{touchette2009the-large}, the perturbative branch (red curve) has to cross to the inactive branch (blue curve), with the change becoming progressively sharp and occurring around $s_c \approx 1/l+1/t$. For $l,t \to \infty$ this corresponds to a discontinuity in the derivative of the CGF at $s=0$, and therefore the model has to undergo an active-inactive first-order phase transition. This transition (cf.\ liquid-vapour in water) gives rise to the dynamic hydrophobicity in the model.

\begin{figure}[t!]
  \centering
   \includegraphics[width=\columnwidth]{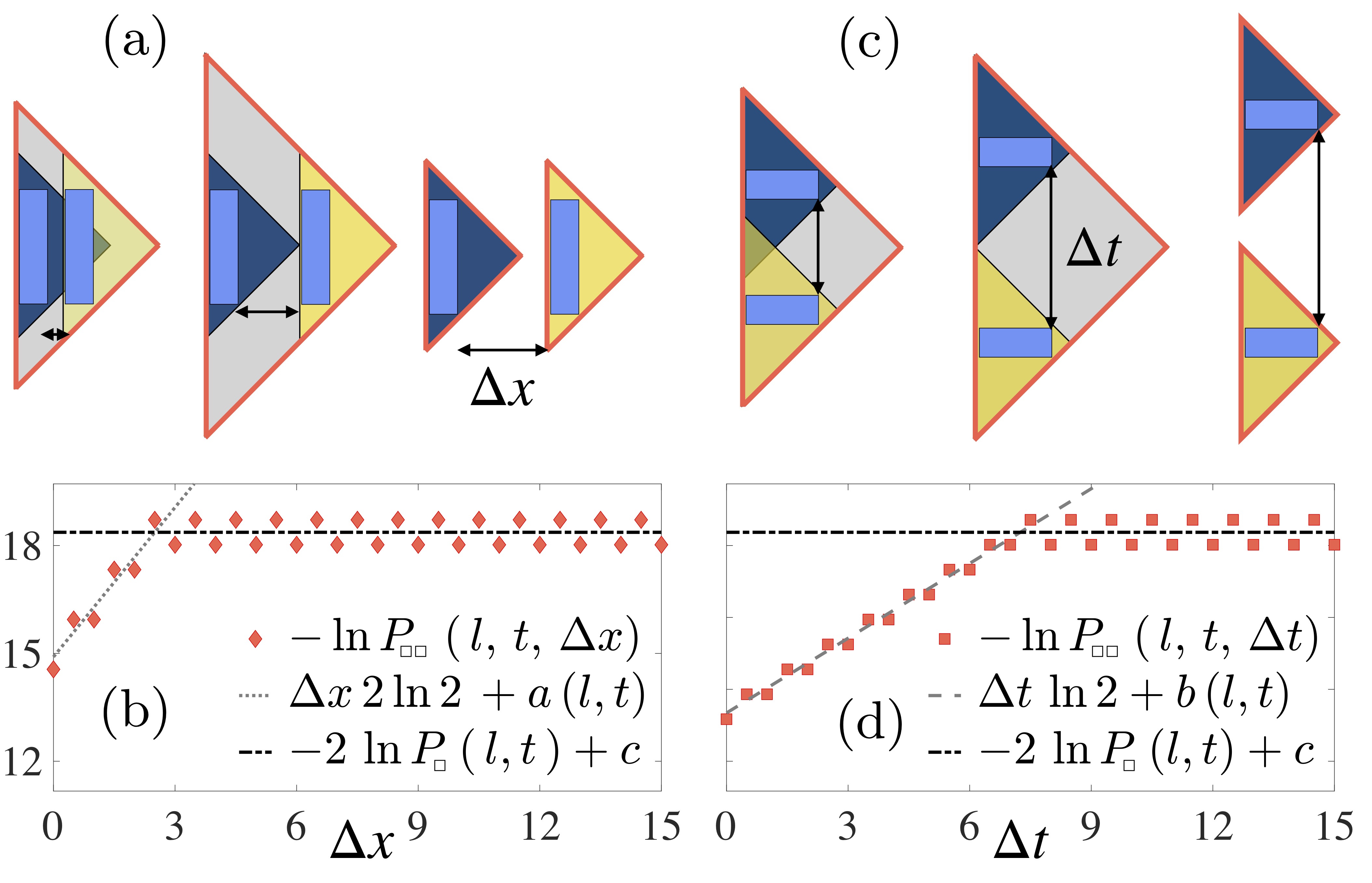}
  \caption{\label{fig4}
      \textbf{Hydrophobic collapse.} 
      (a) Two space-time ``solutes'' (light blue) of size $l \times t$ separated in space by $\Delta x$. Conditions on empty boxes extend to triangular shaped regions (dark blue and yellow). (b) Dynamical free-energy as a function of $\Delta x$. This is given by the perimeter of the optimal triangular bubble enclosing the two boxes (orange contour): for large $\Delta x$ it is given by the perimeters of two triangles enclosing each condition individually, as the probabilities factorise for $\Delta x \geq t/2$; for $\Delta x < t/2$, the free energy is given by the perimeter of a single larger bubble enclosing both conditions, and decreases with decreasing distance giving rise to an effective attraction. The coefficients are $a(l,t)=\ln 2 (4l+t-1/2)$, $c=\ln2 /2$.
      (c,d) Same for a time separation of $\Delta t$. The coefficient in (d) are $b(l,t)=\ln 2 (2l+2t-3/4)$, $c=\ln2 /2$.
      }
\end{figure}

\smallskip
\noindent
{\bf \em Solvation in space-time.---}
To draw an analogy with the case of a solute in water, we consider the probability of a region of space-time with all sites in the zero state. See Fig.~\ref{fig3}(a) for the tensor network representation; given a region $l \times t$ with all spins conditioned to be in the 0 state, we can compute the probability of this event, $P_{\square}(l,t)$, as before, by contracting the outside to the edges of the region using the invariant states. Noting that 
\begin{equation}
  \gatezz = \gateyzz=\gatey,
\end{equation}
and using the relations~\eqref{eq:relInactive} for the inactive gates, we get that for integer $t$  the probability is $\frac{1}{2}$ of the corresponding region being inactive,
\footnote{Note that $P_{\square}(l,t)$ scales with $P_{\rm inact}(l-\frac{1}{2},t-\frac{1}{2})$ due to the way that the sizes are defined in both cases: in the former the size corresponds to the number of unit cells (i.e., twice the number of sites), while in the latter the number of time/space steps. This subtlety arises because of the way that the constraints are applied --- either to sites or gates. See \cite{SM} for an illustration.}
\begin{equation} \label{Psquare}
    P_{\square}(l,t)=2^{-(2l+\lfloor t-\frac{1}{2}\rfloor)} .   
\end{equation}
The key observation is that the free-energy cost for a fluctuation as that of Fig.~\ref{fig3}(a) scales with the perimeter, $2l+t$, of the region, rather than with the total number of sites in the region, $4 l t$. 
This calculation can be repeated for other shapes of the conditioned region. It is easy to show that the optimal one is that of an isosceles right-pointing triangle, see Fig.~\ref{fig3}(b): the probability in this case goes as \cite{SM}
\begin{equation} \label{Ptriangle}
    P_\rhd(t) = 2^{-t}.
\end{equation}
For comparison, a rectangle with the same number of conditioned 0 sites would have a smaller probability of occurring, $P_\square(t+1, t/2) \propto 2^{- 3t/2}$ 
\footnote{
  \change{In the deterministic model, similar perimeter scaling is obtained for the free-energies of space-time regions corresponding to allowed trajectories of the circuit. In contrast, only regions of $0$s show this scaling in the stochastic case. See \cite{SM} for details.} 
}.

The optimality of the right-pointing triangle is shown by computing the average spin density,
$\rho(\Delta x,\Delta t)$, at a distance $(\Delta x, \Delta t)$ from a region conditioned to have all $0$ spins: see Fig.~\ref{fig3}(c,d), where $\projo=[ \begin{smallmatrix} 0 & 0 \\ 0 & 1 \end{smallmatrix} ]$ projects to the state $1$. When the location of the projector is near the solute, 
as in panel (c), all sites within the enclosing triangle (shaded region) are also $0$ at no extra free-energy cost; for the probed site to be $1$ would require a non-existent gate (circled), cf.\ Fig.~\ref{fig1}(a), and as a consequence $\rho(\Delta x,\Delta t)=0$ for any site within the shaded triangle~\cite{SM}. In contrast, for sites outside the triangle, the contraction decouples the site, Fig.~\ref{fig3}(d), which gives the stationary-state value, $\rho(\Delta x,\Delta t)=1/2$. The above reasoning shows that to accommodate any space-time ``solute'', the most efficient fluctuation is to create a triangular ``bubble'', which encloses the solute by locally phase separating the dynamics. The optimal shape also explains the nature of the dynamical fluctuations observed in trajectories such as that of Fig.~\ref{fig1}(c) where the bubbles are all right-pointing triangles.

\smallskip

\noindent
{\bf \em Hydrophobic collapse.---} 
The free-energy of a single solute scaling with the interface of the enclosing bubble leads to an interaction when two or more solutes are present. In water this is known as {\em hydrophobic collapse}~\cite{willard2008the-role}. In the dynamics of the Floquet-East model we have the same in space-time: Fig.~\ref{fig4}(a) shows two space-time regions (light blue rectangles) conditioned to having all their spins in the zero state, with the two regions separated in space by a distance $\Delta x$. For large $\Delta x$ each of these conditions gives rise to a triangular bubble that encloses it, and the (dynamical) free-energy cost is given by the sum of the perimeters of the two triangles. However, as $\Delta x$ decreases, we eventually reach a point beyond which it is more favourable to enclose both solutes within a single triangular bubble, and as $\Delta x$ is reduced further the free-energy also goes down, see Fig.~\ref{fig4}(b). This gives rise to the ``attraction'' between the solutes, a dynamical version of hydrophobic collapse. Fig.~\ref{fig4}(c,d) shows the same for separation in the time direction.

\smallskip

\noindent
{\bf \em Pre-transition effects in the stochastic model.---} 
The stochastic version of the Floquet East model is defined by the following six possible gates \cite{defazio2024exact}:
\begin{equation*}
  \begin{tikzpicture}[baseline={([yshift=-0.85ex]current bounding box.center)},scale=0.5]
    \propS{0}{0}{colU}{black}
    \node at (-0.875,-0.875) {$n_1$};
    \node at (0.875,-0.875) {$n_2$};
    \node at (-0.875,0.875) {$n_1^{\prime}$};
    \node at (0.875,0.875) {$n_2^{\prime}$};
  \end{tikzpicture}\mkern-20mu 
    =
    \left[
    \delta_{n_2,0} \delta_{n_1^{\prime},n_1^{\pprime}}
    +\delta_{n_2,1}
    \left(
      p \, \delta_{n_1^{\prime},1-n_1^{\pprime}}
      + \bar{p} \, \delta_{n_1^{\prime},n_1^{\pprime}}
      \right)
    \right]
    \delta_{n_2^{\prime},n_2^{\pprime}}.
\end{equation*}
Here the constraint is the same as before, but if a flip is possible it only occurs with probability $p$ (with no change with probability $\bar{p}=1-p$). The deterministic model is recovered for $p=1$. The stochastic circuit and the standard (continuous-time) stochastic East model~\cite{garrahan2018aspects} are directly connected through the Trotter-Suzuki decomposition of the integrated time-evolution operator~\cite{trotter1959product,suzuki1991general}.

As proved in Ref.~\cite{defazio2024exact}, the results for the deterministic model ($p=1$) imply similar dynamic hydrophobicity for the stochastic ($p<1$) case. Specifically, one can show that the probability of an inactive space-time region is bounded from below by the deterministic result \cite{defazio2024exact}
\begin{equation*}
  P_{\rm inact}^{(p)}(l,t) 
  \geq
  P_{\rm inact}^{(p=1)}(l,t) = 2^{-( 2l + \lfloor t \rfloor-1)} 
  \;\; (t\ge 1).
\end{equation*}
Similar lower bounds are found for the probabilities of regions of all $0$ spins, cf.\ \era{Psquare}{Ptriangle} \cite{defazio2024exact}. This demonstrates \cite{defazio2024exact} that the stochastic Floquet East has an active/inactive first-order transition, and the space-time perimeter scaling of the dynamical-free energy of inactive regions is analogous to the area (rather than volume) dependence of the free-energy for cavities in water~\cite{lum1999hydrophobicity,chandler2005interfaces}.

\smallskip

\noindent
{\bf \em Conclusions.---} 
Here we studied the Floquet-East model, a circuit version of the kinetically constrained East model. We proved analytically that its trajectories display pre-transition effects, a dynamical analogue of the hydrophobic effect in water. We computed the exact probabilities for inactive space-time fluctuations for the deterministic limit of the model. To our knowledge, this is the first time such pre-transition physics has been computed exactly in any model.
Moreover, as we prove in Ref.~\cite{defazio2024exact} the results for the deterministic circuit lower-bound probability of inactive fluctuations in the stochastic version of the model.
The work here is classical, but since the deterministic Floquet-East is also unitary, our results connect to the ongoing interest in the dynamics of quantum circuits, e.g.\ \cite{osborne2006efficient,nahum2017quantum,nahum2018operator,chan2018solution,vonKeyserlingk2018operator,bertini2019entanglement,bertini2019exact,gopalakrishnan2019unitary,friedman2019spectral,li2019measurement,skinner2019measurement,rakovszky2019sub, zabalo2020critical,claeys2020maximum}.
For example, one could consider (with small alterations, cf.~\cite{klobas2021exact,klobas2021exactrelaxation,klobas2021entanglement}) quantum dynamics, where we expect similar hydrophobic fluctuations to play an important role in entanglement growth and thermalisation \cite{lan2018quantum,bertini2023exact}.

\smallskip
\begin{acknowledgments}
\noindent
We thank K.\ Sfairopoulos for the careful reading of the manuscript, and useful suggestions. We acknowledge financial support from EPSRC Grants No.\ EP/R04421X/1 and EP/V031201/1, and from The Leverhulme Trust through the Early Career Fellowship No.\ ECF-2022-324.
\end{acknowledgments}

\bibliographystyle{apsrev4-2}
\bibliography{bibliography}

\begin{thebibliography}{112}%
\makeatletter
\providecommand \@ifxundefined [1]{%
 \@ifx{#1\undefined}
}%
\providecommand \@ifnum [1]{%
 \ifnum #1\expandafter \@firstoftwo
 \else \expandafter \@secondoftwo
 \fi
}%
\providecommand \@ifx [1]{%
 \ifx #1\expandafter \@firstoftwo
 \else \expandafter \@secondoftwo
 \fi
}%
\providecommand \natexlab [1]{#1}%
\providecommand \enquote  [1]{``#1''}%
\providecommand \bibnamefont  [1]{#1}%
\providecommand \bibfnamefont [1]{#1}%
\providecommand \citenamefont [1]{#1}%
\providecommand \href@noop [0]{\@secondoftwo}%
\providecommand \href [0]{\begingroup \@sanitize@url \@href}%
\providecommand \@href[1]{\@@startlink{#1}\@@href}%
\providecommand \@@href[1]{\endgroup#1\@@endlink}%
\providecommand \@sanitize@url [0]{\catcode `\\12\catcode `\$12\catcode
  `\&12\catcode `\#12\catcode `\^12\catcode `\_12\catcode `\%12\relax}%
\providecommand \@@startlink[1]{}%
\providecommand \@@endlink[0]{}%
\providecommand \url  [0]{\begingroup\@sanitize@url \@url }%
\providecommand \@url [1]{\endgroup\@href {#1}{\urlprefix }}%
\providecommand \urlprefix  [0]{URL }%
\providecommand \Eprint [0]{\href }%
\providecommand \doibase [0]{https://doi.org/}%
\providecommand \selectlanguage [0]{\@gobble}%
\providecommand \bibinfo  [0]{\@secondoftwo}%
\providecommand \bibfield  [0]{\@secondoftwo}%
\providecommand \translation [1]{[#1]}%
\providecommand \BibitemOpen [0]{}%
\providecommand \bibitemStop [0]{}%
\providecommand \bibitemNoStop [0]{.\EOS\space}%
\providecommand \EOS [0]{\spacefactor3000\relax}%
\providecommand \BibitemShut  [1]{\csname bibitem#1\endcsname}%
\let\auto@bib@innerbib\@empty
\bibitem [{\citenamefont {Casimir}(1948)}]{casimir1948attraction}%
  \BibitemOpen
  \bibfield  {author} {\bibinfo {author} {\bibfnamefont {H.~B.}\ \bibnamefont
  {Casimir}},\ }\href@noop {} {\bibfield  {journal} {\bibinfo  {journal} {Proc.
  Kon. Ned. Akad. Wet.}\ }\textbf {\bibinfo {volume} {51}},\ \bibinfo {pages}
  {793} (\bibinfo {year} {1948})}\BibitemShut {NoStop}%
\bibitem [{\citenamefont {Fisher}\ and\ \citenamefont
  {Gennes}(1978)}]{fisher1978wall}%
  \BibitemOpen
  \bibfield  {author} {\bibinfo {author} {\bibfnamefont {M.~E.}\ \bibnamefont
  {Fisher}}\ and\ \bibinfo {author} {\bibfnamefont {P.~G.}\ \bibnamefont
  {Gennes}},\ }\href@noop {} {\bibfield  {journal} {\bibinfo  {journal} {C. R.
  Acad. Sci. Paris}\ }\textbf {\bibinfo {volume} {287}},\ \bibinfo {pages}
  {207} (\bibinfo {year} {1978})}\BibitemShut {NoStop}%
\bibitem [{\citenamefont {Hertlein}\ \emph {et~al.}(2008)\citenamefont
  {Hertlein}, \citenamefont {Helden}, \citenamefont {Gambassi}, \citenamefont
  {Dietrich},\ and\ \citenamefont {Bechinger}}]{hertlein2008direct}%
  \BibitemOpen
  \bibfield  {author} {\bibinfo {author} {\bibfnamefont {C.}~\bibnamefont
  {Hertlein}}, \bibinfo {author} {\bibfnamefont {L.}~\bibnamefont {Helden}},
  \bibinfo {author} {\bibfnamefont {A.}~\bibnamefont {Gambassi}}, \bibinfo
  {author} {\bibfnamefont {S.}~\bibnamefont {Dietrich}},\ and\ \bibinfo
  {author} {\bibfnamefont {C.}~\bibnamefont {Bechinger}},\ }\href
  {https://doi.org/10.1038/nature06443} {\bibfield  {journal} {\bibinfo
  {journal} {Nature}\ }\textbf {\bibinfo {volume} {451}},\ \bibinfo {pages}
  {172} (\bibinfo {year} {2008})}\BibitemShut {NoStop}%
\bibitem [{\citenamefont {Gambassi}(2009)}]{gambassi2009the-casimir}%
  \BibitemOpen
  \bibfield  {author} {\bibinfo {author} {\bibfnamefont {A.}~\bibnamefont
  {Gambassi}},\ }\href {https://dx.doi.org/10.1088/1742-6596/161/1/012037}
  {\bibfield  {journal} {\bibinfo  {journal} {J. Phys.}\ }\textbf {\bibinfo
  {volume} {161}},\ \bibinfo {pages} {012037} (\bibinfo {year}
  {2009})}\BibitemShut {NoStop}%
\bibitem [{\citenamefont {Lipowsky}(1982)}]{lipowsky1982critical}%
  \BibitemOpen
  \bibfield  {author} {\bibinfo {author} {\bibfnamefont {R.}~\bibnamefont
  {Lipowsky}},\ }\href {https://doi.org/10.1103/PhysRevLett.49.1575} {\bibfield
   {journal} {\bibinfo  {journal} {Phys. Rev. Lett.}\ }\textbf {\bibinfo
  {volume} {49}},\ \bibinfo {pages} {1575} (\bibinfo {year}
  {1982})}\BibitemShut {NoStop}%
\bibitem [{\citenamefont {Lipowsky}(1984)}]{lipowsky1984surface-induced}%
  \BibitemOpen
  \bibfield  {author} {\bibinfo {author} {\bibfnamefont {R.}~\bibnamefont
  {Lipowsky}},\ }\href {https://doi.org/10.1063/1.333703} {\bibfield  {journal}
  {\bibinfo  {journal} {J. Appl. Phys.}\ }\textbf {\bibinfo {volume} {55}},\
  \bibinfo {pages} {2485} (\bibinfo {year} {1984})}\BibitemShut {NoStop}%
\bibitem [{\citenamefont {Lum}\ \emph {et~al.}(1999)\citenamefont {Lum},
  \citenamefont {Chandler},\ and\ \citenamefont
  {Weeks}}]{lum1999hydrophobicity}%
  \BibitemOpen
  \bibfield  {author} {\bibinfo {author} {\bibfnamefont {K.}~\bibnamefont
  {Lum}}, \bibinfo {author} {\bibfnamefont {D.}~\bibnamefont {Chandler}},\ and\
  \bibinfo {author} {\bibfnamefont {J.~D.}\ \bibnamefont {Weeks}},\ }\href
  {https://doi.org/10.1021/jp984327m} {\bibfield  {journal} {\bibinfo
  {journal} {J. Phys. Chem. B}\ }\textbf {\bibinfo {volume} {103}},\ \bibinfo
  {pages} {4570} (\bibinfo {year} {1999})}\BibitemShut {NoStop}%
\bibitem [{\citenamefont {Chandler}(2005)}]{chandler2005interfaces}%
  \BibitemOpen
  \bibfield  {author} {\bibinfo {author} {\bibfnamefont {D.}~\bibnamefont
  {Chandler}},\ }\href {https://doi.org/10.1038/nature04162} {\bibfield
  {journal} {\bibinfo  {journal} {Nature}\ }\textbf {\bibinfo {volume} {437}},\
  \bibinfo {pages} {640} (\bibinfo {year} {2005})}\BibitemShut {NoStop}%
\bibitem [{\citenamefont {Katira}\ \emph {et~al.}(2016)\citenamefont {Katira},
  \citenamefont {Mandadapu}, \citenamefont {Vaikuntanathan}, \citenamefont
  {Smit},\ and\ \citenamefont {Chandler}}]{katira2016pre-transition}%
  \BibitemOpen
  \bibfield  {author} {\bibinfo {author} {\bibfnamefont {S.}~\bibnamefont
  {Katira}}, \bibinfo {author} {\bibfnamefont {K.~K.}\ \bibnamefont
  {Mandadapu}}, \bibinfo {author} {\bibfnamefont {S.}~\bibnamefont
  {Vaikuntanathan}}, \bibinfo {author} {\bibfnamefont {B.}~\bibnamefont
  {Smit}},\ and\ \bibinfo {author} {\bibfnamefont {D.}~\bibnamefont
  {Chandler}},\ }\href {https://doi.org/10.7554/eLife.13150} {\bibfield
  {journal} {\bibinfo  {journal} {Elife}\ }\textbf {\bibinfo {volume} {5}},\
  \bibinfo {pages} {e13150} (\bibinfo {year} {2016})}\BibitemShut {NoStop}%
\bibitem [{\citenamefont {Katira}\ \emph {et~al.}(2018)\citenamefont {Katira},
  \citenamefont {Garrahan},\ and\ \citenamefont
  {Mandadapu}}]{katira2018solvation}%
  \BibitemOpen
  \bibfield  {author} {\bibinfo {author} {\bibfnamefont {S.}~\bibnamefont
  {Katira}}, \bibinfo {author} {\bibfnamefont {J.~P.}\ \bibnamefont
  {Garrahan}},\ and\ \bibinfo {author} {\bibfnamefont {K.~K.}\ \bibnamefont
  {Mandadapu}},\ }\href {https://doi.org/10.1103/PhysRevLett.120.260602}
  {\bibfield  {journal} {\bibinfo  {journal} {Phys. Rev. Lett.}\ }\textbf
  {\bibinfo {volume} {120}},\ \bibinfo {pages} {260602} (\bibinfo {year}
  {2018})}\BibitemShut {NoStop}%
\bibitem [{\citenamefont {Fredrickson}\ and\ \citenamefont
  {Andersen}(1984)}]{fredrickson1984kinetic}%
  \BibitemOpen
  \bibfield  {author} {\bibinfo {author} {\bibfnamefont {G.~H.}\ \bibnamefont
  {Fredrickson}}\ and\ \bibinfo {author} {\bibfnamefont {H.~C.}\ \bibnamefont
  {Andersen}},\ }\href {https://doi.org/10.1103/PhysRevLett.53.1244} {\bibfield
   {journal} {\bibinfo  {journal} {Phys. Rev. Lett.}\ }\textbf {\bibinfo
  {volume} {53}},\ \bibinfo {pages} {1244} (\bibinfo {year}
  {1984})}\BibitemShut {NoStop}%
\bibitem [{\citenamefont {J{\"a}ckle}\ and\ \citenamefont
  {Eisinger}(1991)}]{jackle1991a-hierarchically}%
  \BibitemOpen
  \bibfield  {author} {\bibinfo {author} {\bibfnamefont {J.}~\bibnamefont
  {J{\"a}ckle}}\ and\ \bibinfo {author} {\bibfnamefont {S.}~\bibnamefont
  {Eisinger}},\ }\href {https://doi.org/10.1007/BF01453764} {\bibfield
  {journal} {\bibinfo  {journal} {Z. Phys, B}\ }\textbf {\bibinfo {volume}
  {84}},\ \bibinfo {pages} {115} (\bibinfo {year} {1991})}\BibitemShut
  {NoStop}%
\bibitem [{\citenamefont {Ritort}\ and\ \citenamefont
  {Sollich}(2003)}]{ritort2003glassy}%
  \BibitemOpen
  \bibfield  {author} {\bibinfo {author} {\bibfnamefont {F.}~\bibnamefont
  {Ritort}}\ and\ \bibinfo {author} {\bibfnamefont {P.}~\bibnamefont
  {Sollich}},\ }\href {https://doi.org/10.1080/0001873031000093582} {\bibfield
  {journal} {\bibinfo  {journal} {Adv. Phys.}\ }\textbf {\bibinfo {volume}
  {52}},\ \bibinfo {pages} {219} (\bibinfo {year} {2003})}\BibitemShut
  {NoStop}%
\bibitem [{\citenamefont {Chandler}\ and\ \citenamefont
  {Garrahan}(2010)}]{chandler2010dynamics}%
  \BibitemOpen
  \bibfield  {author} {\bibinfo {author} {\bibfnamefont {D.}~\bibnamefont
  {Chandler}}\ and\ \bibinfo {author} {\bibfnamefont {J.~P.}\ \bibnamefont
  {Garrahan}},\ }\href {https://doi.org/10.1146/annurev.physchem.040808.090405}
  {\bibfield  {journal} {\bibinfo  {journal} {Annu. Rev. Phys. Chem.}\ }\textbf
  {\bibinfo {volume} {61}},\ \bibinfo {pages} {191} (\bibinfo {year}
  {2010})}\BibitemShut {NoStop}%
\bibitem [{\citenamefont {Garrahan}(2018)}]{garrahan2018aspects}%
  \BibitemOpen
  \bibfield  {author} {\bibinfo {author} {\bibfnamefont {J.~P.}\ \bibnamefont
  {Garrahan}},\ }\href
  {https://doi.org/https://doi.org/10.1016/j.physa.2017.12.149} {\bibfield
  {journal} {\bibinfo  {journal} {Physica A}\ }\textbf {\bibinfo {volume}
  {504}},\ \bibinfo {pages} {130} (\bibinfo {year} {2018})}\BibitemShut
  {NoStop}%
\bibitem [{\citenamefont {Speck}(2019)}]{speck2019dynamic}%
  \BibitemOpen
  \bibfield  {author} {\bibinfo {author} {\bibfnamefont {T.}~\bibnamefont
  {Speck}},\ }\href {https://doi.org/10.1088/1742-5468/ab2ace} {\bibfield
  {journal} {\bibinfo  {journal} {J. Stat. Mech.: Theory Exp.}\ }\textbf
  {\bibinfo {volume} {2019}}\bibinfo  {number} { (8)},\ \bibinfo {pages}
  {084015}}\BibitemShut {NoStop}%
\bibitem [{\citenamefont {Berthier}\ and\ \citenamefont
  {Biroli}(2011)}]{berthier2011theoretical}%
  \BibitemOpen
\bibfield  {number} {  }\bibfield  {author} {\bibinfo {author} {\bibfnamefont
  {L.}~\bibnamefont {Berthier}}\ and\ \bibinfo {author} {\bibfnamefont
  {G.}~\bibnamefont {Biroli}},\ }\href
  {https://doi.org/10.1103/RevModPhys.83.587} {\bibfield  {journal} {\bibinfo
  {journal} {Rev. Mod. Phys.}\ }\textbf {\bibinfo {volume} {83}},\ \bibinfo
  {pages} {587} (\bibinfo {year} {2011})}\BibitemShut {NoStop}%
\bibitem [{\citenamefont {Biroli}\ and\ \citenamefont
  {Garrahan}(2013)}]{biroli2013perspective}%
  \BibitemOpen
  \bibfield  {author} {\bibinfo {author} {\bibfnamefont {G.}~\bibnamefont
  {Biroli}}\ and\ \bibinfo {author} {\bibfnamefont {J.~P.}\ \bibnamefont
  {Garrahan}},\ }\href {https://doi.org/10.1063/1.4795539} {\bibfield
  {journal} {\bibinfo  {journal} {J. Chem. Phys.}\ }\textbf {\bibinfo {volume}
  {138}},\ \bibinfo {eid} {12A301} (\bibinfo {year} {2013})}\BibitemShut
  {NoStop}%
\bibitem [{\citenamefont {Garrahan}\ and\ \citenamefont
  {Chandler}(2002)}]{garrahan2002geometrical}%
  \BibitemOpen
  \bibfield  {author} {\bibinfo {author} {\bibfnamefont {J.~P.}\ \bibnamefont
  {Garrahan}}\ and\ \bibinfo {author} {\bibfnamefont {D.}~\bibnamefont
  {Chandler}},\ }\href {https://doi.org/10.1103/PhysRevLett.89.035704}
  {\bibfield  {journal} {\bibinfo  {journal} {Phys. Rev. Lett.}\ }\textbf
  {\bibinfo {volume} {89}},\ \bibinfo {pages} {035704} (\bibinfo {year}
  {2002})}\BibitemShut {NoStop}%
\bibitem [{\citenamefont {Merolle}\ \emph {et~al.}(2005)\citenamefont
  {Merolle}, \citenamefont {Garrahan},\ and\ \citenamefont
  {Chandler}}]{merolle2005space-time}%
  \BibitemOpen
  \bibfield  {author} {\bibinfo {author} {\bibfnamefont {M.}~\bibnamefont
  {Merolle}}, \bibinfo {author} {\bibfnamefont {J.~P.}\ \bibnamefont
  {Garrahan}},\ and\ \bibinfo {author} {\bibfnamefont {D.}~\bibnamefont
  {Chandler}},\ }\href
  {https://doi.org/https://doi.org/10.1073/pnas.0504820102} {\bibfield
  {journal} {\bibinfo  {journal} {Proc. Natl. Acad. Sci. USA}\ }\textbf
  {\bibinfo {volume} {102}},\ \bibinfo {pages} {10837} (\bibinfo {year}
  {2005})}\BibitemShut {NoStop}%
\bibitem [{\citenamefont {Garrahan}\ \emph {et~al.}(2007)\citenamefont
  {Garrahan}, \citenamefont {Jack}, \citenamefont {Lecomte}, \citenamefont
  {Pitard}, \citenamefont {van Duijvendijk},\ and\ \citenamefont {van
  Wijland}}]{garrahan2007dynamical}%
  \BibitemOpen
  \bibfield  {author} {\bibinfo {author} {\bibfnamefont {J.~P.}\ \bibnamefont
  {Garrahan}}, \bibinfo {author} {\bibfnamefont {R.~L.}\ \bibnamefont {Jack}},
  \bibinfo {author} {\bibfnamefont {V.}~\bibnamefont {Lecomte}}, \bibinfo
  {author} {\bibfnamefont {E.}~\bibnamefont {Pitard}}, \bibinfo {author}
  {\bibfnamefont {K.}~\bibnamefont {van Duijvendijk}},\ and\ \bibinfo {author}
  {\bibfnamefont {F.}~\bibnamefont {van Wijland}},\ }\href
  {https://doi.org/10.1103/PhysRevLett.98.195702} {\bibfield  {journal}
  {\bibinfo  {journal} {Phys. Rev. Lett.}\ }\textbf {\bibinfo {volume} {98}},\
  \bibinfo {pages} {195702} (\bibinfo {year} {2007})}\BibitemShut {NoStop}%
\bibitem [{\citenamefont {Garrahan}\ \emph {et~al.}(2009)\citenamefont
  {Garrahan}, \citenamefont {Jack}, \citenamefont {Lecomte}, \citenamefont
  {Pitard}, \citenamefont {van Duijvendijk},\ and\ \citenamefont {van
  Wijland}}]{garrahan2009first-order}%
  \BibitemOpen
  \bibfield  {author} {\bibinfo {author} {\bibfnamefont {J.~P.}\ \bibnamefont
  {Garrahan}}, \bibinfo {author} {\bibfnamefont {R.~L.}\ \bibnamefont {Jack}},
  \bibinfo {author} {\bibfnamefont {V.}~\bibnamefont {Lecomte}}, \bibinfo
  {author} {\bibfnamefont {E.}~\bibnamefont {Pitard}}, \bibinfo {author}
  {\bibfnamefont {K.}~\bibnamefont {van Duijvendijk}},\ and\ \bibinfo {author}
  {\bibfnamefont {F.}~\bibnamefont {van Wijland}},\ }\href
  {https://doi.org/10.1088/1751-8113/42/7/075007} {\bibfield  {journal}
  {\bibinfo  {journal} {J. Phys. A}\ }\textbf {\bibinfo {volume} {42}},\
  \bibinfo {pages} {075007} (\bibinfo {year} {2009})}\BibitemShut {NoStop}%
\bibitem [{\citenamefont {Lecomte}\ \emph {et~al.}(2007)\citenamefont
  {Lecomte}, \citenamefont {Appert-Rolland},\ and\ \citenamefont {van
  Wijland}}]{lecomte2007thermodynamic}%
  \BibitemOpen
  \bibfield  {author} {\bibinfo {author} {\bibfnamefont {V.}~\bibnamefont
  {Lecomte}}, \bibinfo {author} {\bibfnamefont {C.}~\bibnamefont
  {Appert-Rolland}},\ and\ \bibinfo {author} {\bibfnamefont {F.}~\bibnamefont
  {van Wijland}},\ }\href {https://doi.org/10.1007/s10955-006-9254-0}
  {\bibfield  {journal} {\bibinfo  {journal} {J. Stat. Phys.}\ }\textbf
  {\bibinfo {volume} {127}},\ \bibinfo {pages} {51} (\bibinfo {year}
  {2007})}\BibitemShut {NoStop}%
\bibitem [{\citenamefont {Touchette}(2009)}]{touchette2009the-large}%
  \BibitemOpen
  \bibfield  {author} {\bibinfo {author} {\bibfnamefont {H.}~\bibnamefont
  {Touchette}},\ }\href
  {https://doi.org/https://doi.org/10.1016/j.physrep.2009.05.002} {\bibfield
  {journal} {\bibinfo  {journal} {Phys. Rep.}\ }\textbf {\bibinfo {volume}
  {478}},\ \bibinfo {pages} {1} (\bibinfo {year} {2009})}\BibitemShut {NoStop}%
\bibitem [{\citenamefont {Jack}(2020)}]{jack2020ergodicity}%
  \BibitemOpen
  \bibfield  {author} {\bibinfo {author} {\bibfnamefont {R.~L.}\ \bibnamefont
  {Jack}},\ }\href {https://doi.org/10.1140/epjb/e2020-100605-3} {\bibfield
  {journal} {\bibinfo  {journal} {Eur. Phys. J. B}\ }\textbf {\bibinfo {volume}
  {93}},\ \bibinfo {pages} {74} (\bibinfo {year} {2020})}\BibitemShut {NoStop}%
\bibitem [{\citenamefont {Esposito}\ \emph {et~al.}(2009)\citenamefont
  {Esposito}, \citenamefont {Harbola},\ and\ \citenamefont
  {Mukamel}}]{esposito2009nonequilibrium}%
  \BibitemOpen
  \bibfield  {author} {\bibinfo {author} {\bibfnamefont {M.}~\bibnamefont
  {Esposito}}, \bibinfo {author} {\bibfnamefont {U.}~\bibnamefont {Harbola}},\
  and\ \bibinfo {author} {\bibfnamefont {S.}~\bibnamefont {Mukamel}},\ }\href
  {https://doi.org/10.1103/RevModPhys.81.1665} {\bibfield  {journal} {\bibinfo
  {journal} {Rev. Mod. Phys.}\ }\textbf {\bibinfo {volume} {81}},\ \bibinfo
  {pages} {1665} (\bibinfo {year} {2009})}\BibitemShut {NoStop}%
\bibitem [{\citenamefont {Wolfram}(1983)}]{wolfram1983statistical}%
  \BibitemOpen
  \bibfield  {author} {\bibinfo {author} {\bibfnamefont {S.}~\bibnamefont
  {Wolfram}},\ }\href {https://doi.org/10.1103/RevModPhys.55.601} {\bibfield
  {journal} {\bibinfo  {journal} {Rev. Mod. Phys.}\ }\textbf {\bibinfo {volume}
  {55}},\ \bibinfo {pages} {601} (\bibinfo {year} {1983})}\BibitemShut
  {NoStop}%
\bibitem [{\citenamefont {Bobenko}\ \emph {et~al.}(1993)\citenamefont
  {Bobenko}, \citenamefont {Bordemann}, \citenamefont {Gunn},\ and\
  \citenamefont {Pinkall}}]{bobenko1993two}%
  \BibitemOpen
  \bibfield  {author} {\bibinfo {author} {\bibfnamefont {A.}~\bibnamefont
  {Bobenko}}, \bibinfo {author} {\bibfnamefont {M.}~\bibnamefont {Bordemann}},
  \bibinfo {author} {\bibfnamefont {C.}~\bibnamefont {Gunn}},\ and\ \bibinfo
  {author} {\bibfnamefont {U.}~\bibnamefont {Pinkall}},\ }\href
  {https://doi.org/10.1007/BF02097234} {\bibfield  {journal} {\bibinfo
  {journal} {Commun. Math. Phys.}\ }\textbf {\bibinfo {volume} {158}},\
  \bibinfo {pages} {127} (\bibinfo {year} {1993})}\BibitemShut {NoStop}%
\bibitem [{\citenamefont {Bu\v{c}a}\ \emph {et~al.}(2021)\citenamefont
  {Bu\v{c}a}, \citenamefont {Klobas},\ and\ \citenamefont
  {Prosen}}]{buca2021rule}%
  \BibitemOpen
  \bibfield  {author} {\bibinfo {author} {\bibfnamefont {B.}~\bibnamefont
  {Bu\v{c}a}}, \bibinfo {author} {\bibfnamefont {K.}~\bibnamefont {Klobas}},\
  and\ \bibinfo {author} {\bibfnamefont {T.}~\bibnamefont {Prosen}},\ }\href
  {https://doi.org/10.1088/1742-5468/ac096b} {\bibfield  {journal} {\bibinfo
  {journal} {J. Stat. Mech.: Theory Exp.}\ }\textbf {\bibinfo {volume}
  {2021}}\bibinfo  {number} { (7)},\ \bibinfo {pages} {074001}}\BibitemShut
  {NoStop}%
\bibitem [{\citenamefont {Lesanovsky}(2011)}]{lesanovsky2011many-body}%
  \BibitemOpen
\bibfield  {number} {  }\bibfield  {author} {\bibinfo {author} {\bibfnamefont
  {I.}~\bibnamefont {Lesanovsky}},\ }\href
  {https://doi.org/10.1103/PhysRevLett.106.025301} {\bibfield  {journal}
  {\bibinfo  {journal} {Phys. Rev. Lett.}\ }\textbf {\bibinfo {volume} {106}},\
  \bibinfo {pages} {025301} (\bibinfo {year} {2011})}\BibitemShut {NoStop}%
\bibitem [{\citenamefont {Lesanovsky}\ and\ \citenamefont
  {Garrahan}(2013)}]{lesanovsky2013kinetic}%
  \BibitemOpen
  \bibfield  {author} {\bibinfo {author} {\bibfnamefont {I.}~\bibnamefont
  {Lesanovsky}}\ and\ \bibinfo {author} {\bibfnamefont {J.~P.}\ \bibnamefont
  {Garrahan}},\ }\href {https://doi.org/10.1103/PhysRevLett.111.215305}
  {\bibfield  {journal} {\bibinfo  {journal} {Phys. Rev. Lett.}\ }\textbf
  {\bibinfo {volume} {111}},\ \bibinfo {pages} {215305} (\bibinfo {year}
  {2013})}\BibitemShut {NoStop}%
\bibitem [{\citenamefont {Naik}\ \emph {et~al.}(2023)\citenamefont {Naik},
  \citenamefont {Trigueros},\ and\ \citenamefont {Heyl}}]{naik2023quantum}%
  \BibitemOpen
  \bibfield  {author} {\bibinfo {author} {\bibfnamefont {V.~D.}\ \bibnamefont
  {Naik}}, \bibinfo {author} {\bibfnamefont {F.~B.}\ \bibnamefont
  {Trigueros}},\ and\ \bibinfo {author} {\bibfnamefont {M.}~\bibnamefont
  {Heyl}},\ }\href {https://arxiv.org/abs/2311.16240} {\bibfield  {journal}
  {\bibinfo  {journal} {arXiv:2311.16240}\ } (\bibinfo {year}
  {2023})}\BibitemShut {NoStop}%
\bibitem [{\citenamefont {Zhang}\ and\ \citenamefont
  {Cai}(2023)}]{zhang2023quantum}%
  \BibitemOpen
  \bibfield  {author} {\bibinfo {author} {\bibfnamefont {T.}~\bibnamefont
  {Zhang}}\ and\ \bibinfo {author} {\bibfnamefont {Z.}~\bibnamefont {Cai}},\
  }\href {https://arxiv.org/abs/2312.04115} {\bibfield  {journal} {\bibinfo
  {journal} {arXiv:2312.04115}\ } (\bibinfo {year} {2023})}\BibitemShut
  {NoStop}%
\bibitem [{\citenamefont {van Horssen}\ \emph {et~al.}(2015)\citenamefont {van
  Horssen}, \citenamefont {Levi},\ and\ \citenamefont
  {Garrahan}}]{horssen2015dynamics}%
  \BibitemOpen
  \bibfield  {author} {\bibinfo {author} {\bibfnamefont {M.}~\bibnamefont {van
  Horssen}}, \bibinfo {author} {\bibfnamefont {E.}~\bibnamefont {Levi}},\ and\
  \bibinfo {author} {\bibfnamefont {J.~P.}\ \bibnamefont {Garrahan}},\ }\href
  {https://doi.org/10.1103/PhysRevB.92.100305} {\bibfield  {journal} {\bibinfo
  {journal} {Phys. Rev. B}\ }\textbf {\bibinfo {volume} {92}},\ \bibinfo
  {pages} {100305} (\bibinfo {year} {2015})}\BibitemShut {NoStop}%
\bibitem [{\citenamefont {Smith}\ \emph {et~al.}(2017)\citenamefont {Smith},
  \citenamefont {Knolle}, \citenamefont {Kovrizhin},\ and\ \citenamefont
  {Moessner}}]{smith2017disorder-free}%
  \BibitemOpen
  \bibfield  {author} {\bibinfo {author} {\bibfnamefont {A.}~\bibnamefont
  {Smith}}, \bibinfo {author} {\bibfnamefont {J.}~\bibnamefont {Knolle}},
  \bibinfo {author} {\bibfnamefont {D.~L.}\ \bibnamefont {Kovrizhin}},\ and\
  \bibinfo {author} {\bibfnamefont {R.}~\bibnamefont {Moessner}},\ }\href
  {https://doi.org/10.1103/PhysRevLett.118.266601} {\bibfield  {journal}
  {\bibinfo  {journal} {Phys. Rev. Lett.}\ }\textbf {\bibinfo {volume} {118}},\
  \bibinfo {pages} {266601} (\bibinfo {year} {2017})}\BibitemShut {NoStop}%
\bibitem [{\citenamefont {Lan}\ \emph {et~al.}(2018)\citenamefont {Lan},
  \citenamefont {van Horssen}, \citenamefont {Powell},\ and\ \citenamefont
  {Garrahan}}]{lan2018quantum}%
  \BibitemOpen
  \bibfield  {author} {\bibinfo {author} {\bibfnamefont {Z.}~\bibnamefont
  {Lan}}, \bibinfo {author} {\bibfnamefont {M.}~\bibnamefont {van Horssen}},
  \bibinfo {author} {\bibfnamefont {S.}~\bibnamefont {Powell}},\ and\ \bibinfo
  {author} {\bibfnamefont {J.~P.}\ \bibnamefont {Garrahan}},\ }\href
  {https://doi.org/10.1103/PhysRevLett.121.040603} {\bibfield  {journal}
  {\bibinfo  {journal} {Phys. Rev. Lett.}\ }\textbf {\bibinfo {volume} {121}},\
  \bibinfo {pages} {040603} (\bibinfo {year} {2018})}\BibitemShut {NoStop}%
\bibitem [{\citenamefont {Pancotti}\ \emph {et~al.}(2020)\citenamefont
  {Pancotti}, \citenamefont {Giudice}, \citenamefont {Cirac}, \citenamefont
  {Garrahan},\ and\ \citenamefont {Ba\~nuls}}]{pancotti2020quantum}%
  \BibitemOpen
  \bibfield  {author} {\bibinfo {author} {\bibfnamefont {N.}~\bibnamefont
  {Pancotti}}, \bibinfo {author} {\bibfnamefont {G.}~\bibnamefont {Giudice}},
  \bibinfo {author} {\bibfnamefont {J.~I.}\ \bibnamefont {Cirac}}, \bibinfo
  {author} {\bibfnamefont {J.~P.}\ \bibnamefont {Garrahan}},\ and\ \bibinfo
  {author} {\bibfnamefont {M.~C.}\ \bibnamefont {Ba\~nuls}},\ }\href
  {https://doi.org/10.1103/PhysRevX.10.021051} {\bibfield  {journal} {\bibinfo
  {journal} {Phys. Rev. X}\ }\textbf {\bibinfo {volume} {10}},\ \bibinfo
  {pages} {021051} (\bibinfo {year} {2020})}\BibitemShut {NoStop}%
\bibitem [{\citenamefont {Roy}\ and\ \citenamefont
  {Lazarides}(2020)}]{roy2020strong}%
  \BibitemOpen
  \bibfield  {author} {\bibinfo {author} {\bibfnamefont {S.}~\bibnamefont
  {Roy}}\ and\ \bibinfo {author} {\bibfnamefont {A.}~\bibnamefont
  {Lazarides}},\ }\href {https://doi.org/10.1103/PhysRevResearch.2.023159}
  {\bibfield  {journal} {\bibinfo  {journal} {Phys. Rev. Res.}\ }\textbf
  {\bibinfo {volume} {2}},\ \bibinfo {pages} {023159} (\bibinfo {year}
  {2020})}\BibitemShut {NoStop}%
\bibitem [{\citenamefont {Valencia-Tortora}\ \emph {et~al.}(2022)\citenamefont
  {Valencia-Tortora}, \citenamefont {Pancotti},\ and\ \citenamefont
  {Marino}}]{valencia-tortora2022kinetically}%
  \BibitemOpen
  \bibfield  {author} {\bibinfo {author} {\bibfnamefont {R.~J.}\ \bibnamefont
  {Valencia-Tortora}}, \bibinfo {author} {\bibfnamefont {N.}~\bibnamefont
  {Pancotti}},\ and\ \bibinfo {author} {\bibfnamefont {J.}~\bibnamefont
  {Marino}},\ }\href {https://doi.org/10.1103/PRXQuantum.3.020346} {\bibfield
  {journal} {\bibinfo  {journal} {PRX Quantum}\ }\textbf {\bibinfo {volume}
  {3}},\ \bibinfo {pages} {020346} (\bibinfo {year} {2022})}\BibitemShut
  {NoStop}%
\bibitem [{\citenamefont {Turner}\ \emph {et~al.}(2018)\citenamefont {Turner},
  \citenamefont {Michailidis}, \citenamefont {Abanin}, \citenamefont {Serbyn},\
  and\ \citenamefont {Papi{\'{c}}}}]{turner2018weak}%
  \BibitemOpen
  \bibfield  {author} {\bibinfo {author} {\bibfnamefont {C.~J.}\ \bibnamefont
  {Turner}}, \bibinfo {author} {\bibfnamefont {A.~A.}\ \bibnamefont
  {Michailidis}}, \bibinfo {author} {\bibfnamefont {D.~A.}\ \bibnamefont
  {Abanin}}, \bibinfo {author} {\bibfnamefont {M.}~\bibnamefont {Serbyn}},\
  and\ \bibinfo {author} {\bibfnamefont {Z.}~\bibnamefont {Papi{\'{c}}}},\
  }\href {https://doi.org/10.1038/s41567-018-0137-5} {\bibfield  {journal}
  {\bibinfo  {journal} {Nature Phys.}\ }\textbf {\bibinfo {volume} {14}},\
  \bibinfo {pages} {745} (\bibinfo {year} {2018})}\BibitemShut {NoStop}%
\bibitem [{\citenamefont {Moudgalya}\ \emph {et~al.}(2022)\citenamefont
  {Moudgalya}, \citenamefont {Bernevig},\ and\ \citenamefont
  {Regnault}}]{moudgalya2022quantum}%
  \BibitemOpen
  \bibfield  {author} {\bibinfo {author} {\bibfnamefont {S.}~\bibnamefont
  {Moudgalya}}, \bibinfo {author} {\bibfnamefont {B.~A.}\ \bibnamefont
  {Bernevig}},\ and\ \bibinfo {author} {\bibfnamefont {N.}~\bibnamefont
  {Regnault}},\ }\href {https://doi.org/10.1088/1361-6633/ac73a0} {\bibfield
  {journal} {\bibinfo  {journal} {Rep. Prog. Phys.}\ }\textbf {\bibinfo
  {volume} {85}},\ \bibinfo {pages} {086501} (\bibinfo {year}
  {2022})}\BibitemShut {NoStop}%
\bibitem [{\citenamefont {Nandkishore}\ and\ \citenamefont
  {Hermele}(2019)}]{nandkishore2019fractons}%
  \BibitemOpen
  \bibfield  {author} {\bibinfo {author} {\bibfnamefont {R.~M.}\ \bibnamefont
  {Nandkishore}}\ and\ \bibinfo {author} {\bibfnamefont {M.}~\bibnamefont
  {Hermele}},\ }\href
  {https://doi.org/10.1146/annurev-conmatphys-031218-013604} {\bibfield
  {journal} {\bibinfo  {journal} {Annu. Rev. Condens. Matt. Phys.}\ }\textbf
  {\bibinfo {volume} {10}},\ \bibinfo {pages} {295} (\bibinfo {year}
  {2019})}\BibitemShut {NoStop}%
\bibitem [{\citenamefont {Pretko}\ \emph {et~al.}(2020)\citenamefont {Pretko},
  \citenamefont {Chen},\ and\ \citenamefont {You}}]{pretko2020fracton}%
  \BibitemOpen
  \bibfield  {author} {\bibinfo {author} {\bibfnamefont {M.}~\bibnamefont
  {Pretko}}, \bibinfo {author} {\bibfnamefont {X.}~\bibnamefont {Chen}},\ and\
  \bibinfo {author} {\bibfnamefont {Y.}~\bibnamefont {You}},\ }\href
  {https://doi.org/10.1142/S0217751X20300033} {\bibfield  {journal} {\bibinfo
  {journal} {Int. J. Mod. Phys. A}\ }\textbf {\bibinfo {volume} {35}},\
  \bibinfo {pages} {2030003} (\bibinfo {year} {2020})}\BibitemShut {NoStop}%
\bibitem [{\citenamefont {De~Fazio}\ \emph {et~al.}(2024)\citenamefont
  {De~Fazio}, \citenamefont {Garrahan},\ and\ \citenamefont
  {Klobas}}]{defazio2024exact}%
  \BibitemOpen
  \bibfield  {author} {\bibinfo {author} {\bibfnamefont {C.}~\bibnamefont
  {De~Fazio}}, \bibinfo {author} {\bibfnamefont {J.~P.}\ \bibnamefont
  {Garrahan}},\ and\ \bibinfo {author} {\bibfnamefont {K.}~\bibnamefont
  {Klobas}},\ }\Eprint {https://arxiv.org/abs/2406.17464} {arXiv:2406.17464}
  (\bibinfo {year} {2024})\BibitemShut {NoStop}%
\bibitem [{SM()}]{SM}%
  \BibitemOpen
  \href@noop {} {\bibinfo {title} {{See Supplemental Material at [URL] for more
  details.}}}\BibitemShut {Stop}%
\bibitem [{Ten()}]{TensorNetwork}%
  \BibitemOpen
  \href@noop {} {\bibinfo {title} {Tensor {N}etwork
  \url{https://tensornetwork.org}}},\ \bibinfo {note} {{A}teccessed:
  21-12-2023}\BibitemShut {NoStop}%
\bibitem [{\citenamefont {Prosen}\ and\ \citenamefont
  {Mej{\'\i}a-Monasterio}(2016)}]{prosen2016integrability}%
  \BibitemOpen
  \bibfield  {author} {\bibinfo {author} {\bibfnamefont {T.}~\bibnamefont
  {Prosen}}\ and\ \bibinfo {author} {\bibfnamefont {C.}~\bibnamefont
  {Mej{\'\i}a-Monasterio}},\ }\href
  {https://doi.org/10.1088/1751-8113/49/18/185003} {\bibfield  {journal}
  {\bibinfo  {journal} {J. Phys. A: Math. Theor.}\ }\textbf {\bibinfo {volume}
  {49}},\ \bibinfo {pages} {185003} (\bibinfo {year} {2016})}\BibitemShut
  {NoStop}%
\bibitem [{\citenamefont {Inoue}\ and\ \citenamefont
  {Takesue}(2018)}]{inoue2018two}%
  \BibitemOpen
  \bibfield  {author} {\bibinfo {author} {\bibfnamefont {A.}~\bibnamefont
  {Inoue}}\ and\ \bibinfo {author} {\bibfnamefont {S.}~\bibnamefont
  {Takesue}},\ }\href {https://doi.org/10.1088/1751-8121/aadc29} {\bibfield
  {journal} {\bibinfo  {journal} {J. Phys. A: Math. Theor.}\ }\textbf {\bibinfo
  {volume} {51}},\ \bibinfo {pages} {425001} (\bibinfo {year}
  {2018})}\BibitemShut {NoStop}%
\bibitem [{\citenamefont {Prosen}\ and\ \citenamefont
  {Bu{\v{c}}a}(2017)}]{prosen2017exact}%
  \BibitemOpen
  \bibfield  {author} {\bibinfo {author} {\bibfnamefont {T.}~\bibnamefont
  {Prosen}}\ and\ \bibinfo {author} {\bibfnamefont {B.}~\bibnamefont
  {Bu{\v{c}}a}},\ }\href {https://doi.org/10.1088/1751-8121/aa85a3} {\bibfield
  {journal} {\bibinfo  {journal} {J. Phys. A: Math. Theor.}\ }\textbf {\bibinfo
  {volume} {50}},\ \bibinfo {pages} {395002} (\bibinfo {year}
  {2017})}\BibitemShut {NoStop}%
\bibitem [{\citenamefont {Gopalakrishnan}(2018)}]{gopalakrishnan2018operator}%
  \BibitemOpen
  \bibfield  {author} {\bibinfo {author} {\bibfnamefont {S.}~\bibnamefont
  {Gopalakrishnan}},\ }\href {https://doi.org/10.1103/PhysRevB.98.060302}
  {\bibfield  {journal} {\bibinfo  {journal} {Phys. Rev. B}\ }\textbf {\bibinfo
  {volume} {98}},\ \bibinfo {pages} {060302(R)} (\bibinfo {year}
  {2018})}\BibitemShut {NoStop}%
\bibitem [{\citenamefont {Bu{\v c}a}\ \emph {et~al.}(2019)\citenamefont {Bu{\v
  c}a}, \citenamefont {Garrahan}, \citenamefont {Prosen},\ and\ \citenamefont
  {Vanicat}}]{buca2019exact}%
  \BibitemOpen
  \bibfield  {author} {\bibinfo {author} {\bibfnamefont {B.}~\bibnamefont
  {Bu{\v c}a}}, \bibinfo {author} {\bibfnamefont {J.~P.}\ \bibnamefont
  {Garrahan}}, \bibinfo {author} {\bibfnamefont {T.}~\bibnamefont {Prosen}},\
  and\ \bibinfo {author} {\bibfnamefont {M.}~\bibnamefont {Vanicat}},\ }\href
  {https://doi.org/10.1103/PhysRevE.100.020103} {\bibfield  {journal} {\bibinfo
   {journal} {Phys. Rev. E}\ }\textbf {\bibinfo {volume} {100}},\ \bibinfo
  {pages} {020103(R)} (\bibinfo {year} {2019})}\BibitemShut {NoStop}%
\bibitem [{\citenamefont {Klobas}\ and\ \citenamefont
  {Prosen}(2020)}]{klobas2020space}%
  \BibitemOpen
  \bibfield  {author} {\bibinfo {author} {\bibfnamefont {K.}~\bibnamefont
  {Klobas}}\ and\ \bibinfo {author} {\bibfnamefont {T.}~\bibnamefont
  {Prosen}},\ }\href {https://doi.org/10.21468/SciPostPhysCore.2.2.010}
  {\bibfield  {journal} {\bibinfo  {journal} {SciPost Phys. Core}\ }\textbf
  {\bibinfo {volume} {2}},\ \bibinfo {pages} {10} (\bibinfo {year}
  {2020})}\BibitemShut {NoStop}%
\bibitem [{\citenamefont {Klobas}\ \emph {et~al.}(2019)\citenamefont {Klobas},
  \citenamefont {Medenjak}, \citenamefont {Prosen},\ and\ \citenamefont
  {Vanicat}}]{klobas2019time}%
  \BibitemOpen
  \bibfield  {author} {\bibinfo {author} {\bibfnamefont {K.}~\bibnamefont
  {Klobas}}, \bibinfo {author} {\bibfnamefont {M.}~\bibnamefont {Medenjak}},
  \bibinfo {author} {\bibfnamefont {T.}~\bibnamefont {Prosen}},\ and\ \bibinfo
  {author} {\bibfnamefont {M.}~\bibnamefont {Vanicat}},\ }\href
  {https://doi.org/10.1007/s00220-019-03494-5} {\bibfield  {journal} {\bibinfo
  {journal} {Commun. Math. Phys.}\ }\textbf {\bibinfo {volume} {371}},\
  \bibinfo {pages} {651} (\bibinfo {year} {2019})}\BibitemShut {NoStop}%
\bibitem [{\citenamefont {Alba}\ \emph {et~al.}(2019)\citenamefont {Alba},
  \citenamefont {Dubail},\ and\ \citenamefont {Medenjak}}]{alba2019operator}%
  \BibitemOpen
  \bibfield  {author} {\bibinfo {author} {\bibfnamefont {V.}~\bibnamefont
  {Alba}}, \bibinfo {author} {\bibfnamefont {J.}~\bibnamefont {Dubail}},\ and\
  \bibinfo {author} {\bibfnamefont {M.}~\bibnamefont {Medenjak}},\ }\href
  {https://doi.org/10.1103/PhysRevLett.122.250603} {\bibfield  {journal}
  {\bibinfo  {journal} {Phys. Rev. Lett.}\ }\textbf {\bibinfo {volume} {122}},\
  \bibinfo {pages} {250603} (\bibinfo {year} {2019})}\BibitemShut {NoStop}%
\bibitem [{\citenamefont {Klobas}\ \emph {et~al.}(2020)\citenamefont {Klobas},
  \citenamefont {Vanicat}, \citenamefont {Garrahan},\ and\ \citenamefont
  {Prosen}}]{klobas2020matrix}%
  \BibitemOpen
  \bibfield  {author} {\bibinfo {author} {\bibfnamefont {K.}~\bibnamefont
  {Klobas}}, \bibinfo {author} {\bibfnamefont {M.}~\bibnamefont {Vanicat}},
  \bibinfo {author} {\bibfnamefont {J.~P.}\ \bibnamefont {Garrahan}},\ and\
  \bibinfo {author} {\bibfnamefont {T.}~\bibnamefont {Prosen}},\ }\href
  {https://doi.org/10.1088/1751-8121/ab8c62} {\bibfield  {journal} {\bibinfo
  {journal} {J. Phys. A: Math. Theor.}\ }\textbf {\bibinfo {volume} {53}},\
  \bibinfo {pages} {335001} (\bibinfo {year} {2020})}\BibitemShut {NoStop}%
\bibitem [{\citenamefont {Wilkinson}\ \emph {et~al.}(2020)\citenamefont
  {Wilkinson}, \citenamefont {Klobas}, \citenamefont {Prosen},\ and\
  \citenamefont {Garrahan}}]{wilkinson2020exact}%
  \BibitemOpen
  \bibfield  {author} {\bibinfo {author} {\bibfnamefont {J.~W.~P.}\
  \bibnamefont {Wilkinson}}, \bibinfo {author} {\bibfnamefont {K.}~\bibnamefont
  {Klobas}}, \bibinfo {author} {\bibfnamefont {T.}~\bibnamefont {Prosen}},\
  and\ \bibinfo {author} {\bibfnamefont {J.~P.}\ \bibnamefont {Garrahan}},\
  }\href {https://doi.org/10.1103/PhysRevE.102.062107} {\bibfield  {journal}
  {\bibinfo  {journal} {Phys. Rev. E}\ }\textbf {\bibinfo {volume} {102}},\
  \bibinfo {pages} {062107} (\bibinfo {year} {2020})}\BibitemShut {NoStop}%
\bibitem [{\citenamefont {Iadecola}\ and\ \citenamefont
  {Vijay}(2020)}]{iadecola2020nonergodic}%
  \BibitemOpen
  \bibfield  {author} {\bibinfo {author} {\bibfnamefont {T.}~\bibnamefont
  {Iadecola}}\ and\ \bibinfo {author} {\bibfnamefont {S.}~\bibnamefont
  {Vijay}},\ }\href {https://doi.org/10.1103/PhysRevB.102.180302} {\bibfield
  {journal} {\bibinfo  {journal} {Phys. Rev. B}\ }\textbf {\bibinfo {volume}
  {102}},\ \bibinfo {pages} {180302} (\bibinfo {year} {2020})}\BibitemShut
  {NoStop}%
\bibitem [{\citenamefont {Gopalakrishnan}\ and\ \citenamefont
  {Zakirov}(2018)}]{gopalakrishnan2018facilitated}%
  \BibitemOpen
  \bibfield  {author} {\bibinfo {author} {\bibfnamefont {S.}~\bibnamefont
  {Gopalakrishnan}}\ and\ \bibinfo {author} {\bibfnamefont {B.}~\bibnamefont
  {Zakirov}},\ }\href {https://doi.org/10.1088/2058-9565/aad759} {\bibfield
  {journal} {\bibinfo  {journal} {Quantum Sci. Technol.}\ }\textbf {\bibinfo
  {volume} {3}},\ \bibinfo {pages} {044004} (\bibinfo {year}
  {2018})}\BibitemShut {NoStop}%
\bibitem [{\citenamefont {Gombor}\ and\ \citenamefont
  {Pozsgay}(2021)}]{gombor2021integrable}%
  \BibitemOpen
  \bibfield  {author} {\bibinfo {author} {\bibfnamefont {T.}~\bibnamefont
  {Gombor}}\ and\ \bibinfo {author} {\bibfnamefont {B.}~\bibnamefont
  {Pozsgay}},\ }\href {https://doi.org/10.1103/PhysRevE.104.054123} {\bibfield
  {journal} {\bibinfo  {journal} {Phys. Rev. E}\ }\textbf {\bibinfo {volume}
  {104}},\ \bibinfo {pages} {054123} (\bibinfo {year} {2021})}\BibitemShut
  {NoStop}%
\bibitem [{\citenamefont {Wilkinson}\ \emph {et~al.}(2022)\citenamefont
  {Wilkinson}, \citenamefont {Prosen},\ and\ \citenamefont
  {Garrahan}}]{wilkinson2022exact}%
  \BibitemOpen
  \bibfield  {author} {\bibinfo {author} {\bibfnamefont {J.~W.~P.}\
  \bibnamefont {Wilkinson}}, \bibinfo {author} {\bibfnamefont {T.}~\bibnamefont
  {Prosen}},\ and\ \bibinfo {author} {\bibfnamefont {J.~P.}\ \bibnamefont
  {Garrahan}},\ }\href {https://doi.org/10.1103/PhysRevE.105.034124} {\bibfield
   {journal} {\bibinfo  {journal} {Phys. Rev. E}\ }\textbf {\bibinfo {volume}
  {105}},\ \bibinfo {pages} {034124} (\bibinfo {year} {2022})}\BibitemShut
  {NoStop}%
\bibitem [{\citenamefont {Deger}\ \emph
  {et~al.}(2022{\natexlab{a}})\citenamefont {Deger}, \citenamefont {Roy},\ and\
  \citenamefont {Lazarides}}]{deger2022arresting}%
  \BibitemOpen
  \bibfield  {author} {\bibinfo {author} {\bibfnamefont {A.}~\bibnamefont
  {Deger}}, \bibinfo {author} {\bibfnamefont {S.}~\bibnamefont {Roy}},\ and\
  \bibinfo {author} {\bibfnamefont {A.}~\bibnamefont {Lazarides}},\ }\href
  {https://doi.org/10.1103/PhysRevLett.129.160601} {\bibfield  {journal}
  {\bibinfo  {journal} {Phys. Rev. Lett.}\ }\textbf {\bibinfo {volume} {129}},\
  \bibinfo {pages} {160601} (\bibinfo {year} {2022}{\natexlab{a}})}\BibitemShut
  {NoStop}%
\bibitem [{\citenamefont {Deger}\ \emph
  {et~al.}(2022{\natexlab{b}})\citenamefont {Deger}, \citenamefont
  {Lazarides},\ and\ \citenamefont {Roy}}]{deger2022constrained}%
  \BibitemOpen
  \bibfield  {author} {\bibinfo {author} {\bibfnamefont {A.}~\bibnamefont
  {Deger}}, \bibinfo {author} {\bibfnamefont {A.}~\bibnamefont {Lazarides}},\
  and\ \bibinfo {author} {\bibfnamefont {S.}~\bibnamefont {Roy}},\ }\href
  {https://doi.org/10.1103/PhysRevLett.129.190601} {\bibfield  {journal}
  {\bibinfo  {journal} {Phys. Rev. Lett.}\ }\textbf {\bibinfo {volume} {129}},\
  \bibinfo {pages} {190601} (\bibinfo {year} {2022}{\natexlab{b}})}\BibitemShut
  {NoStop}%
\bibitem [{Note1()}]{Note1}%
  \BibitemOpen
  \bibinfo {note} {Note that the vector $\bra *{\protect \tilde {P}_x}$ is not
  necessarily a valid probability vector since it does not need to be
  normalized.}\BibitemShut {Stop}%
\bibitem [{Note2()}]{Note2}%
  \BibitemOpen
  \bibinfo {note} {In the quantum setup, analogous objects --- the left and
  right fixed points of the space transfer-matrix --- are referred-to also as
  \protect \emph {influence matrices}~\cite {lerose2021influence}. These have
  recently been understood to provide convenient numerical~\cite
  {banuls2009matrix,muller2012tensor,lerose2021influence,sonner2021influence,lerose2021scaling,friasperez2022lightcone},
  and analytical tools~\cite
  {bertini2019exact,bertini2019entanglement,piroli2020exact,klobas2021exact,klobas2021exactrelaxation,klobas2021entanglement,giudice2022temporal,bertini2022entanglement,bertini2022growth,foligno2023temporal}
  to study quantum many-body dynamics.}\BibitemShut {Stop}%
\bibitem [{\citenamefont {Bertini}\ \emph
  {et~al.}(2019{\natexlab{a}})\citenamefont {Bertini}, \citenamefont {Kos},\
  and\ \citenamefont {Prosen}}]{bertini2019exact}%
  \BibitemOpen
  \bibfield  {author} {\bibinfo {author} {\bibfnamefont {B.}~\bibnamefont
  {Bertini}}, \bibinfo {author} {\bibfnamefont {P.}~\bibnamefont {Kos}},\ and\
  \bibinfo {author} {\bibfnamefont {T.}~\bibnamefont {Prosen}},\ }\href
  {https://doi.org/10.1103/PhysRevLett.123.210601} {\bibfield  {journal}
  {\bibinfo  {journal} {Phys. Rev. Lett.}\ }\textbf {\bibinfo {volume} {123}},\
  \bibinfo {pages} {210601} (\bibinfo {year} {2019}{\natexlab{a}})}\BibitemShut
  {NoStop}%
\bibitem [{\citenamefont {Kos}\ \emph {et~al.}(2021)\citenamefont {Kos},
  \citenamefont {Bertini},\ and\ \citenamefont {Prosen}}]{kos2021correlations}%
  \BibitemOpen
  \bibfield  {author} {\bibinfo {author} {\bibfnamefont {P.}~\bibnamefont
  {Kos}}, \bibinfo {author} {\bibfnamefont {B.}~\bibnamefont {Bertini}},\ and\
  \bibinfo {author} {\bibfnamefont {T.}~\bibnamefont {Prosen}},\ }\href
  {https://doi.org/10.1103/PhysRevX.11.011022} {\bibfield  {journal} {\bibinfo
  {journal} {Phys. Rev. X}\ }\textbf {\bibinfo {volume} {11}},\ \bibinfo
  {pages} {011022} (\bibinfo {year} {2021})}\BibitemShut {NoStop}%
\bibitem [{\citenamefont {Kos}\ and\ \citenamefont
  {Styliaris}(2023)}]{kos2022circuits}%
  \BibitemOpen
  \bibfield  {author} {\bibinfo {author} {\bibfnamefont {P.}~\bibnamefont
  {Kos}}\ and\ \bibinfo {author} {\bibfnamefont {G.}~\bibnamefont
  {Styliaris}},\ }\href
  {https://doi.org/https://doi.org/10.22331/q-2023-05-24-1020} {\bibfield
  {journal} {\bibinfo  {journal} {Quantum}\ }\textbf {\bibinfo {volume} {7}},\
  \bibinfo {pages} {1020} (\bibinfo {year} {2023})}\BibitemShut {NoStop}%
\bibitem [{\citenamefont {Bertini}\ \emph {et~al.}(2018)\citenamefont
  {Bertini}, \citenamefont {Kos},\ and\ \citenamefont
  {Prosen}}]{bertini2018exact}%
  \BibitemOpen
  \bibfield  {author} {\bibinfo {author} {\bibfnamefont {B.}~\bibnamefont
  {Bertini}}, \bibinfo {author} {\bibfnamefont {P.}~\bibnamefont {Kos}},\ and\
  \bibinfo {author} {\bibfnamefont {T.}~\bibnamefont {Prosen}},\ }\href
  {https://doi.org/10.1103/physrevlett.121.264101} {\bibfield  {journal}
  {\bibinfo  {journal} {Phys. Rev. Lett.}\ }\textbf {\bibinfo {volume} {121}},\
  \bibinfo {pages} {264101} (\bibinfo {year} {2018})}\BibitemShut {NoStop}%
\bibitem [{\citenamefont {Bertini}\ \emph
  {et~al.}(2019{\natexlab{b}})\citenamefont {Bertini}, \citenamefont {Kos},\
  and\ \citenamefont {Prosen}}]{bertini2019entanglement}%
  \BibitemOpen
  \bibfield  {author} {\bibinfo {author} {\bibfnamefont {B.}~\bibnamefont
  {Bertini}}, \bibinfo {author} {\bibfnamefont {P.}~\bibnamefont {Kos}},\ and\
  \bibinfo {author} {\bibfnamefont {T.}~\bibnamefont {Prosen}},\ }\href
  {https://doi.org/10.1103/PhysRevX.9.021033} {\bibfield  {journal} {\bibinfo
  {journal} {Phys. Rev. X}\ }\textbf {\bibinfo {volume} {9}},\ \bibinfo {pages}
  {021033} (\bibinfo {year} {2019}{\natexlab{b}})}\BibitemShut {NoStop}%
\bibitem [{\citenamefont {Bertini}\ \emph
  {et~al.}(2020{\natexlab{a}})\citenamefont {Bertini}, \citenamefont {Kos},\
  and\ \citenamefont {Prosen}}]{bertini2020operatorI}%
  \BibitemOpen
  \bibfield  {author} {\bibinfo {author} {\bibfnamefont {B.}~\bibnamefont
  {Bertini}}, \bibinfo {author} {\bibfnamefont {P.}~\bibnamefont {Kos}},\ and\
  \bibinfo {author} {\bibfnamefont {T.}~\bibnamefont {Prosen}},\ }\href
  {https://doi.org/10.21468/SciPostPhys.8.4.067} {\bibfield  {journal}
  {\bibinfo  {journal} {SciPost Phys.}\ }\textbf {\bibinfo {volume} {8}},\
  \bibinfo {pages} {067} (\bibinfo {year} {2020}{\natexlab{a}})}\BibitemShut
  {NoStop}%
\bibitem [{\citenamefont {Bertini}\ \emph
  {et~al.}(2020{\natexlab{b}})\citenamefont {Bertini}, \citenamefont {Kos},\
  and\ \citenamefont {Prosen}}]{bertini2020operatorII}%
  \BibitemOpen
  \bibfield  {author} {\bibinfo {author} {\bibfnamefont {B.}~\bibnamefont
  {Bertini}}, \bibinfo {author} {\bibfnamefont {P.}~\bibnamefont {Kos}},\ and\
  \bibinfo {author} {\bibfnamefont {T.}~\bibnamefont {Prosen}},\ }\href
  {https://doi.org/10.21468/SciPostPhys.8.4.068} {\bibfield  {journal}
  {\bibinfo  {journal} {SciPost Phys.}\ }\textbf {\bibinfo {volume} {8}},\
  \bibinfo {pages} {068} (\bibinfo {year} {2020}{\natexlab{b}})}\BibitemShut
  {NoStop}%
\bibitem [{\citenamefont {Piroli}\ \emph {et~al.}(2020)\citenamefont {Piroli},
  \citenamefont {Bertini}, \citenamefont {Cirac},\ and\ \citenamefont
  {Prosen}}]{piroli2020exact}%
  \BibitemOpen
  \bibfield  {author} {\bibinfo {author} {\bibfnamefont {L.}~\bibnamefont
  {Piroli}}, \bibinfo {author} {\bibfnamefont {B.}~\bibnamefont {Bertini}},
  \bibinfo {author} {\bibfnamefont {J.~I.}\ \bibnamefont {Cirac}},\ and\
  \bibinfo {author} {\bibfnamefont {T.}~\bibnamefont {Prosen}},\ }\href
  {https://doi.org/10.1103/PhysRevB.101.094304} {\bibfield  {journal} {\bibinfo
   {journal} {Phys. Rev. B}\ }\textbf {\bibinfo {volume} {101}},\ \bibinfo
  {pages} {094304} (\bibinfo {year} {2020})}\BibitemShut {NoStop}%
\bibitem [{\citenamefont {Claeys}\ and\ \citenamefont
  {Lamacraft}(2020)}]{claeys2020maximum}%
  \BibitemOpen
  \bibfield  {author} {\bibinfo {author} {\bibfnamefont {P.~W.}\ \bibnamefont
  {Claeys}}\ and\ \bibinfo {author} {\bibfnamefont {A.}~\bibnamefont
  {Lamacraft}},\ }\href {https://doi.org/10.1103/PhysRevResearch.2.033032}
  {\bibfield  {journal} {\bibinfo  {journal} {Phys. Rev. Research}\ }\textbf
  {\bibinfo {volume} {2}},\ \bibinfo {pages} {033032} (\bibinfo {year}
  {2020})}\BibitemShut {NoStop}%
\bibitem [{\citenamefont {Claeys}\ and\ \citenamefont
  {Lamacraft}(2021)}]{claeys2021ergodic}%
  \BibitemOpen
  \bibfield  {author} {\bibinfo {author} {\bibfnamefont {P.~W.}\ \bibnamefont
  {Claeys}}\ and\ \bibinfo {author} {\bibfnamefont {A.}~\bibnamefont
  {Lamacraft}},\ }\href {https://doi.org/10.1103/PhysRevLett.126.100603}
  {\bibfield  {journal} {\bibinfo  {journal} {Phys. Rev. Lett.}\ }\textbf
  {\bibinfo {volume} {126}},\ \bibinfo {pages} {100603} (\bibinfo {year}
  {2021})}\BibitemShut {NoStop}%
\bibitem [{\citenamefont {Bertini}\ \emph {et~al.}(2021)\citenamefont
  {Bertini}, \citenamefont {Kos},\ and\ \citenamefont
  {Prosen}}]{bertini2021random}%
  \BibitemOpen
  \bibfield  {author} {\bibinfo {author} {\bibfnamefont {B.}~\bibnamefont
  {Bertini}}, \bibinfo {author} {\bibfnamefont {P.}~\bibnamefont {Kos}},\ and\
  \bibinfo {author} {\bibfnamefont {T.}~\bibnamefont {Prosen}},\ }\href
  {https://doi.org/10.1007/s00220-021-04139-2} {\bibfield  {journal} {\bibinfo
  {journal} {Commun. Math. Phys.}\ }\textbf {\bibinfo {volume} {387}},\
  \bibinfo {pages} {597} (\bibinfo {year} {2021})}\BibitemShut {NoStop}%
\bibitem [{\citenamefont {Jonay}\ \emph {et~al.}(2021)\citenamefont {Jonay},
  \citenamefont {Khemani},\ and\ \citenamefont
  {Ippoliti}}]{jonay2021triunitary}%
  \BibitemOpen
  \bibfield  {author} {\bibinfo {author} {\bibfnamefont {C.}~\bibnamefont
  {Jonay}}, \bibinfo {author} {\bibfnamefont {V.}~\bibnamefont {Khemani}},\
  and\ \bibinfo {author} {\bibfnamefont {M.}~\bibnamefont {Ippoliti}},\ }\href
  {https://doi.org/10.1103/PhysRevResearch.3.043046} {\bibfield  {journal}
  {\bibinfo  {journal} {Phys. Rev. Res.}\ }\textbf {\bibinfo {volume} {3}},\
  \bibinfo {pages} {043046} (\bibinfo {year} {2021})}\BibitemShut {NoStop}%
\bibitem [{\citenamefont {Kasim}\ and\ \citenamefont
  {Prosen}(2022)}]{kasim2022dual}%
  \BibitemOpen
  \bibfield  {author} {\bibinfo {author} {\bibfnamefont {Y.}~\bibnamefont
  {Kasim}}\ and\ \bibinfo {author} {\bibfnamefont {T.}~\bibnamefont {Prosen}},\
  }\bibfield  {journal} {\bibinfo  {journal} {J. Phys. A: Math. Theor.}\ }\href
  {https://doi.org/10.1088/1751-8121/acb1e0} {10.1088/1751-8121/acb1e0}
  (\bibinfo {year} {2022})\BibitemShut {NoStop}%
\bibitem [{\citenamefont {Suzuki}\ \emph {et~al.}(2022)\citenamefont {Suzuki},
  \citenamefont {Mitarai},\ and\ \citenamefont
  {Fujii}}]{suzuki2022computational}%
  \BibitemOpen
  \bibfield  {author} {\bibinfo {author} {\bibfnamefont {R.}~\bibnamefont
  {Suzuki}}, \bibinfo {author} {\bibfnamefont {K.}~\bibnamefont {Mitarai}},\
  and\ \bibinfo {author} {\bibfnamefont {K.}~\bibnamefont {Fujii}},\ }\href
  {https://doi.org/10.22331/q-2022-01-24-631} {\bibfield  {journal} {\bibinfo
  {journal} {Quantum}\ }\textbf {\bibinfo {volume} {6}},\ \bibinfo {pages}
  {631} (\bibinfo {year} {2022})}\BibitemShut {NoStop}%
\bibitem [{\citenamefont {Foligno}\ and\ \citenamefont
  {Bertini}(2023)}]{foligno2022growth}%
  \BibitemOpen
  \bibfield  {author} {\bibinfo {author} {\bibfnamefont {A.}~\bibnamefont
  {Foligno}}\ and\ \bibinfo {author} {\bibfnamefont {B.}~\bibnamefont
  {Bertini}},\ }\href {https://doi.org/10.1103/PhysRevB.107.174311} {\bibfield
  {journal} {\bibinfo  {journal} {Phys. Rev. B}\ }\textbf {\bibinfo {volume}
  {107}},\ \bibinfo {pages} {174311} (\bibinfo {year} {2023})}\BibitemShut
  {NoStop}%
\bibitem [{\citenamefont {Foligno}\ \emph {et~al.}(2023)\citenamefont
  {Foligno}, \citenamefont {Zhou},\ and\ \citenamefont
  {Bertini}}]{foligno2023temporal}%
  \BibitemOpen
  \bibfield  {author} {\bibinfo {author} {\bibfnamefont {A.}~\bibnamefont
  {Foligno}}, \bibinfo {author} {\bibfnamefont {T.}~\bibnamefont {Zhou}},\ and\
  \bibinfo {author} {\bibfnamefont {B.}~\bibnamefont {Bertini}},\ }\href
  {https://doi.org/10.1103/PhysRevX.13.041008} {\bibfield  {journal} {\bibinfo
  {journal} {Phys. Rev. X}\ }\textbf {\bibinfo {volume} {13}},\ \bibinfo
  {pages} {041008} (\bibinfo {year} {2023})}\BibitemShut {NoStop}%
\bibitem [{\citenamefont {Rampp}\ \emph {et~al.}(2023)\citenamefont {Rampp},
  \citenamefont {Moessner},\ and\ \citenamefont {Claeys}}]{rampp2023dual}%
  \BibitemOpen
  \bibfield  {author} {\bibinfo {author} {\bibfnamefont {M.~A.}\ \bibnamefont
  {Rampp}}, \bibinfo {author} {\bibfnamefont {R.}~\bibnamefont {Moessner}},\
  and\ \bibinfo {author} {\bibfnamefont {P.~W.}\ \bibnamefont {Claeys}},\
  }\href {https://doi.org/10.1103/PhysRevLett.130.130402} {\bibfield  {journal}
  {\bibinfo  {journal} {Phys. Rev. Lett.}\ }\textbf {\bibinfo {volume} {130}},\
  \bibinfo {pages} {130402} (\bibinfo {year} {2023})}\BibitemShut {NoStop}%
\bibitem [{\citenamefont {Lerose}\ \emph
  {et~al.}(2021{\natexlab{a}})\citenamefont {Lerose}, \citenamefont {Sonner},\
  and\ \citenamefont {Abanin}}]{lerose2021influence}%
  \BibitemOpen
  \bibfield  {author} {\bibinfo {author} {\bibfnamefont {A.}~\bibnamefont
  {Lerose}}, \bibinfo {author} {\bibfnamefont {M.}~\bibnamefont {Sonner}},\
  and\ \bibinfo {author} {\bibfnamefont {D.~A.}\ \bibnamefont {Abanin}},\
  }\href {https://doi.org/10.1103/PhysRevX.11.021040} {\bibfield  {journal}
  {\bibinfo  {journal} {Phys. Rev. X}\ }\textbf {\bibinfo {volume} {11}},\
  \bibinfo {pages} {021040} (\bibinfo {year} {2021}{\natexlab{a}})}\BibitemShut
  {NoStop}%
\bibitem [{\citenamefont {Jack}\ \emph {et~al.}(2020)\citenamefont {Jack},
  \citenamefont {Nemoto},\ and\ \citenamefont {Lecomte}}]{jack2020dynamical}%
  \BibitemOpen
  \bibfield  {author} {\bibinfo {author} {\bibfnamefont {R.~L.}\ \bibnamefont
  {Jack}}, \bibinfo {author} {\bibfnamefont {T.}~\bibnamefont {Nemoto}},\ and\
  \bibinfo {author} {\bibfnamefont {V.}~\bibnamefont {Lecomte}},\ }\href
  {https://doi.org/10.1088/1742-5468/ab7af6} {\bibfield  {journal} {\bibinfo
  {journal} {J. Stat. Mech.: Theory Exp.}\ }\textbf {\bibinfo {volume}
  {2020}}\bibinfo  {number} { (5)},\ \bibinfo {pages} {053204}}\BibitemShut
  {NoStop}%
\bibitem [{\citenamefont {Maes}(2020)}]{maes2020frenesy}%
  \BibitemOpen
\bibfield  {number} {  }\bibfield  {author} {\bibinfo {author} {\bibfnamefont
  {C.}~\bibnamefont {Maes}},\ }\href
  {https://doi.org/https://doi.org/10.1016/j.physrep.2020.01.002} {\bibfield
  {journal} {\bibinfo  {journal} {Phys. Rep.}\ }\textbf {\bibinfo {volume}
  {850}},\ \bibinfo {pages} {1} (\bibinfo {year} {2020})}\BibitemShut {NoStop}%
\bibitem [{Note3()}]{Note3}%
  \BibitemOpen
  \bibinfo {note} {Note that $P_{\square }(l,t)$ scales with $P_{\protect \rm
  inact}(l-\protect \frac {1}{2},t-\protect \frac {1}{2})$ due to the way that
  the sizes are defined in both cases: in the former the size corresponds to
  the number of unit cells (i.e., twice the number of sites), while in the
  latter the number of time/space steps. This subtlety arises because of the
  way that the constraints are applied --- either to sites or gates. See \cite
  {SM} for an illustration.}\BibitemShut {Stop}%
\bibitem [{Note4()}]{Note4}%
  \BibitemOpen
  \bibinfo {note} {In the deterministic model, similar perimeter scaling is
  obtained for the free-energies of space-time regions corresponding to allowed
  trajectories of the circuit. In contrast, only regions of $0$s show this
  scaling in the stochastic case. See \cite {SM} for details.}\BibitemShut
  {Stop}%
\bibitem [{\citenamefont {Willard}\ and\ \citenamefont
  {Chandler}(2008)}]{willard2008the-role}%
  \BibitemOpen
  \bibfield  {author} {\bibinfo {author} {\bibfnamefont {A.~P.}\ \bibnamefont
  {Willard}}\ and\ \bibinfo {author} {\bibfnamefont {D.}~\bibnamefont
  {Chandler}},\ }\href {https://doi.org/10.1021/jp077186+} {\bibfield
  {journal} {\bibinfo  {journal} {J. Phys. Chem. B}\ }\textbf {\bibinfo
  {volume} {112}},\ \bibinfo {pages} {6187} (\bibinfo {year}
  {2008})}\BibitemShut {NoStop}%
\bibitem [{\citenamefont {Trotter}(1959)}]{trotter1959product}%
  \BibitemOpen
  \bibfield  {author} {\bibinfo {author} {\bibfnamefont {H.~F.}\ \bibnamefont
  {Trotter}},\ }\href {https://doi.org/10.1090/S0002-9939-1959-0108732-6}
  {\bibfield  {journal} {\bibinfo  {journal} {Proc. Am. Math. Soc.}\ }\textbf
  {\bibinfo {volume} {10}},\ \bibinfo {pages} {545} (\bibinfo {year}
  {1959})}\BibitemShut {NoStop}%
\bibitem [{\citenamefont {Suzuki}(1991)}]{suzuki1991general}%
  \BibitemOpen
  \bibfield  {author} {\bibinfo {author} {\bibfnamefont {M.}~\bibnamefont
  {Suzuki}},\ }\href {https://doi.org/10.1063/1.529425} {\bibfield  {journal}
  {\bibinfo  {journal} {J. Math. Phys.}\ }\textbf {\bibinfo {volume} {32}},\
  \bibinfo {pages} {400} (\bibinfo {year} {1991})}\BibitemShut {NoStop}%
\bibitem [{\citenamefont {Osborne}(2006)}]{osborne2006efficient}%
  \BibitemOpen
  \bibfield  {author} {\bibinfo {author} {\bibfnamefont {T.~J.}\ \bibnamefont
  {Osborne}},\ }\href {https://doi.org/10.1103/PhysRevLett.97.157202}
  {\bibfield  {journal} {\bibinfo  {journal} {Phys. Rev. Lett.}\ }\textbf
  {\bibinfo {volume} {97}},\ \bibinfo {pages} {157202} (\bibinfo {year}
  {2006})}\BibitemShut {NoStop}%
\bibitem [{\citenamefont {Nahum}\ \emph {et~al.}(2017)\citenamefont {Nahum},
  \citenamefont {Ruhman}, \citenamefont {Vijay},\ and\ \citenamefont
  {Haah}}]{nahum2017quantum}%
  \BibitemOpen
  \bibfield  {author} {\bibinfo {author} {\bibfnamefont {A.}~\bibnamefont
  {Nahum}}, \bibinfo {author} {\bibfnamefont {J.}~\bibnamefont {Ruhman}},
  \bibinfo {author} {\bibfnamefont {S.}~\bibnamefont {Vijay}},\ and\ \bibinfo
  {author} {\bibfnamefont {J.}~\bibnamefont {Haah}},\ }\href
  {https://doi.org/10.1103/PhysRevX.7.031016} {\bibfield  {journal} {\bibinfo
  {journal} {Phys. Rev. X}\ }\textbf {\bibinfo {volume} {7}},\ \bibinfo {pages}
  {031016} (\bibinfo {year} {2017})}\BibitemShut {NoStop}%
\bibitem [{\citenamefont {Nahum}\ \emph {et~al.}(2018)\citenamefont {Nahum},
  \citenamefont {Vijay},\ and\ \citenamefont {Haah}}]{nahum2018operator}%
  \BibitemOpen
  \bibfield  {author} {\bibinfo {author} {\bibfnamefont {A.}~\bibnamefont
  {Nahum}}, \bibinfo {author} {\bibfnamefont {S.}~\bibnamefont {Vijay}},\ and\
  \bibinfo {author} {\bibfnamefont {J.}~\bibnamefont {Haah}},\ }\href
  {https://doi.org/10.1103/PhysRevX.8.021014} {\bibfield  {journal} {\bibinfo
  {journal} {Phys. Rev. X}\ }\textbf {\bibinfo {volume} {8}},\ \bibinfo {pages}
  {021014} (\bibinfo {year} {2018})}\BibitemShut {NoStop}%
\bibitem [{\citenamefont {Chan}\ \emph {et~al.}(2018)\citenamefont {Chan},
  \citenamefont {De~Luca},\ and\ \citenamefont {Chalker}}]{chan2018solution}%
  \BibitemOpen
  \bibfield  {author} {\bibinfo {author} {\bibfnamefont {A.}~\bibnamefont
  {Chan}}, \bibinfo {author} {\bibfnamefont {A.}~\bibnamefont {De~Luca}},\ and\
  \bibinfo {author} {\bibfnamefont {J.~T.}\ \bibnamefont {Chalker}},\ }\href
  {https://doi.org/10.1103/PhysRevX.8.041019} {\bibfield  {journal} {\bibinfo
  {journal} {Phys. Rev. X}\ }\textbf {\bibinfo {volume} {8}},\ \bibinfo {pages}
  {041019} (\bibinfo {year} {2018})}\BibitemShut {NoStop}%
\bibitem [{\citenamefont {von Keyserlingk}\ \emph {et~al.}(2018)\citenamefont
  {von Keyserlingk}, \citenamefont {Rakovszky}, \citenamefont {Pollmann},\ and\
  \citenamefont {Sondhi}}]{vonKeyserlingk2018operator}%
  \BibitemOpen
  \bibfield  {author} {\bibinfo {author} {\bibfnamefont {C.~W.}\ \bibnamefont
  {von Keyserlingk}}, \bibinfo {author} {\bibfnamefont {T.}~\bibnamefont
  {Rakovszky}}, \bibinfo {author} {\bibfnamefont {F.}~\bibnamefont
  {Pollmann}},\ and\ \bibinfo {author} {\bibfnamefont {S.~L.}\ \bibnamefont
  {Sondhi}},\ }\href {https://doi.org/10.1103/PhysRevX.8.021013} {\bibfield
  {journal} {\bibinfo  {journal} {Phys. Rev. X}\ }\textbf {\bibinfo {volume}
  {8}},\ \bibinfo {pages} {021013} (\bibinfo {year} {2018})}\BibitemShut
  {NoStop}%
\bibitem [{\citenamefont {Gopalakrishnan}\ and\ \citenamefont
  {Lamacraft}(2019)}]{gopalakrishnan2019unitary}%
  \BibitemOpen
  \bibfield  {author} {\bibinfo {author} {\bibfnamefont {S.}~\bibnamefont
  {Gopalakrishnan}}\ and\ \bibinfo {author} {\bibfnamefont {A.}~\bibnamefont
  {Lamacraft}},\ }\href {https://doi.org/10.1103/PhysRevB.100.064309}
  {\bibfield  {journal} {\bibinfo  {journal} {Phys. Rev. B}\ }\textbf {\bibinfo
  {volume} {100}},\ \bibinfo {pages} {064309} (\bibinfo {year}
  {2019})}\BibitemShut {NoStop}%
\bibitem [{\citenamefont {Friedman}\ \emph {et~al.}(2019)\citenamefont
  {Friedman}, \citenamefont {Chan}, \citenamefont {De~Luca},\ and\
  \citenamefont {Chalker}}]{friedman2019spectral}%
  \BibitemOpen
  \bibfield  {author} {\bibinfo {author} {\bibfnamefont {A.~J.}\ \bibnamefont
  {Friedman}}, \bibinfo {author} {\bibfnamefont {A.}~\bibnamefont {Chan}},
  \bibinfo {author} {\bibfnamefont {A.}~\bibnamefont {De~Luca}},\ and\ \bibinfo
  {author} {\bibfnamefont {J.~T.}\ \bibnamefont {Chalker}},\ }\href
  {https://doi.org/10.1103/PhysRevLett.123.210603} {\bibfield  {journal}
  {\bibinfo  {journal} {Phys. Rev. Lett.}\ }\textbf {\bibinfo {volume} {123}},\
  \bibinfo {pages} {210603} (\bibinfo {year} {2019})}\BibitemShut {NoStop}%
\bibitem [{\citenamefont {Li}\ \emph {et~al.}(2019)\citenamefont {Li},
  \citenamefont {Chen},\ and\ \citenamefont {Fisher}}]{li2019measurement}%
  \BibitemOpen
  \bibfield  {author} {\bibinfo {author} {\bibfnamefont {Y.}~\bibnamefont
  {Li}}, \bibinfo {author} {\bibfnamefont {X.}~\bibnamefont {Chen}},\ and\
  \bibinfo {author} {\bibfnamefont {M.~P.~A.}\ \bibnamefont {Fisher}},\ }\href
  {https://doi.org/10.1103/PhysRevB.100.134306} {\bibfield  {journal} {\bibinfo
   {journal} {Phys. Rev. B}\ }\textbf {\bibinfo {volume} {100}},\ \bibinfo
  {pages} {134306} (\bibinfo {year} {2019})}\BibitemShut {NoStop}%
\bibitem [{\citenamefont {Skinner}\ \emph {et~al.}(2019)\citenamefont
  {Skinner}, \citenamefont {Ruhman},\ and\ \citenamefont
  {Nahum}}]{skinner2019measurement}%
  \BibitemOpen
  \bibfield  {author} {\bibinfo {author} {\bibfnamefont {B.}~\bibnamefont
  {Skinner}}, \bibinfo {author} {\bibfnamefont {J.}~\bibnamefont {Ruhman}},\
  and\ \bibinfo {author} {\bibfnamefont {A.}~\bibnamefont {Nahum}},\ }\href
  {https://doi.org/10.1103/PhysRevX.9.031009} {\bibfield  {journal} {\bibinfo
  {journal} {Phys. Rev. X}\ }\textbf {\bibinfo {volume} {9}},\ \bibinfo {pages}
  {031009} (\bibinfo {year} {2019})}\BibitemShut {NoStop}%
\bibitem [{\citenamefont {Rakovszky}\ \emph {et~al.}(2019)\citenamefont
  {Rakovszky}, \citenamefont {Pollmann},\ and\ \citenamefont {von
  Keyserlingk}}]{rakovszky2019sub}%
  \BibitemOpen
  \bibfield  {author} {\bibinfo {author} {\bibfnamefont {T.}~\bibnamefont
  {Rakovszky}}, \bibinfo {author} {\bibfnamefont {F.}~\bibnamefont
  {Pollmann}},\ and\ \bibinfo {author} {\bibfnamefont {C.~W.}\ \bibnamefont
  {von Keyserlingk}},\ }\href {https://doi.org/10.1103/PhysRevLett.122.250602}
  {\bibfield  {journal} {\bibinfo  {journal} {Phys. Rev. Lett.}\ }\textbf
  {\bibinfo {volume} {122}},\ \bibinfo {pages} {250602} (\bibinfo {year}
  {2019})}\BibitemShut {NoStop}%
\bibitem [{\citenamefont {Zabalo}\ \emph {et~al.}(2020)\citenamefont {Zabalo},
  \citenamefont {Gullans}, \citenamefont {Wilson}, \citenamefont
  {Gopalakrishnan}, \citenamefont {Huse},\ and\ \citenamefont
  {Pixley}}]{zabalo2020critical}%
  \BibitemOpen
  \bibfield  {author} {\bibinfo {author} {\bibfnamefont {A.}~\bibnamefont
  {Zabalo}}, \bibinfo {author} {\bibfnamefont {M.~J.}\ \bibnamefont {Gullans}},
  \bibinfo {author} {\bibfnamefont {J.~H.}\ \bibnamefont {Wilson}}, \bibinfo
  {author} {\bibfnamefont {S.}~\bibnamefont {Gopalakrishnan}}, \bibinfo
  {author} {\bibfnamefont {D.~A.}\ \bibnamefont {Huse}},\ and\ \bibinfo
  {author} {\bibfnamefont {J.~H.}\ \bibnamefont {Pixley}},\ }\href
  {https://doi.org/10.1103/PhysRevB.101.060301} {\bibfield  {journal} {\bibinfo
   {journal} {Phys. Rev. B}\ }\textbf {\bibinfo {volume} {101}},\ \bibinfo
  {pages} {060301(R)} (\bibinfo {year} {2020})}\BibitemShut {NoStop}%
\bibitem [{\citenamefont {Klobas}\ \emph {et~al.}(2021)\citenamefont {Klobas},
  \citenamefont {Bertini},\ and\ \citenamefont {Piroli}}]{klobas2021exact}%
  \BibitemOpen
  \bibfield  {author} {\bibinfo {author} {\bibfnamefont {K.}~\bibnamefont
  {Klobas}}, \bibinfo {author} {\bibfnamefont {B.}~\bibnamefont {Bertini}},\
  and\ \bibinfo {author} {\bibfnamefont {L.}~\bibnamefont {Piroli}},\ }\href
  {https://doi.org/10.1103/PhysRevLett.126.160602} {\bibfield  {journal}
  {\bibinfo  {journal} {Phys. Rev. Lett.}\ }\textbf {\bibinfo {volume} {126}},\
  \bibinfo {pages} {160602} (\bibinfo {year} {2021})}\BibitemShut {NoStop}%
\bibitem [{\citenamefont {Klobas}\ and\ \citenamefont
  {Bertini}(2021{\natexlab{a}})}]{klobas2021exactrelaxation}%
  \BibitemOpen
  \bibfield  {author} {\bibinfo {author} {\bibfnamefont {K.}~\bibnamefont
  {Klobas}}\ and\ \bibinfo {author} {\bibfnamefont {B.}~\bibnamefont
  {Bertini}},\ }\href {https://doi.org/10.21468/SciPostPhys.11.6.106}
  {\bibfield  {journal} {\bibinfo  {journal} {SciPost Phys.}\ }\textbf
  {\bibinfo {volume} {11}},\ \bibinfo {pages} {106} (\bibinfo {year}
  {2021}{\natexlab{a}})}\BibitemShut {NoStop}%
\bibitem [{\citenamefont {Klobas}\ and\ \citenamefont
  {Bertini}(2021{\natexlab{b}})}]{klobas2021entanglement}%
  \BibitemOpen
  \bibfield  {author} {\bibinfo {author} {\bibfnamefont {K.}~\bibnamefont
  {Klobas}}\ and\ \bibinfo {author} {\bibfnamefont {B.}~\bibnamefont
  {Bertini}},\ }\href {https://doi.org/10.21468/SciPostPhys.11.6.107}
  {\bibfield  {journal} {\bibinfo  {journal} {SciPost Phys.}\ }\textbf
  {\bibinfo {volume} {11}},\ \bibinfo {pages} {107} (\bibinfo {year}
  {2021}{\natexlab{b}})}\BibitemShut {NoStop}%
\bibitem [{\citenamefont {Bertini}\ \emph {et~al.}(2024)\citenamefont
  {Bertini}, \citenamefont {De~Fazio}, \citenamefont {Garrahan},\ and\
  \citenamefont {Klobas}}]{bertini2023exact}%
  \BibitemOpen
  \bibfield  {author} {\bibinfo {author} {\bibfnamefont {B.}~\bibnamefont
  {Bertini}}, \bibinfo {author} {\bibfnamefont {C.}~\bibnamefont {De~Fazio}},
  \bibinfo {author} {\bibfnamefont {J.~P.}\ \bibnamefont {Garrahan}},\ and\
  \bibinfo {author} {\bibfnamefont {K.}~\bibnamefont {Klobas}},\ }\href
  {https://doi.org/10.1103/PhysRevLett.132.120402} {\bibfield  {journal}
  {\bibinfo  {journal} {Phys. Rev. Lett.}\ }\textbf {\bibinfo {volume} {132}},\
  \bibinfo {pages} {120402} (\bibinfo {year} {2024})}\BibitemShut {NoStop}%
\bibitem [{\citenamefont {Ba\~nuls}\ \emph {et~al.}(2009)\citenamefont
  {Ba\~nuls}, \citenamefont {Hastings}, \citenamefont {Verstraete},\ and\
  \citenamefont {Cirac}}]{banuls2009matrix}%
  \BibitemOpen
  \bibfield  {author} {\bibinfo {author} {\bibfnamefont {M.~C.}\ \bibnamefont
  {Ba\~nuls}}, \bibinfo {author} {\bibfnamefont {M.~B.}\ \bibnamefont
  {Hastings}}, \bibinfo {author} {\bibfnamefont {F.}~\bibnamefont
  {Verstraete}},\ and\ \bibinfo {author} {\bibfnamefont {J.~I.}\ \bibnamefont
  {Cirac}},\ }\href {https://doi.org/10.1103/PhysRevLett.102.240603} {\bibfield
   {journal} {\bibinfo  {journal} {Phys. Rev. Lett.}\ }\textbf {\bibinfo
  {volume} {102}},\ \bibinfo {pages} {240603} (\bibinfo {year}
  {2009})}\BibitemShut {NoStop}%
\bibitem [{\citenamefont {M{\"u}ller-Hermes}\ \emph {et~al.}(2012)\citenamefont
  {M{\"u}ller-Hermes}, \citenamefont {Cirac},\ and\ \citenamefont
  {Ba{\~n}uls}}]{muller2012tensor}%
  \BibitemOpen
  \bibfield  {author} {\bibinfo {author} {\bibfnamefont {A.}~\bibnamefont
  {M{\"u}ller-Hermes}}, \bibinfo {author} {\bibfnamefont {J.~I.}\ \bibnamefont
  {Cirac}},\ and\ \bibinfo {author} {\bibfnamefont {M.~C.}\ \bibnamefont
  {Ba{\~n}uls}},\ }\href {https://doi.org/10.1088/1367-2630/14/7/075003}
  {\bibfield  {journal} {\bibinfo  {journal} {New J. Phys.}\ }\textbf {\bibinfo
  {volume} {14}},\ \bibinfo {pages} {075003} (\bibinfo {year}
  {2012})}\BibitemShut {NoStop}%
\bibitem [{\citenamefont {Sonner}\ \emph {et~al.}(2021)\citenamefont {Sonner},
  \citenamefont {Lerose},\ and\ \citenamefont {Abanin}}]{sonner2021influence}%
  \BibitemOpen
  \bibfield  {author} {\bibinfo {author} {\bibfnamefont {M.}~\bibnamefont
  {Sonner}}, \bibinfo {author} {\bibfnamefont {A.}~\bibnamefont {Lerose}},\
  and\ \bibinfo {author} {\bibfnamefont {D.~A.}\ \bibnamefont {Abanin}},\
  }\href {https://doi.org/10.1016/j.aop.2021.168677} {\bibfield  {journal}
  {\bibinfo  {journal} {Ann. Physics}\ }\textbf {\bibinfo {volume} {435}},\
  \bibinfo {pages} {168677} (\bibinfo {year} {2021})}\BibitemShut {NoStop}%
\bibitem [{\citenamefont {Lerose}\ \emph
  {et~al.}(2021{\natexlab{b}})\citenamefont {Lerose}, \citenamefont {Sonner},\
  and\ \citenamefont {Abanin}}]{lerose2021scaling}%
  \BibitemOpen
  \bibfield  {author} {\bibinfo {author} {\bibfnamefont {A.}~\bibnamefont
  {Lerose}}, \bibinfo {author} {\bibfnamefont {M.}~\bibnamefont {Sonner}},\
  and\ \bibinfo {author} {\bibfnamefont {D.~A.}\ \bibnamefont {Abanin}},\
  }\href {https://doi.org/10.1103/PhysRevB.104.035137} {\bibfield  {journal}
  {\bibinfo  {journal} {Phys. Rev. B}\ }\textbf {\bibinfo {volume} {104}},\
  \bibinfo {pages} {035137} (\bibinfo {year} {2021}{\natexlab{b}})}\BibitemShut
  {NoStop}%
\bibitem [{\citenamefont {Fr\'{\i}as-P\'erez}\ and\ \citenamefont
  {Ba\~nuls}(2022)}]{friasperez2022lightcone}%
  \BibitemOpen
  \bibfield  {author} {\bibinfo {author} {\bibfnamefont {M.}~\bibnamefont
  {Fr\'{\i}as-P\'erez}}\ and\ \bibinfo {author} {\bibfnamefont {M.~C.}\
  \bibnamefont {Ba\~nuls}},\ }\href
  {https://doi.org/10.1103/PhysRevB.106.115117} {\bibfield  {journal} {\bibinfo
   {journal} {Phys. Rev. B}\ }\textbf {\bibinfo {volume} {106}},\ \bibinfo
  {pages} {115117} (\bibinfo {year} {2022})}\BibitemShut {NoStop}%
\bibitem [{\citenamefont {Giudice}\ \emph {et~al.}(2022)\citenamefont
  {Giudice}, \citenamefont {Giudici}, \citenamefont {Sonner}, \citenamefont
  {Thoenniss}, \citenamefont {Lerose}, \citenamefont {Abanin},\ and\
  \citenamefont {Piroli}}]{giudice2022temporal}%
  \BibitemOpen
  \bibfield  {author} {\bibinfo {author} {\bibfnamefont {G.}~\bibnamefont
  {Giudice}}, \bibinfo {author} {\bibfnamefont {G.}~\bibnamefont {Giudici}},
  \bibinfo {author} {\bibfnamefont {M.}~\bibnamefont {Sonner}}, \bibinfo
  {author} {\bibfnamefont {J.}~\bibnamefont {Thoenniss}}, \bibinfo {author}
  {\bibfnamefont {A.}~\bibnamefont {Lerose}}, \bibinfo {author} {\bibfnamefont
  {D.~A.}\ \bibnamefont {Abanin}},\ and\ \bibinfo {author} {\bibfnamefont
  {L.}~\bibnamefont {Piroli}},\ }\href
  {https://doi.org/10.1103/PhysRevLett.128.220401} {\bibfield  {journal}
  {\bibinfo  {journal} {Phys. Rev. Lett.}\ }\textbf {\bibinfo {volume} {128}},\
  \bibinfo {pages} {220401} (\bibinfo {year} {2022})}\BibitemShut {NoStop}%
\bibitem [{\citenamefont {Bertini}\ \emph
  {et~al.}(2022{\natexlab{a}})\citenamefont {Bertini}, \citenamefont {Klobas},\
  and\ \citenamefont {Lu}}]{bertini2022entanglement}%
  \BibitemOpen
  \bibfield  {author} {\bibinfo {author} {\bibfnamefont {B.}~\bibnamefont
  {Bertini}}, \bibinfo {author} {\bibfnamefont {K.}~\bibnamefont {Klobas}},\
  and\ \bibinfo {author} {\bibfnamefont {T.-C.}\ \bibnamefont {Lu}},\ }\href
  {https://doi.org/10.1103/PhysRevLett.129.140503} {\bibfield  {journal}
  {\bibinfo  {journal} {Phys. Rev. Lett.}\ }\textbf {\bibinfo {volume} {129}},\
  \bibinfo {pages} {140503} (\bibinfo {year} {2022}{\natexlab{a}})}\BibitemShut
  {NoStop}%
\bibitem [{\citenamefont {Bertini}\ \emph
  {et~al.}(2022{\natexlab{b}})\citenamefont {Bertini}, \citenamefont {Klobas},
  \citenamefont {Alba}, \citenamefont {Lagnese},\ and\ \citenamefont
  {Calabrese}}]{bertini2022growth}%
  \BibitemOpen
  \bibfield  {author} {\bibinfo {author} {\bibfnamefont {B.}~\bibnamefont
  {Bertini}}, \bibinfo {author} {\bibfnamefont {K.}~\bibnamefont {Klobas}},
  \bibinfo {author} {\bibfnamefont {V.}~\bibnamefont {Alba}}, \bibinfo {author}
  {\bibfnamefont {G.}~\bibnamefont {Lagnese}},\ and\ \bibinfo {author}
  {\bibfnamefont {P.}~\bibnamefont {Calabrese}},\ }\href
  {https://doi.org/10.1103/PhysRevX.12.031016} {\bibfield  {journal} {\bibinfo
  {journal} {Phys. Rev. X}\ }\textbf {\bibinfo {volume} {12}},\ \bibinfo
  {pages} {031016} (\bibinfo {year} {2022}{\natexlab{b}})}\BibitemShut
  {NoStop}%
\end{thebibliography}%

\bigskip

\bigskip

\bigskip

\onecolumngrid
\setcounter{equation}{0}
\setcounter{figure}{0}
\setcounter{table}{0}
\setcounter{section}{0}
\renewcommand{\theequation}{\textsc{sm}-\arabic{equation}}
\renewcommand{\theHfigure}{sm-figure-\arabic{figure}}
\renewcommand{\thefigure}{\textsc{sm}-\arabic{figure}}
\renewcommand{\thetable}{\textsc{sm}-\arabic{table}}



\begin{center}
  {\large \bf
Supplemental Material for\\ \emph{Exact pre-transition effects in kinetically constrained circuits: dynamical fluctuations in the Floquet-East model}}
\end{center}
Here we report some useful information complementing the main text. In particular
\begin{itemize}
  \item[-] In Sec.~\ref{sec:definitions} we introduce the graphical notation and summarize all
    the local properties of the gates that are needed for the other calculations.
  \item[-] In Sec.~\ref{sec:hydroCross} we provide additional details on the ``hydrophobic crossover'' calculation.
  \item[-] In Sec.~\ref{sec:dynLDs} we discuss the phase transition in dynamical large deviations.
  \item[-] In Sec.~\ref{sec:probHoles} we provide additional details on the calculations of solvation probabilities and discussion of ``hydrophobic collapse''.
  \item[-] In Sec.~\ref{sec:bipartite} we find the density profile after starting from an inhomogeneous initial state where half of the system is empty and the other one prepared in the stationary state.
    
\end{itemize}

\section{Summary of graphical notation}\label{sec:definitions}
Here we briefly define the graphical notation used throughout the paper. 
\change{We follow the now well established standards for tensor diagrams: (i) tensors are indicated by shapes (squares, circles, etc.) and their indices are indicated by lines coming out of the shapes; (ii) connecting two lines implies a contraction (i.e., sum over the indices). For a pedagogical introduction see the TensorNetwork website Ref.~\cite{TensorNetwork}.
}

We start by defining the 
local deterministic gate for the Floquet East model,
\begin{equation}
  \begin{tikzpicture}[baseline={([yshift=-0.85ex]current bounding box.center)},scale=0.5]
    \prop{0}{0}{colU}
    \node at (-0.875,-0.875) {$n_1$};
    \node at (0.875,-0.875) {$n_2$};
    \node at (-0.875,0.875) {$n_1^{\prime}$};
    \node at (0.875,0.875) {$n_2^{\prime}$};
  \end{tikzpicture}\mkern-8mu =
  \delta_{n_2^{\prime},n_2^{\pprime}}\left(
  \delta_{n_2,0}\delta_{n_1^{\prime},n_1^{\pprime}}+
  \delta_{n_2,1}\delta_{n_1^{\prime},1-n_1^{\pprime}}
  \right),
\end{equation}
which can be shown to satisfy the following set of relations
\begin{equation}\label{eq:SMRelU}
   \gatefsd = \fsd~\fsd\,,
 \qquad
  \gatefsu = \fsu~\fsu\,,
  \qquad
   \gatefsr = \begin{matrix} \fsr \\[-0.25em] \fsr \end{matrix}\,,\qquad
  \begin{tikzpicture}[baseline={([yshift=-0.6ex]current bounding box.center)},scale=0.5]
    \prop{0}{0}{colU}
    \MEld{-0.5}{0.5}
    \MErd{-0.5}{-0.5}
  \end{tikzpicture}=
 2 \, \ds
  ,\qquad
  \begin{tikzpicture}[baseline={([yshift=-0.6ex]current bounding box.center)},scale=0.5]
    \prop{0}{0}{colU}
    \MErd{-0.5}{-0.5}
    \MErd{0.5}{0.5}
  \end{tikzpicture}=
  \begin{tikzpicture}[baseline={([yshift=-0.6ex]current bounding box.center)},scale=0.5]
    \nctgridLine{0.625}{-0.625}{0.125}{-0.125}
    \nctgridLine{-0.625}{0.625}{-0.125}{0.125}
    \MEld{-0.125}{0.125}
    \MEld{0.125}{-0.125}
  \end{tikzpicture},\qquad
  \begin{tikzpicture}[baseline={([yshift=-0.6ex]current bounding box.center)},scale=0.5]
    \prop{0}{0}{colU}
    \MEld{0.5}{-0.5}
    \MEld{-0.5}{0.5}
  \end{tikzpicture}=
  \begin{tikzpicture}[baseline={([yshift=-0.6ex]current bounding box.center)},scale=0.5]
    \nctgridLine{-0.625}{-0.625}{-0.125}{-0.125}
    \nctgridLine{0.625}{0.625}{0.125}{0.125}
    \MErd{0.125}{0.125}
    \MErd{-0.125}{-0.125}
  \end{tikzpicture}.
\end{equation}
Here
$\begin{tikzpicture}[baseline={([yshift=-0.6ex]current bounding box.center)},scale=0.5]
    \nctgridLine{0}{0}{0}{0.5}
    \MEh{0}{0}
\end{tikzpicture}$ denotes the flat state $[1\,1]$ as either the row or column matrix.

The projectors to empty and full state are $2\times2$ diagonal matrices,
\begin{equation}
  \begin{tikzpicture}[baseline={([yshift=-0.6ex]current bounding box.center)},scale=0.5]
    \nctgridLine{0}{-0.5}{0}{0.5}
    \obsZero{0}{0}
  \end{tikzpicture}=
  \begin{tikzpicture}[baseline={([yshift=-0.6ex]current bounding box.center)},scale=0.5]
    \nctgridLine{0}{-0.75}{0}{0.75}
    \obsZero{0}{0.25}
    \obsZero{0}{-0.25}
  \end{tikzpicture}=
  \begin{bmatrix} 1& 0 \\ 0 & 0 \end{bmatrix},\qquad
  \begin{tikzpicture}[baseline={([yshift=-0.6ex]current bounding box.center)},scale=0.5]
    \nctgridLine{0}{-0.5}{0}{0.5}
    \obsOne{0}{0}
  \end{tikzpicture}=
  \begin{tikzpicture}[baseline={([yshift=-0.6ex]current bounding box.center)},scale=0.5]
    \nctgridLine{0}{-0.75}{0}{0.75}
    \obsOne{0}{0.25}
    \obsOne{0}{-0.25}
  \end{tikzpicture}=
  \begin{bmatrix} 0& 0 \\ 0 & 1 \end{bmatrix},\qquad
  \begin{tikzpicture}[baseline={([yshift=-0.6ex]current bounding box.center)},scale=0.5]
    \nctgridLine{0}{-0.75}{0}{0.75}
    \obsZero{0}{0.25}
    \obsOne{0}{-0.25}
  \end{tikzpicture}=
  \begin{tikzpicture}[baseline={([yshift=-0.6ex]current bounding box.center)},scale=0.5]
    \nctgridLine{0}{-0.75}{0}{0.75}
    \obsOne{0}{0.25}
    \obsZero{0}{-0.25}
  \end{tikzpicture}=
  0,
\end{equation}
in terms of which we can define the active and inactive gates as,
\begin{equation}\label{eq:defactInact}
  \begin{tikzpicture}[baseline={([yshift=-0.6]current bounding box.center)},scale=0.5]
    \prop{0}{0}{colUinactive}
  \end{tikzpicture} :=
  \begin{tikzpicture}[baseline={([yshift=-0.6]current bounding box.center)},scale=0.5]
    \nctgridLine{0}{0}{0.75}{0.75}
    \nctgridLine{0}{0}{0.75}{-0.75}
    \prop{0}{0}{colU}
    \obsZero{0.5}{0.5}
    \obsZero{0.5}{-0.5}
  \end{tikzpicture},\qquad
  \begin{tikzpicture}[baseline={([yshift=-0.6]current bounding box.center)},scale=0.5]
    \prop{0}{0}{colUactive}
  \end{tikzpicture} :=
  \begin{tikzpicture}[baseline={([yshift=-0.6]current bounding box.center)},scale=0.5]
    \nctgridLine{0}{0}{0.75}{0.75}
    \nctgridLine{0}{0}{0.75}{-0.75}
    \prop{0}{0}{colU}
    \obsOne{0.5}{0.5}
    \obsOne{0.5}{-0.5}
  \end{tikzpicture},
\end{equation}
where we note that the deterministic gate is the sum of the two,
\begin{equation}
  \begin{tikzpicture}[baseline={([yshift=-0.6ex]current bounding box.center)},scale=0.5]
    \prop{0}{0}{colU}
  \end{tikzpicture}=
  \begin{tikzpicture}[baseline={([yshift=-0.6ex]current bounding box.center)},scale=0.5]
    \prop{0}{0}{colUinactive}
  \end{tikzpicture}+
  \begin{tikzpicture}[baseline={([yshift=-0.6ex]current bounding box.center)},scale=0.5]
    \prop{0}{0}{colUactive}
  \end{tikzpicture}.
\end{equation}
These reduced gates satisfy the following two sets of relations that will be used later,
\begin{equation}\label{eq:SMRelInactive}
  \gateyfs{1} = \dsr~\projzfs{1},\quad \gateyfs{-1} = \dsr~\projzfs{-1},
\end{equation}
and
\begin{equation}\label{eq:SMRelActive}
  \begin{tikzpicture}[baseline={([yshift=-0.6ex]current bounding box.center)},scale=0.5]
    \prop{0}{0}{colUactive}
    \MErd{-0.5}{-0.5}
    \MErd{0.5}{0.5}
  \end{tikzpicture}=
  \begin{tikzpicture}[baseline={([yshift=-0.6ex]current bounding box.center)},scale=0.5]
    \nctgridLine{0.625}{-0.625}{0.125}{-0.125}
    \nctgridLine{-0.625}{0.625}{-0.125}{0.125}
    \MEld{-0.125}{0.125}
    \MEld{0.125}{-0.125}
    \obsOne{0.35}{-0.35}
  \end{tikzpicture},\qquad
  \begin{tikzpicture}[baseline={([yshift=-0.6ex]current bounding box.center)},scale=0.5]
    \prop{0}{0}{colUactive}
    \MEld{0.5}{-0.5}
    \MEld{-0.5}{0.5}
  \end{tikzpicture}=
  \begin{tikzpicture}[baseline={([yshift=-0.6ex]current bounding box.center)},scale=0.5]
    \nctgridLine{-0.625}{-0.625}{-0.125}{-0.125}
    \nctgridLine{0.625}{0.625}{0.125}{0.125}
    \MErd{0.125}{0.125}
    \MErd{-0.125}{-0.125}
    \obsOne{0.35}{0.35}
  \end{tikzpicture}.
\end{equation}
We finish this section by reporting a number of other useful graphical
identities that can be expressed using the projectors introduced above,
\begin{equation} \label{eq:IdentitiesForTriangles}
  \begin{tikzpicture}[baseline={([yshift=-0.6]current bounding box.center)},scale=0.5]
    \nctgridLine{0}{0}{-0.75}{-0.75}
    \nctgridLine{0}{0}{0.75}{-0.75}
    \prop{0}{0}{colU}
    \obsZero{-0.5}{-0.5}
    \obsZero{0.5}{-0.5}
    \node at (1.25,0) {$=$};
    \begin{scope}[shift={(2.5,0)}]
      \nctgridLine{0.75}{0.75}{0}{0}
      \nctgridLine{-0.75}{0.75}{0}{0}
      \prop{0}{0}{colU}
      \obsZero{-0.5}{0.5}
      \obsZero{0.5}{0.5}
    \node at (1.25,0) {$=$};
    \end{scope}
    \begin{scope}[shift={(5,0)}]
      \nctgridLine{0.75}{0.75}{-0.75}{-0.75}
      \nctgridLine{-0.75}{0.75}{0.75}{-0.75}
      \prop{0}{0}{colU}
      \obsZero{-0.5}{0.5}
      \obsZero{0.5}{0.5}
      \obsZero{-0.5}{-0.5}
      \obsZero{0.5}{-0.5}
    \end{scope}
  \end{tikzpicture},\qquad
  \begin{tikzpicture}[baseline={([yshift=-0.6]current bounding box.center)},scale=0.5]
    \nctgridLine{0}{0}{0.75}{-0.75}
    \prop{0}{0}{colU}
    \obsZero{0.5}{-0.5}
    \node at (1.25,0) {$=$};
    \begin{scope}[shift={(2.5,0)}]
      \nctgridLine{0}{0}{0.75}{0.75}
      \prop{0}{0}{colU}
      \obsZero{0.5}{0.5}
      \node at (1.25,0) {$=$};
    \end{scope}
    \begin{scope}[shift={(5,0)}]
      \nctgridLine{0.75}{0.75}{0}{0}
      \nctgridLine{0.75}{-0.75}{0}{0}
      \prop{0}{0}{colU}
      \obsZero{0.5}{0.5}
      \obsZero{0.5}{-0.5}
    \end{scope}
  \end{tikzpicture},\qquad
  \begin{tikzpicture}[baseline={([yshift=-0.6]current bounding box.center)},scale=0.5]
    \nctgridLine{0}{0}{-0.75}{-0.75}
    \nctgridLine{0}{0}{-0.75}{0.75}
    \prop{0}{0}{colU}
    \obsZero{-0.5}{-0.5}
    \obsZero{-0.5}{0.5}
    \node at (1.25,0) {$=$};
    \begin{scope}[shift={(2.5,0)}]
      \nctgridLine{-0.75}{-0.75}{0.75}{0.75}
      \nctgridLine{-0.75}{0.75}{0.75}{-0.75}
      \prop{0}{0}{colU}
      \obsZero{0.5}{-0.5}
      \obsZero{-0.5}{-0.5}
      \obsZero{0.5}{0.5}
      \obsZero{-0.5}{0.5}
    \end{scope}
  \end{tikzpicture},\qquad
  \begin{tikzpicture}[baseline={([yshift=-0.6]current bounding box.center)},scale=0.5]
    \nctgridLine{0}{0}{-0.75}{-0.75}
    \nctgridLine{0}{0}{0.75}{-0.75}
    \prop{0}{0}{colU}
    \obsZero{-0.5}{-0.5}
    \obsOne{0.5}{-0.5}
    \node at (1.25,0) {$=$};
    \begin{scope}[shift={(2.5,0)}]
      \nctgridLine{0.75}{0.75}{-0.75}{-0.75}
      \nctgridLine{-0.75}{0.75}{0.75}{-0.75}
      \prop{0}{0}{colU}
      \obsOne{-0.5}{0.5}
      \obsOne{0.5}{0.5}
      \obsZero{-0.5}{-0.5}
      \obsOne{0.5}{-0.5}
    \end{scope}
  \end{tikzpicture}.
\end{equation}

\section{Probabilities of inactive space-time regions}\label{sec:hydroCross}

\begin{figure}
  \begin{center}
    \includegraphics[width=0.9\textwidth]{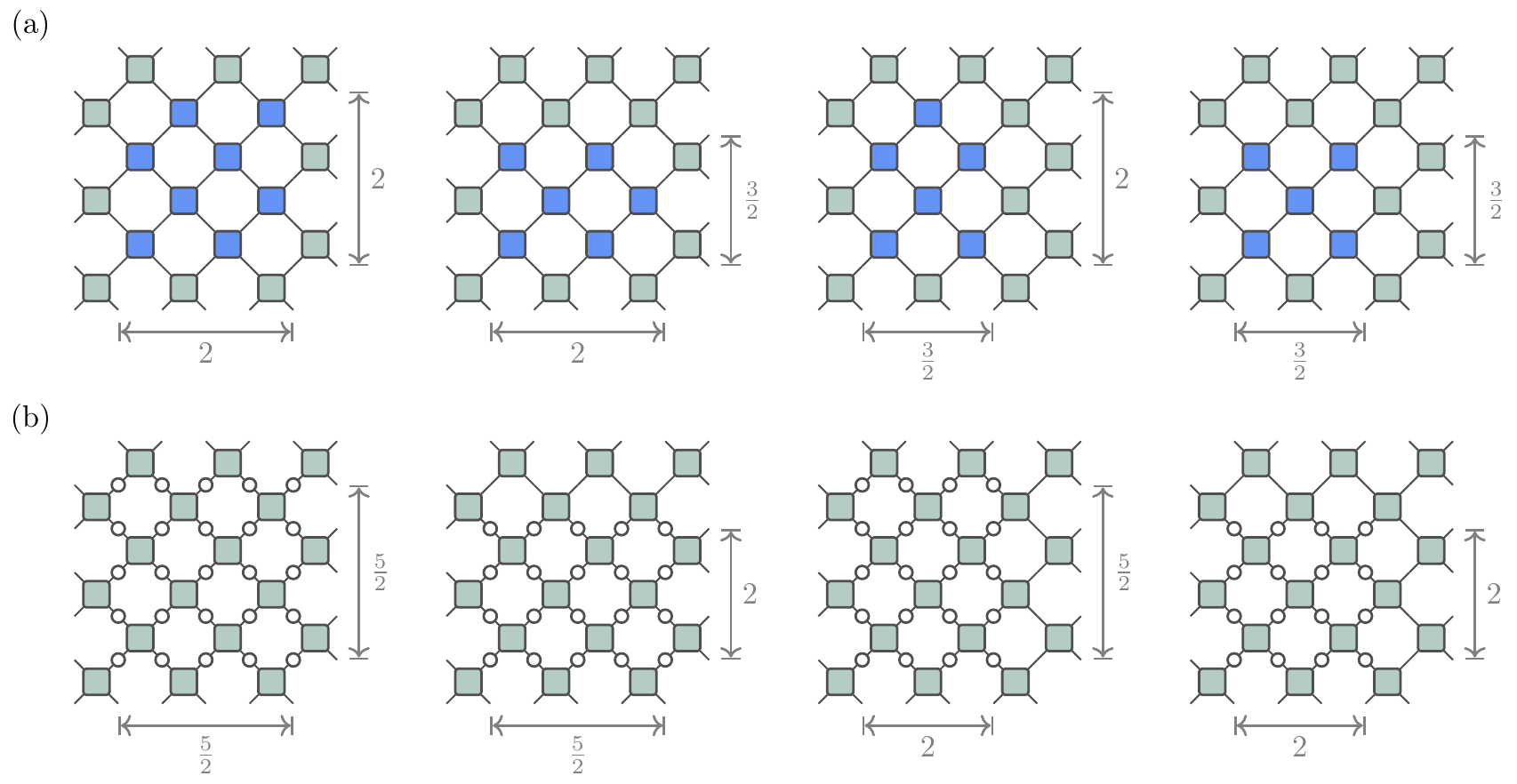}
  \end{center}
  \caption{\label{fig:defSubsystemsize}
    Example of regions with integer and half-integer sizes $l\times t$. For
    simplicity we always assume rectangular regions to have a gate in the
    bottom-left corner. The two conventions differ because of the different
    nature of conditioning. If we condition on the gates (a), we define the
    size of the region to be the number of time and space steps needed to get
    out of the region, so that the region with $l=0$ or $t=0$ has no gates in
    it. Conversely, if the conditioning is on sites (b), the size is defined as
    the number of unit cells, with each one of them containing two sites, so
    that the area contains $2l\times 2t$ sites.
  }
\end{figure}

The first problem we consider is the probability of a space-time rectangular region to be
inactive (i.e., no spin occurring in this region). This probability can be graphically
represented as,
\begin{equation} \label{eq:defProbNoact}
  P_{\rm inact}(l,t)=
  2^{-2L}
  \begin{tikzpicture}[baseline={([yshift=-0.6ex]current bounding box.center)},scale=0.5]
    \def\X{10}
    \def\Y{10}
    \foreach \x in {0,2,...,\X}{
      \foreach \y in {0,2,...,\Y}{
        \prop{\x+0.5}{\y}{colU}
        \prop{\x+1.5}{\y+1}{colU}
      }
    }
    \foreach \x in {2,4,6}{
      \foreach \y in {2,4,6,8}{
        \prop{\x+0.5}{\y}{colUinactive}
        \prop{\x+1.5}{\y+1}{colUinactive}
      }
    }
    \foreach \x in {0,2,...,\X}{
      \MErd{\x}{-0.5}
      \MEld{\x+1}{-0.5}
      \MEld{\x+1}{\Y+1.5}
      \MErd{\x+2}{\Y+1.5}
    }
    \draw[thick,|<->|,gray] (2,-1) -- (8,-1) node[midway,below] {$l$};
    \draw[thick,|<->|,gray] (\Y+2.5,1.5) -- (\Y+2.5,9.5) node[midway,right] {$t$};
  \end{tikzpicture}
\end{equation}
where we assume the full system size $L$ to be much larger than the subsystem size $l$,
and the blue gates are the inactive gates as defined in~\eqref{eq:defactInact}.
The sizes $l\times t$ are measured in the number of full time-steps and number of 
sites with integer label, so that they can take integer or half-integer values as shown
in Fig.~\ref{fig:defSubsystemsize}. Assuming first that the sizes take integer values,
we can use the fixed-point relations~\eqref{ist}, and~\eqref{iss} to contract the
tensor network~\eqref{eq:defProbNoact} to only contain the inactive region, and then
repeatedly apply~\eqref{eq:relInactive} to further simplify the inactive tensor-network
and calculate the probability as
\begin{equation}
  \left. P_{\rm inact}(l,t)\right|_{l,t\in\mathbb{N}}=
    2^{-(2l+t)}
    \begin{tikzpicture}[baseline={([yshift=-0.6ex]current bounding box.center)},scale=0.5]
      \foreach \x in {2,4,6}{
        \foreach \y in {2,4,6,8}{
          \prop{\x+0.5}{\y}{colUinactive}
          \prop{\x+1.5}{\y+1}{colUinactive}
        }
        \MErd{\x}{1.5}
        \MEld{\x+1}{1.5}
        \MEld{\x+1}{9.5}
        \MErd{\x+2}{9.5}
      }
      \foreach \y in {2,4,6}{
        \MEld{8}{\y+0.5}
        \MErd{8}{\y+1.5}
        \bendLud{2}{\y+0.5}{\y+1.5}
      }
      \MEld{8}{8.5}
      \MEld{2}{8.5}
    \end{tikzpicture}=
    2^{-(2l+t)}
    \begin{tikzpicture}[baseline={([yshift=-0.6ex]current bounding box.center)},scale=0.5]
      \MEld{2}{8.5}
      \MErd{2}{1.5}
      \foreach \y in {2,4,6}{\bendLud{2}{\y+0.5}{\y+1.5}}
      \foreach \y in {2,4,6,8}{\bendRud{2}{\y-0.5}{\y+0.5}}
    \end{tikzpicture}=
    2^{-(2l+t-1)}.
\end{equation}
Whenever $t\ge 1$ this expression can be straightforwardly generalized to half-integer
values of subsystem sizes,
\begin{equation}
  \left. P_{\rm inact}(l,t)\right|_{t\ge 1}=
  2^{-(2l+\lfloor t +l-\lfloor l\rfloor\rfloor-1)}.
\end{equation}
Note that the overall scaling with $t$ and $l$ stays the same, but depending on which
of the two sizes is (half)integer we can obtain a $\pm 1$ correction to the exponent.
If $t=\frac{1}{2}$, the probability scales differently,
\begin{equation}
  P_{\rm inact}(l,\frac{1}{2})=
  2^{-2\lceil l \rceil}
  \begin{tikzpicture}[baseline={([yshift=-0.6ex]current bounding box.center)},scale=0.5]
    \foreach \x in {2,4,6}{
      \foreach \y in {2}{
        \prop{\x+0.5}{\y}{colUinactive}
        \MEld{\x}{\y+0.5}
        \MErd{\x+1}{\y+0.5}
        \MErd{\x}{\y-0.5}
        \MEld{\x+1}{\y-0.5}
      }
    }
  \end{tikzpicture}=
  \left(
    \frac{1}{4}
    \begin{tikzpicture}[baseline={([yshift=-0.6ex]current bounding box.center)},scale=0.5]
      \foreach \x in {2}{
        \foreach \y in {2}{
          \bendLud{\x+1}{\y-0.5}{\y+0.5}
          \bendRud{\x}{\y-0.5}{\y+0.5}
          \obsZero{\x+0.8}{\y}
          \MEld{\x}{\y+0.5}
          \MErd{\x+1}{\y+0.5}
          \MErd{\x}{\y-0.5}
          \MEld{\x+1}{\y-0.5}
        }
      }
    \end{tikzpicture}
  \right)^{\lceil l \rceil}
  = 2^{-\lceil l \rceil}.
\end{equation}
Combining everything together gives us
\begin{equation}
  P_{\rm inact}(l,t) =
  \begin{cases}
    2^{-( 2l + \lfloor t + l -\lfloor l\rfloor \rfloor-1)},& t\ge 1,\\
    2^{-\lceil l \rceil},& t=\frac{1}{2}.
  \end{cases}
\end{equation}
Since we are interested in the scaling with $t$ for fixed $l$ it makes sense to
express this probability by assuming $l$ to be either integer or half-integer,
in which case the expression above takes a simplified form,
\begin{equation}
  \left. P_{\rm inact}(l,t)\right|_{l\in\mathbb{Z}} =
  \begin{cases}
    2^{-( 2l + \lfloor t \rfloor-1)},& t\ge 1,\\
    2^{- l },& t=\frac{1}{2},
  \end{cases}\qquad
  \left. P_{\rm inact}(l,t)\right|_{l-\frac{1}{2}\in\mathbb{Z}} =
  \begin{cases}
    2^{-( 2l + \lceil t \rceil-1)},& t\ge 1,\\
    2^{- l -1 },& t=\frac{1}{2}.
  \end{cases}
\end{equation}
\section{Dynamical large deviations}\label{sec:dynLDs}
We now consider a generalization of the quantity considered in Sec.~\ref{sec:hydroCross}
by replacing the inactive region with a region of tilted gates for which every-time
a spin flip occurs, we obtain a factor $e^{-s}$, $s\in\mathbb{R}$,
\begin{equation}
  \begin{tikzpicture}[baseline={([yshift=-0.6ex]current bounding box.center)},scale=0.5]
    \prop{0}{0}{colUtilted} 
  \end{tikzpicture}=
  \begin{tikzpicture}[baseline={([yshift=-0.6ex]current bounding box.center)},scale=0.5]
    \prop{0}{0}{colUinactive} 
  \end{tikzpicture}
  +
  e^{-s}
  \begin{tikzpicture}[baseline={([yshift=-0.6ex]current bounding box.center)},scale=0.5]
    \prop{0}{0}{colUactive} 
  \end{tikzpicture}.
\end{equation}
We are interested in the partition function for a tilted space-time region of
size $l\times t$ embedded in a much larger system of size $L\times T$ with
deterministic dynamics,
\begin{equation} \label{eq:defPartitionFunctionDeterministic}
  Z_{l,t}(s)=
  2^{-2L}
  \begin{tikzpicture}[baseline={([yshift=-0.6ex]current bounding box.center)},scale=0.5]
    \def\X{10}
    \def\Y{10}
    \foreach \x in {0,2,...,\X}{
      \foreach \y in {0,2,...,\Y}{
        \prop{\x+0.5}{\y}{colU}
        \prop{\x+1.5}{\y+1}{colU}
      }
    }
    \foreach \x in {2,4,6}{
      \foreach \y in {2,4,6,8}{
        \prop{\x+0.5}{\y}{colUtilted}
        \prop{\x+1.5}{\y+1}{colUtilted}
      }
    }
    \foreach \x in {0,2,...,\X}{
      \MErd{\x}{-0.5}
      \MEld{\x+1}{-0.5}
      \MEld{\x+1}{\Y+1.5}
      \MErd{\x+2}{\Y+1.5}
    }
    \draw[thick,|<->|,gray] (2,-1) -- (8,-1) node[midway,below] {$l$};
    \draw[thick,|<->|,gray] (\Y+2.5,1.5) -- (\Y+2.5,9.5) node[midway,right] {$t$};
  \end{tikzpicture}=
  2^{-(2l+t)}
  \begin{tikzpicture}[baseline={([yshift=-0.6ex]current bounding box.center)},scale=0.5]
    \foreach \x in {2,4,6}{
      \foreach \y in {2,4,6,8}{
        \prop{\x+0.5}{\y}{colUtilted}
        \prop{\x+1.5}{\y+1}{colUtilted}
      }
    }
    \foreach \x in {2,4,6}{
      \MErd{\x}{1.5}
      \MEld{\x+1}{1.5}
      \MEld{\x+1}{9.5}
      \MErd{\x+2}{9.5}
    }
    \foreach \t in {2,4,6}{
      \MEld{8}{\t+0.5}
      \MErd{8}{\t+1.5}
      \bendLud{2}{\t+0.5}{\t+1.5}
    }
    \MEld{8}{8.5}
    \MEld{2}{8.5}
  \end{tikzpicture},
\end{equation}
where for simplicity we assume $l,t\in\mathbb{Z}$, and we first take the limit
$L,T\to\infty$ leaving $l$ and $t$ finite.

The moment-generating function (MGF) $Z_{l,t}(s)$ is hard to calculate in full
generality, but we are able to obtain some special points. We start by noting
that in the limit $s\to\infty$ the active gates get completely suppressed, and
the expression is reduced to the probability of the region being inactive,
\begin{equation}
  \lim_{s\to\infty}Z_{l,t}(s)=P_{\rm inact}(l,t)=2^{-(2l+t-1)}.
\end{equation}
The other simple point is $s=0$, in which case the MGF reduces to the partition
function of the usual (non-tilted) model, and is $1$ by construction,
\begin{equation}
  Z_{l,t}(0)=1.
\end{equation}
Furthermore, for small $s$ we are able to find perturbative corrections to $1$,
\begin{equation}
  Z_{l,t}(s)\approx 1 - s Z_{l,t}^{(1)} + s^2 Z_{l,t}^{(2)} + \ldots.
\end{equation}
The first order is given by one-point correlation functions,
\begin{equation}
  Z_{l,t}^{(1)} = \sum_{\{x,y\}}
  2^{-(2l+t)}
  \begin{tikzpicture}[baseline={([yshift=-0.6ex]current bounding box.center)},scale=0.5]
    \foreach \x in {2,4,6}{
      \foreach \y in {2,4,6,8}{
        \prop{\x+0.5}{\y}{colU}
        \prop{\x+1.5}{\y+1}{colU}
      }
    }
    \prop{5.5}{3}{colUactive}
    \foreach \x in {2,4,6}{
      \MErd{\x}{1.5}
      \MEld{\x+1}{1.5}
      \MEld{\x+1}{9.5}
      \MErd{\x+2}{9.5}
    }
    \foreach \t in {2,4,6}{
      \MEld{8}{\t+0.5}
      \MErd{8}{\t+1.5}
      \bendLud{2}{\t+0.5}{\t+1.5}
    }
    \MEld{8}{8.5}
    \MEld{2}{8.5}
    \draw [thick,gray,|<->|] (2.5,1) -- (5.5,1) node[midway,below]{$x$};
    \draw [thick,gray,|<->|] (8.5,2) -- (8.5,3) node[midway,right]{$y$};
  \end{tikzpicture}
  =2 l t \times\frac{1}{4}
  \begin{tikzpicture}[baseline={([yshift=-0.6ex]current bounding box.center)},scale=0.5]
    \prop{5.5}{3}{colUactive}
    \MErd{5}{2.5}
    \MEld{5}{3.5}
    \MEld{6}{2.5}
    \MErd{6}{3.5}
  \end{tikzpicture}= l t,
\end{equation}
where the first equality follows directly from the definition, the second
equality by using the local fixed-point relations \eqref{eq:SMRelU}, and the
last one from~\eqref{eq:SMRelActive}.

Similarly, we can express the second order as
\begin{equation}\label{eq:deterministicSecond}
Z_{l,t}^{(2)}=\frac{1}{2} Z_{l,t}^{(1)}
+\frac{1}{2}\mkern16mu \smashoperator{\sum_{\{x_1,y_1\}\neq\{x_2,y_2\}}}
\underbrace{
2^{-(2l+t)}
  \begin{tikzpicture}[baseline={([yshift=-0.6ex]current bounding box.center)},scale=0.5]
  \foreach \x in {2,4,6}{
    \foreach \y in {2,4,6,8}{
      \prop{\x+0.5}{\y}{colU}
      \prop{\x+1.5}{\y+1}{colU}
    }
  }
  \prop{5.5}{3}{colUactive}
  \prop{3.5}{7}{colUactive}
  \foreach \x in {2,4,6}{
    \MErd{\x}{1.5}
    \MEld{\x+1}{1.5}
    \MEld{\x+1}{9.5}
    \MErd{\x+2}{9.5}
  }
  \foreach \t in {2,4,6}{
    \MEld{8}{\t+0.5}
    \MErd{8}{\t+1.5}
    \bendLud{2}{\t+0.5}{\t+1.5}
  }
  \MEld{8}{8.5}
  \MEld{2}{8.5}
  \draw [thick,gray,|<->|] (2.5,1) -- (5.5,1) node[midway,below]{$x_1$};
  \draw [thick,gray,|<->|] (8.5,2) -- (8.5,3) node[midway,right]{$y_1$};
  \draw [thick,gray,|<->|] (2.5,10) -- (3.5,10) node[midway,above]{$x_2$};
  \draw [thick,gray,|<->|] (1.5,2) -- (1.5,7) node[midway,left]{$y_2$};
\end{tikzpicture}}_{C_{2}(\{x_1,y_1\},\{x_2,y_2\})},
\end{equation}
where the prefactor $\frac{1}{2}$ in front of the sum takes care of double-counting.
Whenever the two active gates are not diagonal neighbours,
the coefficient $C_{2}(\{x_1,y_1\},\{x_2,y_2\})$ factorizes upon the repeated
application of local relations~\eqref{eq:SMRelU},
\begin{equation}
  \left.C_{2}(\{x_1,y_1\},\{x_2,y_2\}) \right|_{\{x_2,y_2\}\neq \{x_1\pm\frac{1}{2},y_1\pm\frac{1}{2}\}}
  =
  \left(\frac{1}{4}
    \begin{tikzpicture}[baseline={([yshift=-0.6ex]current bounding box.center)},scale=0.5]
      \prop{0}{0}{colUactive}
      \MErd{-0.5}{-0.5}
      \MErd{0.5}{0.5}
      \MEld{0.5}{-0.5}
      \MEld{-0.5}{0.5}
  \end{tikzpicture}\right)^2=\frac{1}{4}.
\end{equation}
The rest of the coefficients do not immediately factorize, but nonetheless, using
relations~\eqref{eq:SMRelU} and~\eqref{eq:SMRelActive}, we can show that they give
the same contribution,
\begin{equation}
  C_{2}(\{x_1,y_1\},\{x_1,y_1\}\pm\{\frac{1}{2},\frac{1}{2}\})
  =
  \frac{1}{8}
  \begin{tikzpicture}[baseline={([yshift=-0.6ex]current bounding box.center)},scale=0.5]
    \prop{0}{0}{colUactive}
    \prop{1}{1}{colUactive}
    \MErd{-0.5}{-0.5}
    \MErd{1.5}{1.5}
    \MEld{0.5}{-0.5}
    \MEld{-0.5}{0.5}
    \MEld{1.5}{0.5}
    \MEld{1.5}{0.5}
    \MEld{0.5}{1.5}
  \end{tikzpicture}
  =\frac{1}{4},\qquad
  C_{2}(\{x_1,y_1\},\{x_1,y_1\}\pm\{\frac{1}{2},-\frac{1}{2}\})
  =
  \frac{1}{8}
  \begin{tikzpicture}[baseline={([yshift=-0.6ex]current bounding box.center)},scale=0.5]
    \prop{0}{0}{colUactive}
    \prop{1}{-1}{colUactive}
    \MEld{-0.5}{0.5}
    \MEld{1.5}{-1.5}
    \MErd{0.5}{0.5}
    \MErd{-0.5}{-0.5}
    \MErd{1.5}{-0.5}
    \MErd{1.5}{-0.5}
    \MErd{0.5}{-1.5}
  \end{tikzpicture}
  =\frac{1}{4}.
\end{equation}
This finally gives us the second-order in the expansion as
\begin{equation}
  Z_{l,t}^{(2)}=\frac{1}{2} Z_{l,t}^{(1)}
  +\frac{1}{2}\mkern16mu\smashoperator{\sum_{\{x_1,y_1\}\neq\{x_2,y_2\}}}
  \mkern16mu \frac{1}{4}=
  \frac{l t}{2} + \frac{1}{8} \left((2lt)^2-2lt\right)
  =\frac{(l t)^2}{2}+\frac{l t}{4}.
\end{equation}

\section{Emptiness formation probabilities}\label{sec:probHoles}
Probabilities of empty rectangular regions are closely connected to the
probabilities of inactive space-time regions considered in
Sec.~\ref{sec:hydroCross}.  In particular, the probability of an empty
rectangular region with sizes $l$ and $t$ scales as
$P_{\rm inact}(l-\frac{1}{2},t-\frac{1}{2})$, 
\begin{equation}
  \left. P_{\square}(l,t)\right|_{l,t\in\mathbb{N}}=
  2^{-(2l+t-1)}
  \begin{tikzpicture}[baseline={([yshift=-0.6ex]current bounding box.center)},scale=0.5]
    \foreach \x in {2,4,6}{
      \nctgridLine{\x-0.25}{1.25}{\x+0.5}{2}
      \nctgridLine{\x+1.25}{1.25}{\x+0.5}{2}
      \MErd{\x-0.25}{1.25}
      \MEld{\x+1.25}{1.25}
    }
    \foreach \y in {2,4,6}{
      \nctgridLine{7.25}{\y+0.75}{6.5}{\y}
      \nctgridLine{7.25}{\y+1.25}{6.5}{\y+2}
      \MErd{7.25}{\y+0.75}
      \MEld{7.25}{\y+1.25}
      \bendLud{2}{\y+0.5}{\y+1.5}
    }
    \foreach \x in {2,4,6}{
      \nctgridLine{\x+0.5}{8}{\x+1.25}{8.75}
      \nctgridLine{\x+0.5}{8}{\x-0.25}{8.75}
      \MEld{\x-0.25}{8.75}
      \MErd{\x+1.25}{8.75}
    }
    \foreach \x in {2,4}{
      \foreach \y in {2,4,6}{
        \prop{\x+0.5}{\y}{colU}
        \prop{\x+1.5}{\y+1}{colU}
      }
    }
    \foreach \x in {2,4,6}{\prop{\x+0.5}{8}{colU}}
    \foreach \x in {2,4,6}{\prop{6.5}{\x}{colU}}
    \foreach \x in {2,...,7}
    {\foreach \y in {1,...,8}{\obsZero{\x}{\y+0.5}}}
  \end{tikzpicture}=
  2^{-(2l+t-1)}=
  \frac{1}{2} P_{\rm inact}(l-\frac{1}{2},t-\frac{1}{2}).
\end{equation}
Here the factor of $\frac{1}{2}$ with respect to $P_{\rm inact}$ is due
to the fact that requiring the area to be inactive allows for the left-most
column to be either in the state $1$ or state $0$, while here we force it
to be $0$. Analogous expressions can be obtained for all combinations of
parities of $2l$ and $2t$, and for sufficiently large $t$ and $l$ we obtain
(up to $\mathcal{O}(1)$ corrections in the exponent)
the same scaling with $P_{\rm inact}(l-\frac{1}{2},t-\frac{1}{2})$.
In particular, taking into account all the possible combinations of parities
and finite-size effects for small $t$ and $l$, we obtain the following general
expression,
\begin{equation}
  \left. P_{\square}(l,t)\right|_{l\in\mathbb{Z}}=2^{-(2l+\lfloor t-\frac{1}{2}\rfloor)},
  \qquad
  \left. P_{\square}(l,t)\right|_{l+\frac{1}{2}\in\mathbb{Z}}=
    2^{-(2l+\lceil t-\frac{1}{2}\rceil+\delta_{t,1}(1-\delta_{l,\frac{1}{2}}))}.
\end{equation}
Note that the cases with $l,t\le 1$ do not scale as $P_{\rm inact}(l-\frac{1}{2},t-\frac{1}{2})$, and therefore $P_{\square}(l,t)$ does not exhibit the transition in
Fig.~\ref{fig2}(b).

A shape that is more favourable than a rectangle is that of a right-pointing triangle,
\begin{equation}
  P_{\rhd}(t)=
  2^{-L}
  \begin{tikzpicture}[baseline={([yshift=-0.6ex]current bounding box.center)},scale=0.5]
    \foreach \x in {0,2,4,6}{
      \foreach \y in {0,2,4,6,8,10,12}
      {
        \prop{\x}{\y+1}{colU}
        \prop{\x+1}{\y}{colU}
      }
    }
    \foreach \x in {0,2,4,6}{
      \MEld{\x+1.5}{-0.5}
      \MErd{\x+0.5}{-0.5}
      \MErd{\x+0.5}{13.5}
      \MEld{\x-0.5}{13.5}
    }
    \foreach \y in {0,...,11}{\obsZero{0.5}{\y+0.5}}
    \foreach \y in {1,...,10}{\obsZero{1.5}{\y+0.5}}
    \foreach \y in {2,...,9}{\obsZero{2.5}{\y+0.5}}
    \foreach \y in {3,...,8}{\obsZero{3.5}{\y+0.5}}
    \foreach \y in {4,...,7}{\obsZero{4.5}{\y+0.5}}
    \foreach \y in {5,...,6}{\obsZero{5.5}{\y+0.5}}
  \end{tikzpicture}=
  2^{-t}
  \begin{tikzpicture}[baseline={([yshift=-0.6ex]current bounding box.center)},scale=0.5]
    \foreach \x in {0,...,5}{\nctgridLine{\x+0.5}{0.5+\x}{\x+0.75}{0.25+\x}}
    \foreach \x in {0,...,5}{\nctgridLine{\x+0.5}{11+0.5-\x}{\x+0.75}{11-\x+0.75}}
    \foreach \x in {0,...,5}{\MErd{\x+0.75}{11.75-\x}}
    \foreach \x in {0,...,5}{\MEld{\x+0.75}{\x+0.25}}
    \foreach \x in {0,2,...,10}{\bendLud{0.5}{\x+0.5}{\x+1.5}}
    \foreach \y in {2,4,...,10}{\prop{1}{\y}{colU}}
    \foreach \y in {3,5,...,9}{\prop{2}{\y}{colU}}
    \foreach \y in {4,6,8}{\prop{3}{\y}{colU}}
    \foreach \y in {5,7}{\prop{4}{\y}{colU}}
    \foreach \y in {6}{\prop{5}{\y}{colU}}
    \foreach \y in {0,...,11}{\obsZero{0.5}{\y+0.5}}
    \foreach \y in {1,...,10}{\obsZero{1.5}{\y+0.5}}
    \foreach \y in {2,...,9}{\obsZero{2.5}{\y+0.5}}
    \foreach \y in {3,...,8}{\obsZero{3.5}{\y+0.5}}
    \foreach \y in {4,...,7}{\obsZero{4.5}{\y+0.5}}
    \foreach \y in {5,...,6}{\obsZero{5.5}{\y+0.5}}
  \end{tikzpicture}=
  2^{-t}.
\end{equation}
This scales much more favourably compared to other shapes. For example,
the probability of an optimal rectangle with the same area as the triangle of height
$t$ scales as 
\begin{equation}
  P_{\square}(\frac{t}{2\sqrt{2}},\frac{t}{\sqrt{2}})
  \approx
  2^{-t\sqrt{2}}.
\end{equation}

In fact, the right-pointing triangle is not only the most favourable shape, but
it is the only shape that voids take: assuming a certain area of the space-time
lattice is empty, it follows that the smallest right-pointing triangle
enclosing that shape is also empty. To see this, it is enough to consider two
limiting cases, an empty horizontal and vertical line.

We start with the former: the horizontal empty line of length $l$ implies that
a triangle with horizontal length $l$ and vertical length $2l$ is also empty,
which follows from
\begin{equation}
\label{eq:emptytriangles}
  \begin{tikzpicture}[baseline={([yshift=-0.6ex]current bounding box.center)},scale=0.5]
    \foreach \x in {0,2,4}{
      \foreach \y in {0,2,4,6,8,10}
      {
        \prop{\x}{\y+1}{colU}
        \prop{\x+1}{\y}{colU}
      }
    }
    \foreach \x in {1,...,5}{\obsZero{\x-0.5}{5.5}}
    \draw[thick,gray,|<->|] (0.5,-0.875) --(4.5,-0.875) node [midway,below] {$l$};
  \end{tikzpicture}=
  \begin{tikzpicture}[baseline={([yshift=-0.6ex]current bounding box.center)},scale=0.5]
    \foreach \x in {0,2,4}{
      \foreach \y in {0,2,4,6,8,10}
      {
        \prop{\x}{\y+1}{colU}
        \prop{\x+1}{\y}{colU}
      }
    }
    \foreach \x in {1,...,1}{\obsZero{\x-0.5}{9.5}}
    \foreach \x in {1,...,2}{\obsZero{\x-0.5}{8.5}}
    \foreach \x in {1,...,3}{\obsZero{\x-0.5}{7.5}}
    \foreach \x in {1,...,4}{\obsZero{\x-0.5}{6.5}}
    \foreach \x in {1,...,5}{\obsZero{\x-0.5}{5.5}}
    \foreach \x in {1,...,5}{\obsZero{\x-0.5}{4.5}}
    \foreach \x in {1,...,4}{\obsZero{\x-0.5}{3.5}}
    \foreach \x in {1,...,3}{\obsZero{\x-0.5}{2.5}}
    \foreach \x in {1,...,2}{\obsZero{\x-0.5}{1.5}}
    \foreach \x in {1,...,1}{\obsZero{\x-0.5}{0.5}}
    \draw[thick,gray,|<->|] (0.5,-0.875) --(4.5,-0.875) node [midway,below] {$l$};
    \draw[thick,gray,|<->|] (-0.875,0.5) --(-0.875,9.5) node [midway,left] {$2l$};
  \end{tikzpicture}=
  \begin{tikzpicture}[baseline={([yshift=-0.6ex]current bounding box.center)},scale=0.5]
    \foreach \x in {0,2,4}{
      \foreach \y in {0,2,4,6,8,10}
      {
        \prop{\x}{\y+1}{colU}
        \prop{\x+1}{\y}{colU}
      }
    }
    \foreach \x in {1,...,5}{\obsZero{\x-0.5}{4.5}}
    \draw[thick,gray,|<->|] (0.5,-0.875) --(4.5,-0.875) node [midway,below] {$l$};
  \end{tikzpicture}.
\end{equation}
Here the middle diagram is obtained by repeated application
of~\eqref{eq:IdentitiesForTriangles} on both the right and left diagrams. Note that
depending on the parity of the right-most point, the precise vertical position
of the triangle shifts slightly. Indeed, both left and right diagrams result in the
same triangle (centred diagram), despite the starting void on the right being
vertically shifted with respect to the left one.

Similarly, starting from a vertical empty line implies an empty triangle with the line
being the base,
\begin{equation}
  \begin{tikzpicture}[baseline={([yshift=-0.6ex]current bounding box.center)},scale=0.5]
    \foreach \x in {0,2,4}{
      \foreach \y in {0,2,4,6,8,10}
      {
        \prop{\x}{\y+1}{colU}
        \prop{\x+1}{\y}{colU}
      }
    }
    \foreach \t in {0,...,9}{\obsZero{0.5}{\t+0.5}}
    \draw[red,thick] (0.5,9.5) circle (0.45);
    \draw[red,thick] (0.5,0.5) circle (0.45);
    \draw[thick,gray,|<->|] (-0.875,0.5) --(-0.875,9.5) node [midway,left] {$t$};
  \end{tikzpicture}=
  \begin{tikzpicture}[baseline={([yshift=-0.6ex]current bounding box.center)},scale=0.5]
    \foreach \x in {0,2,4}{
      \foreach \y in {0,2,4,6,8,10}
      {
        \prop{\x}{\y+1}{colU}
        \prop{\x+1}{\y}{colU}
      }
    }
    \foreach \t in {0,...,9}{\obsZero{0.5}{\t+0.5}}
    \foreach \t in {1,...,8}{\obsZero{1.5}{\t+0.5}}
    \foreach \t in {2,...,7}{\obsZero{2.5}{\t+0.5}}
    \foreach \t in {3,...,6}{\obsZero{3.5}{\t+0.5}}
    \foreach \t in {4,...,5}{\obsZero{4.5}{\t+0.5}}
    \draw[thick,gray,|<->|] (-0.875,0.5) --(-0.875,9.5) node [midway,left] {$t$};
    \draw[thick,gray,|<->|] (0.5,-0.875) --(4.5,-0.875) node [midway,below] {$\frac{t}{2}$};
    \draw[thick,white,|<->|] (0.5,11.875) --(4.5,11.875) node [midway,above] {$\frac{t}{2}$};
  \end{tikzpicture},
\end{equation}
where the red circles denote the boundary sites that can be removed from the left-hand side
and produce the same result on the right-hand side.

This reasoning implies more generally: any shape of the void necessarily
implies a larger triangular void around it, as illustrated in Fig.~\ref{fig4}.
This is also consistent with the scaling of the probabilities of rectangular voids,
which (up to $\mathcal{O}(1)$ corrections in the exponent) obeys
\begin{equation}
  P_{\square}(l,t)\sim P_{\rhd}(2l+t).
\end{equation}

\section{Bipartitioning protocol}\label{sec:bipartite}
As the last example, we wish to study the properties of the interface between
an inactive and active phase. We probe this by considering an initial value
problem, where we assume that at time $t=0$ we have a very large system
prepared in an inhomogeneous state $\ket{P_{\rm inh}}$ so that the left half is
empty, one site to the right of it is full, and the rest is in a maximally
mixed state,
\begin{equation}
  \ket{P_{\rm inh}}=
  \frac{1}{2^L}
  \begin{bmatrix} 1 \\ 0 \end{bmatrix}^{\otimes L-1} 
  \otimes
  \begin{bmatrix} 0 \\ 1 \end{bmatrix} \otimes
  \begin{bmatrix} 1 \\ 1 \end{bmatrix}^{\otimes L}.
\end{equation}
We are interested in the density of particles $\rho_{x,t}$ at time $t$ and
position $x$ relative to the interface, which is graphically expressed as
\begin{equation}
  \rho_{x,t}=
  \frac{1}{2^{L}}
  \begin{tikzpicture}[baseline={([yshift=-0.6ex]current bounding box.center)},scale=0.5]
    \def\X{16}
    \def\Y{8}
    \foreach \x in {1,3,...,\X}{
      \nctgridLine{\x+0.25}{\Y+0.75}{\x-0.5}{\Y}
      \nctgridLine{\x+0.75}{\Y+0.75}{\x+1.5}{\Y}
      \nctgridLine{\x+1.75}{0.25}{\x+2.5}{1}
      \nctgridLine{\x+1.25}{0.25}{\x+0.5}{1}
    }
    \draw[ultra thick,white] (\X+1,0.5) --(\X+1.5,1);
    \draw[ultra thick,white] (1-0.5*0.125,\Y+0.5-0.5*.125) --(0.5,\Y);
    \foreach \y in {0,2,...,\Y}{\nctgridLine{\X+1+0.3}{0.5+\y+0.3}{\X+1}{0.5+\y}}
    \foreach \y in {1,3,...,\Y}{\nctgridLine{\X+1+0.3}{0.5+\y-0.3}{\X+1}{0.5+\y}}
    \foreach \x in {1,3,...,\X}{
      \foreach \y in {1,3,...,\Y}{
        \prop{\x+0.5}{\y}{colU}
        \prop{\x+1.5}{\y+1}{colU}
      }
      \MEld{\x+1.25}{0.25}
      \MErd{\x+1.75}{0.25}

      \MEld{\x+0.75}{\Y+0.75}
      \MErd{\x+0.25}{\Y+0.75}
    }
    \foreach \x in {2,...,13}{\obsZero{\x}{0.5}}
    \obsOne{14}{0.5}
    \obsOne{13}{8.5}
    \draw[thick,gray,|->|] (14,9.25) -- (13,9.25) node[midway,yshift=5pt] {$x$};
    \draw[thick,gray,|->|] (17.5,0.5) -- (17.5,8.5) node[midway,xshift=3pt] {$t$};
  \end{tikzpicture}.
\end{equation}
At $t=0$ the expectation value can be immediately evaluated to be equal to
\begin{equation}
  \rho_{x,0}=\begin{cases}
    \frac{1}{2},& x\ge\frac{1}{2},\\
    1,& x=0,\\
    0,& x\le-\frac{1}{2}.
  \end{cases}
\end{equation}
For $t>0$, we first note that relations~\eqref{eq:IdentitiesForTriangles}
allow us to rewrite the time-evolved state $\ket{P_{{\rm inh},t}}$ as 
\begin{equation}
  \ket{P_{{\rm inh},t}}=
  \frac{1}{2^{L}}
  \begin{tikzpicture}[baseline={([yshift=-0.6ex]current bounding box.center)},scale=0.5]
    \def\X{16}
    \def\Y{8}
    \foreach \x in {1,3,...,\X}{
      \nctgridLine{\x+0.25}{\Y+0.75}{\x-0.5}{\Y}
      \nctgridLine{\x+0.75}{\Y+0.75}{\x+1.5}{\Y}
      \nctgridLine{\x+1.75}{0.25}{\x+2.5}{1}
      \nctgridLine{\x+1.25}{0.25}{\x+0.5}{1}
    }
    \draw[ultra thick,white] (\X+1,0.5) --(\X+1.5,1);
    \draw[ultra thick,white] (1-0.5*0.125,\Y+0.5-0.5*.125) --(0.5,\Y);
    \foreach \y in {0,2,...,\Y}{\nctgridLine{\X+1+0.3}{0.5+\y+0.3}{\X+1}{0.5+\y}}
    \foreach \y in {1,3,...,\Y}{\nctgridLine{\X+1+0.3}{0.5+\y-0.3}{\X+1}{0.5+\y}}
    \foreach \x in {1,3,...,\X}{
      \foreach \y in {1,3,...,\Y}{
        \prop{\x+0.5}{\y}{colU}
        \prop{\x+1.5}{\y+1}{colU}
      }
      \MEld{\x+1.25}{0.25}
      \MErd{\x+1.75}{0.25}
    }
    \foreach \x in {2,...,13}{\obsZero{\x}{0.5}}
    \foreach \x in {2,...,12}{\obsZero{\x}{1.5}}
    \foreach \x in {2,...,11}{\obsZero{\x}{2.5}}
    \foreach \x in {2,...,10}{\obsZero{\x}{3.5}}
    \foreach \x in {2,...,9}{\obsZero{\x}{4.5}}
    \foreach \x in {2,...,8}{\obsZero{\x}{5.5}}
    \foreach \x in {2,...,7}{\obsZero{\x}{6.5}}
    \foreach \x in {2,...,6}{\obsZero{\x}{7.5}}
    \foreach \x in {2,...,5}{\obsZero{\x}{8.5}}
    \foreach \y in {0,...,\Y}{\obsOne{14-\y}{0.5+\y}}
    \foreach \y in {1,...,\Y}{\obsOne{15-\y}{0.5+\y}}
    \draw[ultra thick,white] (0.75,\Y+0.25) -- (1.25,\Y+0.75);
  \end{tikzpicture}.
\end{equation}
From here it immediately follows that for $x\le-t+\frac{1}{2}$ the expectation
value is given as
\begin{equation}
  \rho_{-t+\frac{1}{2},t}=\rho_{-t,t}=1,\qquad
  \left.\rho_{x,t}\right|_{x\le-t-\frac{1}{2}}=0.
\end{equation}
Let us now assume that $x\ge -t+1$. The cases $x+t\in\mathbb{Z}$ and 
$x+t\in\mathbb{Z}+\frac{1}{2}$ need to be considered separately, so we
start with the former. Using the above expression for $\ket{P_{{\rm inh},t}}$,
and using the first two of~\eqref{eq:SMRelU}, the expectation value of the density
can be expressed as the following tilted rectangle,
\begin{equation}\label{eq:bipartition2}
  \left.\rho_{x,t}\right|_{x+t\in\mathbb{Z}+\frac{1}{2}}=
    2^{-L}
    \mkern-18mu
    \begin{tikzpicture}[baseline={([yshift=-0.6ex]current bounding box.center)},scale=0.5]
      \def\X{16}
      \def\Y{8}
      \foreach \y in {1,...,7}{
        \nctgridLine{15-\y-0.25}{0.25+\y}{15-\y+0.5}{1+\y}
      }
      \foreach{\x} in {9,11,15,17,19}{\MErd{\x}{8.5}}
      \foreach{\x} in {8,10,12,14,16,18}{\MEld{\x}{8.5}}
      \nctgridLine{12.5}{8}{13.25}{8.75}
      \nctgridLine{19}{0.5}{19.255}{0.75}
      \MErd{13.25}{8.75}
      \MErd{15}{0.5}
      \MEld{16}{0.5}
      \MErd{17}{0.5}
      \MEld{18}{0.5}
      \MErd{19}{0.5}
      \prop{15.5}{1}{colU}
      \prop{17.5}{1}{colU}
      \prop{14.5}{2}{colU}
      \prop{16.5}{2}{colU}
      \prop{18.5}{2}{colU}
      \prop{13.5}{3}{colU}
      \prop{15.5}{3}{colU}
      \prop{17.5}{3}{colU}
      \prop{12.5}{4}{colU}
      \prop{14.5}{4}{colU}
      \prop{16.5}{4}{colU}
      \prop{18.5}{4}{colU}
      \prop{11.5}{5}{colU}
      \prop{13.5}{5}{colU}
      \prop{15.5}{5}{colU}
      \prop{17.5}{5}{colU}
      \prop{10.5}{6}{colU}
      \prop{12.5}{6}{colU}
      \prop{14.5}{6}{colU}
      \prop{16.5}{6}{colU}
      \prop{18.5}{6}{colU}
      \prop{9.5}{7}{colU}
      \prop{11.5}{7}{colU}
      \prop{13.5}{7}{colU}
      \prop{15.5}{7}{colU}
      \prop{17.5}{7}{colU}
      \prop{8.5}{8}{colU}
      \prop{10.5}{8}{colU}
      \prop{12.5}{8}{colU}
      \prop{14.5}{8}{colU}
      \prop{16.5}{8}{colU}
      \prop{18.5}{8}{colU}
      \foreach \y in {1,...,7}{
        \MErd{15-\y-0.25}{0.5+\y-0.25}
        \obsOne{15-\y}{0.5+\y}
      }
      \obsOne{13}{8.5}
      \draw[thick,gray,|->|] (14,9.25) -- (13,9.25) node[midway,above] {$x$};
      \draw[thick,gray,|->|] (19.5,0.5) -- (19.5,8.5) node[midway,right] {$t$};
  \end{tikzpicture}=
  2^{-(x+t-\frac{1}{2})}
  \begin{tikzpicture}[baseline={([yshift=-0.6ex]current bounding box.center)},scale=0.5]
    \def\X{16}
    \def\Y{8}
    \foreach \y in {1,2,3,4,5}{
      \nctgridLine{15-\y-0.25}{0.25+\y}{15-\y+0.5}{1+\y}
    }
    \foreach{\x} in {0,1,2}{\MEld{12-\x}{8.5-\x}}
    \foreach{\x} in {1,2,3,4}{\MErd{13+\x}{8.5-\x}}
    \foreach{\x} in {0,1,2}{\MEld{15+\x}{1.5+\x}}
    \nctgridLine{12.5}{8}{13.25}{8.75}
    \MErd{13.25}{8.75}
    \prop{14.5}{2}{colU}
    \prop{13.5}{3}{colU}
    \prop{15.5}{3}{colU}
    \prop{12.5}{4}{colU}
    \prop{14.5}{4}{colU}
    \prop{16.5}{4}{colU}
    \prop{11.5}{5}{colU}
    \prop{13.5}{5}{colU}
    \prop{15.5}{5}{colU}
    \prop{10.5}{6}{colU}
    \prop{12.5}{6}{colU}
    \prop{14.5}{6}{colU}
    \prop{11.5}{7}{colU}
    \prop{13.5}{7}{colU}
    \prop{12.5}{8}{colU}
    \foreach \y in {1,2,3,4,5}{
      \MErd{15-\y-0.25}{0.5+\y-0.25}
      \obsOne{15-\y}{0.5+\y}
    }
    \obsOne{13}{8.5}
    \draw[thick,gray,|->|] (14,9.25) -- (13,9.25) node[midway,above] {$x$};
    \draw[thick,gray,|->|] (17.5,1.5) -- (17.5,8.5) node[midway,rotate=90,below] {$t-\frac{1}{2}$};
    \draw[thick,gray,|<->|] (9.25,6.25) -- (12.25,9.25) node[midway,rotate=45,above] {$x+t-\frac{1}{2}$};
  \end{tikzpicture}.
\end{equation}
Using now the third relation of~\eqref{eq:SMRelU} the rectangle is transformed into
a line, which can be fully contracted using the last of~\eqref{eq:SMRelU},
\begin{equation}\label{eq:bipartition3}
  \left.\rho_{x,t}\right|_{x+t\in\mathbb{Z}+\frac{1}{2}}=
    2^{-(x+t-\frac{1}{2})}
  \begin{tikzpicture}[baseline={([yshift=-0.6ex]current bounding box.center)},scale=0.5]
    \def\X{16}
    \def\Y{8}
    \foreach{\x} in {0,1,2}{\MEld{12-\x}{8.5-\x}}
    \foreach{\x} in {0,1,2}{\MEld{13-\x}{7.5-\x}}
    \nctgridLine{9.75}{5.25}{13.25}{8.75}
    \MErd{13.25}{8.75}
    \MErd{9.75}{5.25}
    \prop{10.5}{6}{colU}
    \prop{11.5}{7}{colU}
    \prop{12.5}{8}{colU}
    \obsOne{10}{5.5}
    \obsOne{13}{8.5}
    \draw[thick,gray,|<->|] (9.25,6.25) -- (12.25,9.25) node[midway,rotate=45,above] {$x+t-\frac{1}{2}$};
  \end{tikzpicture}=\frac{1}{2}.
\end{equation}
Similarly, for $x+t\in\mathbb{Z}$,
\begin{equation}\label{eq:bipartition4}
  \left.\rho_{x,t}\right|_{x+t\in\mathbb{Z}}=
    2^{-(x+t)}
  \begin{tikzpicture}[baseline={([yshift=-0.6ex]current bounding box.center)},scale=0.5]
    \def\X{16}
    \def\Y{8}
    \foreach \y in {1,...,4}{
      \nctgridLine{15-\y-0.25}{0.25+\y}{15-\y+0.5}{1+\y}
    }
    \nctgridLine{13}{7.5}{12.75}{7.75}
    \prop{14.5}{2}{colU}
    \prop{13.5}{3}{colU}
    \prop{15.5}{3}{colU}
    \prop{12.5}{4}{colU}
    \prop{14.5}{4}{colU}
    \prop{16.5}{4}{colU}
    \prop{11.5}{5}{colU}
    \prop{13.5}{5}{colU}
    \prop{15.5}{5}{colU}
    \prop{12.5}{6}{colU}
    \prop{14.5}{6}{colU}
    \prop{13.5}{7}{colU}
    \foreach \y in {1,...,4}{
      \MErd{15-\y-0.25}{0.5+\y-0.25}
      \obsOne{15-\y}{0.5+\y}
      \MErd{13+\y}{8.5-\y}
    }
    \foreach \y in {0,...,2}{
      \MEld{15+\y}{1.5+\y}
    }
    \MEld{12.75}{7.75}
    \MEld{12}{6.5}
    \MEld{11}{5.5}
    \obsOne{13}{7.5}
    \draw[thick,gray,|->|] (14,8.25) -- (13,8.25) node[midway,above] {$x$};
    \draw[thick,gray,|->|] (17.5,1.5) -- (17.5,7.5) node[midway,right] {$t-\frac{1}{2}$};
    \draw[thick,gray,|<->|] (9.75,5.75) -- (12.75,8.75) node[midway,rotate=45,above] {$x+t$};
  \end{tikzpicture}=
  2^{-(x+t)}
  \begin{tikzpicture}[baseline={([yshift=-0.6ex]current bounding box.center)},scale=0.5]
    \def\X{16}
    \def\Y{8}
    \foreach{\x} in {1,2}{\MEld{12-\x}{8.5-\x}}
    \foreach{\x} in {0,1,2}{\MEld{13-\x}{7.5-\x}}
    \nctgridLine{9.75}{5.25}{10.25}{5.75}
    \nctgridLine{11.75}{8.75}{12.5}{8}
    \MEld{11.75}{8.75}
    \MErd{9.75}{5.25}
    \MErd{13}{8.5}
    \prop{10.5}{6}{colU}
    \prop{11.5}{7}{colU}
    \prop{12.5}{8}{colU}
    \obsOne{10}{5.5}
    \obsOne{12}{8.5}
    \draw[thick,gray,|<->|] (9,6.5) -- (12,9.5) node[midway,rotate=45,above] {$x+t$};
  \end{tikzpicture}=\frac{1}{2}.
\end{equation}
We note that to obtain the tilted rectangles in~\eqref{eq:bipartition2},
and~\eqref{eq:bipartition4}, we implicitly assumed $x\le t-\frac{1}{2}$. If
this is not the case, the string of projectors on the left and the observable
on the right immediately factorize and instead of the rectangle we immediately
obtain (i.e., without the step in~\eqref{eq:bipartition3}),
\begin{equation}
  \left.\rho_{x,t}\right|_{x\ge t}= \frac{1}{2}.
\end{equation}
Finally, combining these together gives us
\begin{equation}
  \rho_{x,t} =
  \begin{cases}
    0,& x\le -t -\frac{1}{2},\\
    1,& -t \le x \le -t+\frac{1}{2},\\
    \frac{1}{2},& x \ge -t+1.
  \end{cases}
\end{equation}

\end{document}